\documentclass[10pt]{llncs}

\usepackage{amsmath}
\usepackage{latexsym}
\usepackage{proof}
\usepackage{amsxtra} 
\usepackage{amssymb}
\usepackage{paralist}
\usepackage{tikz}
\usetikzlibrary{automata,positioning, calc}
\usepackage[inline,shortlabels]{enumitem}

\usepackage{algorithm}
\usepackage{algorithmicx}
\usepackage[noend]{algpseudocode}

\usepackage{wrapfig}
\usepackage{adjustbox}

\newcommand{\np}{$\mathsf{NP}$}

\newcommand{\nexptime}{$\mathsf{NEXPTIME}$}
\newcommand{\pspace}{$\mathsf{PSPACE}$~}
\newcommand{\sigmatwop}{$\mathsf{\Sigma}_2^{\scriptscriptstyle{\mathrm{P}}}$}
\newcommand{\pitwop}{$\mathsf{\Pi}_2^{\scriptscriptstyle{\mathrm{P}}}$}
\newcommand{\expspace}{$\mathsf{EXPSPACE}$}

\newcommand{\rbr}{{\bf ]\!]}}
\newcommand{\lbr}{{\bf [\![}}
\newcommand{\sem}[1]{\lbr #1 \rbr}

\newcommand{\set}[1]{\left\{ #1 \right\}}
\newcommand{\tuple}[1]{\left\langle #1 \right\rangle}
\renewcommand{\vec}[1]{\mathbf #1}

\newcommand{\len}[1]{{|{#1}|}}
\newcommand{\card}[1]{{|\!|{#1}|\!|}}

\newcommand{\arrow}[2]{\xrightarrow{{\scriptscriptstyle #1}}_{{\scriptscriptstyle #2}}}

\newcommand{\nat}{{\bf \mathbb{N}}}

\renewcommand{\paragraph}[1]{\noindent{\bf #1}}

\let\Asterisk\undefined
\newcommand{\Asterisk}{\mathop{\scalebox{1.7}{\raisebox{-0.2ex}{$\ast$}}}}%

\newcommand{\Coloneqq}{::=}

\newif\ifLongVersion\LongVersiontrue

\newcommand{\ssorts}[1]{#1^\mathrm{s}}
\newcommand{\sfuns}[1]{#1^\mathrm{f}}

\newcommand{\pow}[1]{\mathcal{P}({#1})}
\newcommand{\finpow}[1]{\mathcal{P}_{\!\!\!\scriptscriptstyle{\mathit{fin}}}({#1})}

\newcommand{\eqset}{\mathsf{eq}}
\newcommand{\mayalloc}{\mathsf{alloc}^{+}}
\newcommand{\mustalloc}{\mathsf{alloc}^{-}}
\newcommand{\symhp}{\Pi}
\newcommand{\symhs}{\Theta}
\newcommand{\locsi}{\mathsf{L}}
\newcommand{\heaps}{\mathsf{Heaps}}
\newcommand{\trees}{\mathsf{Cover}}
\newcommand{\teq}{\approx}

\newcommand{\I}{\mathcal{I}}

\newcommand{\herb}{\mathcal{H}}
\newcommand{\tinyherb}{{\scriptscriptstyle\mathcal{H}}}
\newcommand{\locs}{\mathsf{Loc}}

\newcommand{\emp}{\mathsf{emp}}
\newcommand{\wand}{
 \mathrel{\mbox{$\hspace*{-0.03em}\mathord{-}\hspace*{-0.66em}
 \mathord{-}\hspace*{-0.36em}\mathord{*}$\hspace*{-0.005em}}}} 
\newcommand{\seplog}{\mathsf{SL}}
\newcommand{\tinyseplog}{\mathsf{\scriptscriptstyle{sl}}}

\newcommand{\fv}[1]{\mathrm{FV}(#1)}

\newcommand{\dom}{\mathrm{dom}}
\newcommand{\img}{\mathrm{img}}
\newcommand{\fr}{\mathrm{fr}}

\newcommand{\vars}{\mathsf{Var}}
\newcommand{\preds}{\mathsf{Pred}}
\newcommand{\T}{\mathcal{T}}
\newcommand{\X}{\mathcal{X}}
\newcommand{\utree}{\mathsf{\scriptscriptstyle{u}}}

\newcommand{\Y}{\mathcal{Y}}
\newcommand{\sys}{\mathcal{S}}
\newcommand{\lseq}{\Gamma}
\newcommand{\rseq}{\Delta}
\newcommand{\typelabel}{\Lambda}

\renewcommand{\int}{\mathsf{\scriptscriptstyle{i}}}

\newcommand{\cf}{\mathcal{F}}
\newcommand{\sw}{\leq_{\mathrm{sw}}}
\newcommand{\subst}[1]{\mathrm{Sk}(#1)}
\newcommand{\DS}{\mathfrak{D}}
\newcommand{\DSSL}{\mathfrak{D}^\tinyseplog}

\newcommand{\infname}[1]{\begin{array}{c}\text{ ($#1$) }\\\\\end{array}}

\newcommand{\sub}[2]{{#1}_{|_{#2}}}
\newcommand{\context}[2]{{#1}_{[#2]}}

\newcommand{\subtree}{\sqsubset}
\newcommand{\subtreeq}{\sqsubseteq}

\newcommand{\LU}{\mathrm{LU}}
\newcommand{\RU}{\mathrm{RU}}
\newcommand{\RD}{\mathrm{RD}}
\newcommand{\RI}{\wedge\mathrm{R}}
\newcommand{\SP}{\mathrm{SP}}
\newcommand{\ID}{\mathrm{ID}}
\newcommand{\AX}{\mathrm{AX}}

\newcommand{\Bool}{\mathsf{Bool}}

\newcommand{\predno}{\mathsf{p}^{\scriptscriptstyle \#}}
\newcommand{\ruleno}{\mathsf{r}^{\scriptscriptstyle \#}}
\newcommand{\subgno}{\mathsf{s}^{\scriptscriptstyle \#}}
\newcommand{\baseno}{\mathsf{b}^{\scriptscriptstyle \#}}

\newcommand{\proofrule}[4]{\begin{array}[t]{@{}l@{\;\,}l@{}} #1 & \vcenter{\infer[\text{$#4$}]{#2}{#3}}\end{array}}

\newcommand{\proofrulepivot}[6]{\begin{array}[t]{@{}l@{\;\,}l@{}} \begin{array}{@{}l@{}}#1\\\\\\\end{array} & \vcenter{\infer*[#6]{#2}{\infer[\text{$#5$}]{#3}{#4}}}\end{array}}

\newcommand{\RInd}{\mathcal{R}_\mathsf{Id}}

\newcommand{\IR}{\mathsf{R}}
\newcommand{\ir}{\mathsf{r}}
\newcommand{\nextval}{\gtrdot}
\newcommand{\p}{$\mathsf{P}$}
\newcommand{\RIndsl}{\mathcal{R}_\mathsf{Id}^\tinyseplog}
\newcommand{\RUsl}{\mathrm{RU}_\tinyseplog}
\newcommand{\RDsl}{\mathrm{RD}_\tinyseplog}
\newcommand{\SPsl}{\mathrm{SP}_\tinyseplog}
\newcommand{\AXsl}{\mathrm{AX}_\tinyseplog}

\begin{document}

\title{Complete Cyclic Proof Systems for Inductive Entailments}         



\author{Radu Iosif \and Cristina Serban} 
\institute{CNRS/VERIMAG/Universit\'e Grenoble Alpes \\ 
\email{\{Radu.Iosif,Cristina.Serban\}@univ-grenoble-alpes.fr}
}

\maketitle
\begin{abstract}
In this paper we develop cyclic proof systems for the problem of
inclusion between the least sets of models of mutually recursive
predicates, when the ground constraints in the inductive definitions
belong to the quantifier-free fragments of~\begin{inparaenum}[(i)]
\item First Order Logic with the canonical Herbrand interpretation and
\item Separation Logic, respectively.
\end{inparaenum}
Inspired by classical automata-theoretic techniques of proving
language inclusion between tree automata, we give a small set of
inference rules, that are proved to be sound and complete, under
certain semantic restrictions, involving the set of constraints in the
inductive system. Moreover, we investigate the decidability and
computational complexity of these restrictions for all the logical
fragments considered and provide a proof search semi-algorithm that
becomes a decision procedure for the entailment problem, for those
systems that fulfill the restrictions. 
\end{abstract}





\section{Introduction}

Inductive definitions play an important role in computing, being an
essential component of the syntax and semantics of programming
languages, databases, automated reasoning and program verification
systems. The main advantage of using inductive definitions is the
ability of reasoning about sets of logical objects, by means of
recursion. The semantics of these definitions is defined in terms of
least fixed points of higher-order functions on assignments of
predicates to sets of models. A natural problem is the
\emph{entailment}, that asks whether the least solution of one
predicate is included in the least solution of another. Examples of
entailments are language inclusion between finite-state (tree)
automata or context-free grammars, or verification conditions
generated by shape analysis tools using specifications of recursive
data structures as contracts of program correctness.

The principle of \emph{Infinite Descent} \cite{Bussey18}, initially
formalized by Fermat, has become an important tool for reasoning about
entailments between inductively defined predicates. In a nutshell, a
proof by infinite descent is a particular proof by contradiction, in
which we assume the existence of a counterexample from a well-founded
domain and show that this leads to the existence of a strictly smaller
counterexample for the same entailment problem.  By repeating this
step, we obtain an infinite descending chain of counterexamples,
which is impossible because we assumed the domain of interpretation to
be well-founded. Hence, there is no counterexample to start with and
the entailment holds. This principle is found in all \emph{cyclic
proof systems}, where branches of a sequent calculus proof are
closed by specific infinite descent rules \cite{BrotherstonS11}.

The interest for automatic proof generation is two-fold. On the one
hand, machine-checkable proofs are certificates for the correctness
of the answer given by an automated checker, that increase our trust
in the reliability of a particular implementation. For 
instance, the need for proof generation as means of certifying SMT
solvers has been widely recognized, see e.g.\ \cite{Stump13}. In this
paper we provide means to certify implementations of automata-theoretic
algorithms, such as language inclusion and emptiness for (tree) automata
\cite{deWulfDoyenHenzingerRaskin06,HolikLSV11}.


On the other hand, the existence of a sound and complete proof system
provides a (theoretical) decision procedure for the entailment
problem, based on the following argument. Assuming that the sets of
models and derivations are both recursively enumerable, one can
interleave the enumeration of counter-models with the enumeration of
derivations; if the entailment holds one finds a finite proof (provided
that the proof system is complete), or a  finite counterexample,
otherwise. Moreover, proof generation can be made effective by
providing suitable strategies that limit the possibilities of applying
the inference rules and guide the search towards finding a proof or a
counterexample.

In this paper we give a set of inference rules that are shown to be
sound and complete for entailments in an inductive system $\sys$,
provided that the set of constraints of $\sys$ meets several
restrictions.  In general, most authors define these restrictions by
the syntax of the logical fragment in which the constraints of the
system are written. In contrast, here we consider a rather general 
logic, namely First Order Logic with the canonical Herbrand 
interpretation, and define the restrictions necessary for soundness 
and completeness by a number of decidable semantic conditions, which 
can be checked by existing decision procedures. We then adapt these
restrictions to Separation Logic. Additionally, we investigate the
computational complexity of checking whether a given system complies
with these restrictions. 

The set of inference rules given in this paper is used by a proof
generation semi-algorithm. Our goal is to describe inference rules and
a proof search method that are general enough so they can be adapted
to other types of logic. We provide a prototype implementation of this
semi-algorithm \cite{Inductor}, that uses a variant of the inference
rules adjusted for proving entailments between inductive predicates
written in Separation Logic \cite{Reynolds02}.

\subsubsection*{\bf Related Work}

The problem of generating cyclic proofs for entailments between
inductively defined predicates in First Order Logic has been
considered by Brotherston, who gave a first sound proof system
\cite{Brotherston05}, based on Gentzen-style natural deduction. There
soundness requires that each infinite trace in the proof goes through
infinitely many progress points (left unfoldings), which can be
reduced to an inclusion between B\"uchi $\omega$-automata. In our
case, the requirement that the pivot (companion) of an infinite
descent rule is placed on the path from the root to the consequent
(bud) is instrumental in giving a sufficient local progress condition
that can be checked easily. Completeness in \cite{Brotherston05} is
relative to proofs using an induction rule and uses a cut rule for the
introduction of inductive invariants. In our case, completeness is
subject to a number of (decidable) semantic requirements on the set of
constraints of the inductive system. 

Brotherston and Simpson\cite{BrotherstonS11} describe the proof system
$\mathsf{CLKID}^\omega$, in which the inductive arguments are
discovered during proof construction. The $\mathsf{CLKID}^\omega$
system is shown to be sound and complete w.r.t. Henkin
models. Soundness is guaranteed by a global trace condition ensuring
that each infinite trace visits infinitely many progress points, while
relative completeness is established by translation of proofs written
using Martin-L\"{o}f's complete induction system, called
$\mathsf{LKID}$, into $\mathsf{CLKID}^\omega$ proofs. Furthermore, the
equivalence of $\mathsf{LKID}$ and $\mathsf{CLKID}^\omega$ is
conjectured, but left as an open question. More recently, Berardi and
Tatsuta \cite{Berardi2017CSML} have disproved the equivalence by
providing a statement, called $2$-Hydra, expressed by means of
first-order logic with the constant $0$, the successor function, the
natural number predicate and an additional binary predicate, and
showing how it is provable in $\mathsf{CLKID}^\omega$, but not in
$\mathsf{LKID}$.

The relation between cyclic proof systems and automata is
long-standing, one of the main results being a complete proof system
for the modal $\mu$-calculus proposed by Walukiewicz
\cite{Walukiewicz00}. Rosu et al. \cite{Rosu13} also describe
reachability logic, a framework tailored towards reasoning about
reachability properties of systems. A proof system containing a
circularity rule is presented.  Similarly to proofs using infinite
descent, a claim holds if it can be proved using itself as a
circularity, and requiring progress before circularities are used
ensures soundness.

\cite{GalmicheMery17} present a labeled sequent calculus that supports
arbitrary inductive predicates, with Separation Logic constraints
using the separating implication connective \cite{Reynolds02}. To our
knowledge this is the first proof system that supports all connectives
of Separation Logic, but unfortunately, no proof of completeness is
provided. We chose not to include the separating implication (magic
wand) in our systems because the Bernays-Sch{\"{o}}nfinkel-Ramsey
fragment of Separation Logic with this connective appears to be
undecidable \cite{Vmcai17}, which prevents us from effectively
checking our sufficient conditions for completeness.

\cite{ChuJT14} propose a proof system for Separation Logic that
extends the basic cyclic proof method with a cut rule type that uses
previously encountered sequents as inductive hypothesis and applies
them by matching and replacing the left with the right-hand side of
such a hypothesis. This method can prove entailments between
predicates whose coverage trees differ, but again, only soundness is
guaranteed. It remains an open question for which class of entailment
problems this type of cut rules yields a complete proof system. An
automata-based decision procedure that tackles such entailments is
given in \cite{Atva14}. This method translates the entailment problem
to a language inclusion between tree automata and uses a closure
operation on automata to match divergent predicates. Unlike proof
search, this method uses existing tree automata inclusion algorithms,
which do not produce proof witnesses. 

In a different vein, \cite{EneaSW15} give a proof system that uses
automatic generation of concatenation lemmas for inductively defined
Separation Logic predicates. Their system is sound and moreover, most
concatenation lemmas can be shown to have a cyclic proof in our
system, whereas completeness remains still an open question. Further,
\cite{TaLKC16} describes {\it mutual explicit induction proofs}, an
induction method based on a well-founded order on Separation Logic
models. Akin to \cite{EneaSW15}, this method considers symbolic heaps
extended with constraints on the data values stored within the heap
structures. In a similar fashion with \cite{ChuJT14}, they keep a
vault with hypotheses, which are marked as {\it valid} or {\it
unknown}. Valid hypotheses can be freely applied, but unknown
hypotheses are only applied if certain side conditions, which ensure a
decrease in the size of the heap model, are satisfied. This approach
allows the hypotheses to be used anywhere in the proof tree, and are
not restricted to the branch from which they originated. This method
is sound but no completeness arguments are given.

\section{Preliminaries}

For two integers $0 \leq i \leq j$, we denote by $[i,j]$ the set
$\set{i,i+1,\ldots,j}$ and by $[i]$ the set $[1,i]$, where $[0]$ is
the empty set. Given a finite set $S$, $\card{S}$ denotes its
cardinality, $\pow{S}$ the powerset and $\finpow{S}$ the set of finite
subsets of $S$.

A \emph{signature} $\Sigma = (\ssorts{\Sigma},\sfuns{\Sigma})$
consists of a set $\ssorts{\Sigma}$ of \emph{sort symbols} and a set
$\sfuns{\Sigma}$ of \emph{function symbols} $f^{\sigma_1 \cdots
  \sigma_n \sigma}$, where $n \geq 0$ is its arity, $\sigma_1, \ldots,
\sigma_n \in \ssorts{\Sigma}$ are the sorts of its arguments and
$\sigma \in \ssorts{\Sigma}$ is the sort of its result. If $n=0$, we
call $f^\sigma$ a \emph{constant symbol}. We assume that every
signature contains the boolean sort, and write $\top$ and $\bot$ for
the boolean constants \emph{true} and \emph{false}. Let $\vars$ be a
countable set of \emph{first-order variables}, each variable $x^\sigma
\in \vars$ having an associated sort $\sigma$. We omit specifying the
sorts of the function symbols and variables, whenever they are not
important. We write $\vec{x}, \vec{y}, \ldots$ for both sets and
ordered tuples of variables, when no confusion arises. 

Terms are defined recursively: any constant symbol or variable is a
term, and if $t_1,\ldots,t_n$ are terms of sorts
$\sigma_1,\ldots,\sigma_n$, respectively, and $f^{\sigma_1 \cdots
  \sigma_n \sigma} \in \sfuns{\Sigma}$, then $t=f(t_1,\ldots,t_n)$ is
a term of sort $\sigma$, denoted $t^\sigma$. We denote by
$\T_\Sigma(\vec{x})$ the set of terms with function symbols in
$\sfuns{\Sigma}$ and variables in $\vec{x}$, and we write $\T_\Sigma$
for the set $\T_\Sigma(\emptyset)$ of \emph{ground} terms, in which no
variable occurs. Given ground terms $u,t \in \T_\Sigma$, we denote by
$u \subtreeq t$ the fact that $u$ is a subterm of $t$. Observe that
$(\T_\Sigma,\subtreeq)$ is a wfqo, because $\T_\Sigma$ consists of
finite trees only.

Formulae are also defined recursively: a term of boolean sort, an
equality $t \teq u$ (where $t$ and $u$ are terms of the same sort) are
formulae, and a quantified boolean combination of formulae is a
formula. For a formula $\phi$ (set of formulae $F$), we
denote by $\fv{\phi}$ ($\bigcup_{\phi \in F}\fv{\phi}$) the
set of variables not occurring under a quantifier scope, and
$\phi(\vec{x})$ ($F(\vec{x})$) means that $\vec{x}
\subseteq \fv{\phi}$ ($\vec{x} \subseteq \bigcup_{\phi \in
F}\fv{\phi}$). The \emph{size} of a formula is the total
number of variables, function symbols and logical connectives occurring
in it.

Given sets of variables $\vec{x}$ and $\vec{y}$, a \emph{substitution}
$\theta : \vec{x} \rightarrow \T_\Sigma(\vec{y})$ is a mapping of the
variables in $\vec{x}$ to terms in $\T_\Sigma(\vec{y})$. For a set of
variables $\vec{x}$ we denote by $\vec{x}\theta = \set{\theta(x) \mid
x \in \vec{x}}$ its image under the substitution $\theta$. A
substitution $\theta$ is \emph{flat} if $\vars\theta \subseteq \vars$,
i.e.\ each variable is mapped to a variable. For a formula
$\phi(\vec{x})$, we denote by $\phi\theta$ the formula obtained by
replacing each occurrence of $x\in\vec{x}$ with the term $\theta(x)$,
and lift this notation to sets as $F\theta = \set{\phi\theta
\mid \phi \in F}$. Observe that $\theta$ is always a
surjective mapping between $\fv{\phi}$ and $\fv{\phi\theta}$.


An \emph{interpretation} $\I$ maps each sort symbol $\sigma \in
\ssorts{\Sigma}$ to a non-empty set $\sigma^\I$, each function symbol
$f^{\sigma_1 \ldots \sigma_n \sigma} \in \Sigma$ to a total function
$f^\I : \sigma^\I_1 \times \ldots \times \sigma^\I_n \rightarrow
\sigma^\I$ where $n > 0$, and to an element of $\sigma^\I$ when $n=0$.
A distinguished example is the \emph{Herbrand (canonical)
interpretation} $\herb$, which maps each sort $\sigma \in
\ssorts{\Sigma}$ into $\T_\Sigma$, each constant symbol $c$
into the term $c$ and each function symbol $f^{\sigma_1 \ldots
\sigma_n \sigma}$ into the function carrying the tuple of terms
$t^{\sigma_1}_1,\ldots,t^{\sigma_n}_n$ into the term
$f(t_1,\ldots,t_n)$.

Given an interpretation $\I$, a \emph{valuation} $\nu$ maps each
variable $x^\sigma \in \vars$ to an element of $\sigma^\I$. Given a
tuple of variables $\vec{x} = (x_1,\ldots,x_k)$, we write
$\nu(\vec{x})$ for $(\nu(x_1),\ldots,\nu(x_k))$. For a term $t$,
$t^\I_\nu$ is the value obtained by replacing each function symbol $f$
by its interpretation $f^\I$ and each variable $x$ by its valuation
$\nu(x)$.  For a quantifier-free formula $\phi$, we write $\I,\nu
\models \phi$ if the formula obtained by replacing each term $t$ in
$\phi$ by the value $t^\I_\nu$ is equivalent to true. The semantics of
first order quantifiers is defined as: $\I,\nu \models \exists
x^\sigma \,.\, \phi(x)$ iff $\I,\nu[x \leftarrow \alpha] \models
\phi$, for some value $\alpha \in \sigma^\I$, where $\nu[x \leftarrow
\alpha]$ is the same as $\nu$, except for $\nu[x \leftarrow
\alpha](x)=\alpha$.

A formula $\phi$ is \emph{satisfiable} in the interpretation $\I$ if
there exists a valuation $\nu$ such that $\I,\nu \models \phi$. Given
formulae $\phi$ and $\psi$, we say that \emph{$\phi$ entails $\psi$}
in the interpretation $\I$, denoted $\phi \models^\I \psi$, iff $\I,\nu
\models \phi$ implies $\I,\nu \models \psi$, for each valuation $\nu$.

\subsection{Systems of Inductive Definitions} \label{sec:systems}

Let $\preds$ be a countable set of \emph{predicates}, each
$p^{\sigma_1 \ldots \sigma_n} \in \preds$ having an associated tuple
of argument sorts. Given $p^{\sigma_1 \ldots \sigma_n} \in \preds$ and
a tuple of terms $(t_1^{\sigma_1}, \ldots,
t_n^{\sigma_n})$, we call $p(t_1,\ldots,t_n)$ a \emph{predicate atom}.
A \emph{predicate rule} is a pair
$\langle \{\phi(\vec{x},\vec{x}_1,\ldots,\vec{x}_n),
q_1(\vec{x}_1), \ldots, q_n(\vec{x}_n)\},p(\vec{x})\rangle$, where
$\vec{x}, \vec{x}_1,\ldots,\vec{x}_n$ are pairwise disjoint sets of
variables, $\phi$ is a formula, called the \emph{constraint},
$p(\vec{x})$ is a predicate atom called the \emph{goal} and
$q_1(\vec{x}_1),\ldots,q_n(\vec{x}_n)$ are predicate atoms called
\emph{subgoals}. The variables $\vec{x}$ are the \emph{goal
  variables}, whereas the ones in $\bigcup_{i=1}^n\vec{x}_i$ are the
\emph{subgoal variables} of the rule.

An \emph{inductive system} $\sys$ (system, for short) is a finite set
of predicate rules. We assume w.l.o.g. that there are no goals with the same
predicate and different goal variables, and write \(p(\vec{x})
\leftarrow_\sys R_1 \mid \ldots \mid R_m\) when
$\set{\tuple{R_1,p(\vec{x})},\ldots,\tuple{R_m,p(\vec{x})}}$ is the
set of all predicate rules in $\sys$ with goal $p(\vec{x})$. The size
of $\sys$ is the sum of the sizes of all constraints occurring in the
rules of $\sys$.

We assume that each constraint is a quantifier-free formula in which
no disjunction occurs positively and no conjunction occurs negatively
and, moreover, that the set of constraints of a system has a decidable
satisfiability problem. Observe that disjunctions can be eliminated
w.l.o.g. from quantifier-free constraints, by splitting each rule
$\langle\{\phi_1 \vee \ldots \vee \phi_m,$ $q_1(\vec{x}_1),\ldots,
q_n(\vec{x}_n)\},p(\vec{x})\rangle$ into $m$ rules $\tuple{\set{\phi_i,
        q_1(\vec{x}_1),\ldots,q_n(\vec{x}_n)},p(\vec{x})}$, one for each
$i\in[m]$. Finally, we assume that each predicate $p \in \preds$ is
the goal of at least one rule of $\sys$.

\begin{example}\label{ex:fol-sys}
	Consider the following inductive system of predicates:
	\begin{center}
		\(\begin{array}{l@{\;}c@{\;}l@{\quad}l@{\;}c@{\;}l}
		p(x) & \leftarrow_\sys & x \teq f(x_1, x_2), p_1(x_1), p_2(x_2) &
		q(x) & \leftarrow_\sys & x \teq f(x_1, x_2), q_1(x_1), q_2(x_2) \\	
		p_1(x) & \leftarrow_\sys & x \teq g(x_1), p_1(x_1) \mid x \teq a &
		& \mid & x \teq f(x_1, x_2), q_2(x_1), q_1(x_2) \\
		p_2(x) & \leftarrow_\sys & x \teq g(x_1), p_2(x_1) \mid x \teq b &
		q_1(x) & \leftarrow_\sys & x \teq g(x_1), q_1(x_1) \mid x \teq a \\	
		& & & q_2(x) & \leftarrow_\sys & x \teq g(x_1), q_2(x_1) \mid x \teq b
		\end{array}\)
	\end{center}
Intuitively, $\sys$ models two tree automata with final states given by the predicates $p$ and $q$, where $p$ accepts trees of the form $f(g^m(a), g^n(b))$, while $q$ accepts trees of the form $f(g^m(a), g^n(b))$ and $f(g^m(b), g^n(a))$, where $m, n \leq 0$. \hfill\qed
\end{example}

Given an inductive system $\sys$ and an interpretation $\I$, an
\emph{assignment} $\X$ maps each predicate $p^{\sigma_1 \ldots
\sigma_n} \in \preds$ to a set $\X(p) \subseteq \sigma_1^\I \times
\ldots \times \sigma_n^\I$. By a slight abuse of notation, we lift
assignments from predicates to a set $F = \{\phi, q_1(\vec{x}_1),
\ldots, q_m(\vec{x}_m)\}$, where $\phi$ is a first order formula
and $q_i(\vec{x}_i)$ are predicate atoms, we define $\X(\bigwedge F)
= \{\nu \mid \I,\nu \models \phi, \nu(\vec{x}_i) \in \X(q_i),
\forall i \in [m]\}$.

The system $\sys$ and interpretation $\I$ induce a function
$\mathbb{F}^\I_\sys(\X)$ on assignments, which maps each predicate $p
\in \preds$ into the set $\bigcup_{i=1}^m \{\nu(\vec{x}) \mid \nu \in
\X(\bigwedge R_i)\}$, where $p(\vec{x}) \leftarrow_\sys R_{1} \mid
\ldots \mid R_{m}$ and, for a tuple of variables $\vec{x} =
(x_1,\ldots,x_k)$, we write $\nu(\vec{x})$ for $(\nu(x_1),\ldots,
\nu(x_k))$. A \emph{solution} of $\sys$ is an assignment $\X$
such that $\mathbb{F}^\I_\sys(\X) \subseteq \X$, where inclusion
between assignments is defined pointwise. It can easily be shown that
the set of all assignments, together with the $\subseteq$ relation, is
a complete lattice, since any power set equipped with the subset
relation is a complete lattice. Because $\mathbb{F}^\I_\sys$ is
monotone, it follows from Tarski's theorem \cite{Tarski1955LTF} that
$\mu\sys^\I = \bigcap\{\X \mid \mathbb{F}^\I_\sys(\X) \subseteq \X\}$
is the least fixed point of $\mathbb{F}^\I_\sys$ and the least solution
of $\sys$.

Note that predicates $p \in \preds$ that are not the goal of any predicate
rule in an inductive system $\sys$ will have empty least solutions, i.e.\
$\mu\sys^\I(p) = \emptyset$, thus each predicate rule containing such
a predicate as a subgoal can be safely removed from the system. This
justifies our earlier assumption that each predicate $p \in \preds$ is
the goal of at least one predicate rule of $\sys$.

\begin{example}\label{ex:fo-solution}
	For the inductive system in Example \ref{ex:fol-sys}:
	\[\begin{array}{l}
		\mu\sys^\I(p) =\{f(g^n(a), g^m(b)) \mid n, m \geq 0\} \\[2pt]
		\mu\sys^\I(p_1) = \mu\sys^\I(q_1) = \{g^n(a) \mid n\geq 0\} \\[2pt]
		\mu\sys^\I(p_2) = \mu\sys^\I(q_2) = \{g^n(b) \mid n \geq 0\} \\[2pt]
		\mu\sys^\I(q) = \{f(g^n(a), g^m(b)) \mid n, m \geq 0\} \cup \{f(g^n(b), g^m(a)) \mid n, m \geq 0\}
	\end{array}\]
	Since $\mu\sys^\I(p) \subseteq \mu\sys^\I(q)$, it follows that the entailment $p \models_\sys^\I q$ holds. In other words, the language accepted by the state represented as the predicate $p$ is included in the language of the state represented by $q$.
\end{example}

We are concerned with the following \emph{entailment problem}: given
an inductive system $\sys$, an interpretation $\I$, and predicates
$p^{\sigma_1 \ldots \sigma_m}$, $q_1^{\sigma_1 \ldots \sigma_m},
\ldots, q_n^{\sigma_1 \ldots \sigma_m}$, having the same tuple of
argument sorts, is it true that $\mu\sys^\I(p) \subseteq
\bigcup_{i=1}^n \mu\sys^\I(q_i)$? We denote entailment problems as $p
\models^\I_\sys q_1, \ldots, q_n$.

\subsection{Well Quasi-Orders} \label{sec:wqo}

Given a set $D$, a \emph{quasi-order} (qo) is a reflexive
and transitive relation $\preceq~ \subseteq \mathcal{D} \times
\mathcal{D}$. An infinite sequence $d_1, d_2, \ldots$ from $\mathcal{D}$
is \emph{saturating} if $d_i \preceq d_j$ for some $i < j$. A quasi-order
$\preceq$ is a \emph{well-quasi-order} (wqo) if every infinite sequence
is saturating. A quasi-order $\preceq$ is \emph{well-founded} (wfqo) iff
there are no infinite decreasing sequences $d_1 \succ d_2 \succ \ldots$
Every wqo is well-founded, but not viceversa.

We extend any wqo $(\mathcal{D},\preceq)$ to the following order on
the set of finite subsets of $\mathcal{D}$. For all finite sets $S,T
\in \finpow{\mathcal{D}}$, we have $S \preceq^{\forall\exists} T$ if
and only if for all $a \in S$ there exists $b \in T$ such that $a
\preceq b$. The following is a consequence of Higman's Lemma
\cite{Higman52}:

\begin{lemma}\label{lemma:finite-sets-wqo}
  If $\mathcal{D}$ is countable and $(\mathcal{D},\preceq)$ is a wqo,
  then $(\finpow{\mathcal{D}}, \preceq^{\forall\exists})$ is a wqo.
\end{lemma}

A \emph{multiset} over $\mathcal{D}$ is a mapping $M : \mathcal{D}
\rightarrow \nat$. The multiset $M$ is finite if $M(d) > 0$ for a
finite number of elements $d \in \mathcal{D}$. We denote by
$\mathcal{M}(\mathcal{D})$ the set of finite multisets over
$\mathcal{D}$, and lift the operations of subset, union, intersection
and difference to multisets, as usual. The \emph{multiset order}
induced by $\preceq$ is defined as in \cite{MannaDershowitz79}. We 
write $N \preceq^\dagger M$ if and only if either $M = N$, or there
exists a non-empty finite multiset $X \subseteq M$ and a (possibly
empty) multiset $Y$, where for all $y \in Y$ there exists $x \in X$
such that $y \prec x$ and $N = (M \setminus X) \cup Y$. Roughly, 
$N$ is obtained by replacing a non-empty submultiset of $M$ with a
possibly empty multiset of strictly smaller elements. The following
theorem was proved in \cite{MannaDershowitz79}:

\begin{theorem}\label{thm:manna-dershowitz}  
  $(\mathcal{M}(\mathcal{D}),\preceq^\dagger)$ is a wfqo if and only
  if $(\mathcal{D},\preceq)$ is a wfqo.
\end{theorem}

\subsection{Canonical Interpretation} \label{sec:herb}

Let $\nat^*$ be the set of sequences of natural numbers, $\varepsilon
\in \nat^*$ be the empty sequence, and $p \cdot q$ denote the
concatenation of two sequences $p,q\in \nat^*$. We say that $p$ is a
prefix of $q$ iff $p \cdot r = q$, for some $r \in \nat^*$. A set 
$X \subseteq \nat^*$ is \emph{prefix-closed} if $p \in X$ implies that
every prefix of $p$ is in $X$. A \emph{tree} over the signature
$\Sigma = (\ssorts{\Sigma}, \sfuns{\Sigma})$ is a ground term
$t \in \T_\Sigma$, viewed as a finite partial function
$t : \nat^* \rightharpoonup_{fin} \sfuns{\Sigma}$, where $\dom(t)$
is prefix-closed and, for all $\alpha \in \dom(t)$ such that
$t(\alpha) = f^{\sigma_1 \ldots \sigma_n \sigma}$, we have
$\set{i \in \nat \mid pi \in \dom(t)} = [n]$. We denote by
$\fr(t) = \set{\alpha \in \dom(t) \mid \alpha \cdot 1 \not\in \dom(t)}$
the frontier of $t$. Given a tree $t$ and a position $\alpha \in \dom(t)$,
we denote by $\sub{t}{\alpha}$ the \emph{subtree} of $t$ rooted at
$\alpha$, where, for each $\beta \in \nat^*$, we have $\sub{t}{\alpha}(\beta)
= t(\alpha \cdot \beta)$. The subtree order is defined by $u \subtreeq t$
iff $u = \sub{t}{\alpha}$, for some $\alpha \in \dom(t)$. It is easy to see
that $(\T_\Sigma,\subtreeq)$ is a wfqo, because $\T_\Sigma$ consists only of
finite trees, making it impossible to build an infinite strictly decreasing
sequence of subtrees.


For a function symbol $f^{\sigma_1 \ldots \sigma_n} \in
\sfuns{\Sigma}$ and trees $t_1, \ldots, t_n \in \T_\Sigma$, let
$\tau_n(f,t_1,\ldots,t_n)$ be the tree $t$ such that $t(\varepsilon) =
f$ and $\sub{t}{i} = t_i$, for all $i \in [n]$. The \emph{Herbrand
(canonical) interpretation} $\herb$ maps each sort $\sigma \in
\ssorts{\Sigma}$ into $\T_\Sigma$, each constant symbol $c$ into the
tree $c^\tinyherb = \set{(\varepsilon,c)}$ consisting of a leaf which
is also the root, and each function symbol
$f^{\sigma_1 \ldots \sigma_n \sigma}$ into the function $f^\tinyherb$
mapping each tuple of trees $t_1,\ldots,t_n$ into
$\tau_n(f,t_1,\ldots,t_n)$. Even in this simple case, where function
symbols do not have any equational properties (e.g.\ commutativity,
associativity, etc.) entailment problems are undecidable, as stated by
the following theorem.

\begin{theorem}\label{thm:entailment-undec}
  The entailment problem is undecidable for First Order Logic, under
the Herbrand interpretation.
\end{theorem}

This negative result excludes the possibility of having a complete
proof system for solving entailments between predicates of inductive
systems using (unrestricted) first-order logic constraints, under the
canonical interpretation. A possible workaround is to restrict the
class of systems considered, by imposing several semantic restrictions
on the set of constraints that occur within the rules of the system
(\S\ref{sec:restrictions}).

\section{Cyclic Proofs for Inductive Predicate Entailments}
\label{sec:canonical-proof-system}

In this section we present our set of inference rules (proof system)
for proving entailments between predicates in a given inductive
system. In a nutshell, our proof system is based on a classical
\emph{unfold-and-match} scheme, extended with a \emph{split} rule
(explained next), and whose termination relies on an infinite descent
inductive argument. The proof system is inspired by an antichain-based
language inclusion algorithm for top-down tree automata
\cite{HolikLSV11}, briefly explained in \S\ref{sec:ta}. The
presentation of the inference rules uses a comparison with the
internals of the language inclusion algorithm. Because, moreover,
language inclusion is decidable for tree automata, we extract several
sufficient conditions that guarantee completeness of the proof system,
and thus the existence of a decision procedure for the entailment
problem. 

\subsection{Tree Automata Inclusion as Cyclic Proof Search}
\label{sec:ta}

We assume basic knowledge of tree automata \cite{Tata05} and consider
top-down nondeterministic finite tree automata (NFTA), whose actions
are described by transition rules $q \arrow{f}{} (q_1, \ldots, q_n)$,
with the following meaning: if the automaton is in state $q$ and the
input is a ground term $f(t_1,\ldots,t_n)$, then it moves
simultaneously on each $t_i$ changing its state to $q_i$, for all $i
\in [n]$. A ground term is \emph{accepted} by an automaton $A$ in
state $q$ if each constant subterm (leaf) can be eventually read by a
rule of the form $q \arrow{a}{} ()$. The language of a state $q$ in
$A$, denoted $\mathcal{L}(A,q)$, is the set of ground terms accepted
by $A$ starting with $q$.

An NFTA can be naturally viewed as an inductive system, where
predicates represent states and predicate rules are obtained directly
from transition rules, as follows. For instance, the transition rule
$q \arrow{f}{} (q_1, \ldots, q_n)$ can be written as $\langle\{x \teq
f(x_1,\ldots,x_n),$ $q_1(x_1), \ldots, q_n(x_n)\}, q(x) \rangle$,
where variables range over ground terms and the function symbols are
interpreted in the canonical (Herbrand) sense. Then $\mathcal{L}(A,q)
= \mu\sys^\tinyherb(q)$, for any state (predicate) $q$. Then a
language inclusion problem $\mathcal{L}(A,p) \subseteq
\bigcup_{i=1}^k\mathcal{L}(A,q_i)$ is equivalent to the entailment
problem $p \models^\tinyherb_\sys q_1,\ldots,q_k$, where $\sys$ is the
inductive system corresponding to $A$.

\begin{example}\label{ex:ta-sys}
    Consider the following inductive system:
    {\[\begin{array}{@{}l@{\;}c@{\;}l@{\qquad}l@{\;}c@{\;}l@{}}
        p(x) & \leftarrow_\sys & x \teq f(x_1, x_2), p_1(x_1), p_2(x_2) &
        q(x) & \leftarrow_\sys & x \teq f(x_1, x_2), q_1(x_1), q_2(x_2) \\[2pt] 
        p_1(x) & \leftarrow_\sys & x \teq g(x_1), p_1(x_1) \mid x \teq a &
        & \mid & x \teq f(x_1, x_2), q_2(x_1), q_1(x_2) \\[2pt]
        p_2(x) & \leftarrow_\sys & x \teq g(x_1), p_2(x_1) \mid x \teq b &
        q_1(x) & \leftarrow_\sys & x \teq g(x_1), q_1(x_1) \mid x \teq a \\[2pt]
        & & & q_2(x) & \leftarrow_\sys & x \teq g(x_1), q_2(x_1) \mid x \teq b
        \end{array}\]}
    Intuitively, $p$ accepts terms of the form $f(g^n(a), g^m(b))$ and
    $q$ accepts both terms of the form $f(g^n(a), g^m(b))$ and
    $f(g^n(b), g^m(a))$ with $n, m \geq 0$. Then $p
    \models_\sys^\tinyherb q$ holds, while $q \not\models_\sys^\tinyherb
    p$. \hfill$\blacksquare$
\end{example}

Since language inclusion is decidable for NFTA\footnote{See,
    e.g.\ \cite[Corollary 1.7.9]{Tata05}.}, we leverage from an existing
algorithm for this problem by Hol{\'{\i}}k et al. \cite{HolikLSV11} to
build a complete set of inference rules and derive a proof search
technique. This algorithm searches for counterexamples of the
inclusion problem $\mathcal{L}(A,p) \subseteq \bigcup_{i=1}^k
\mathcal{L}(A,q_i)$ by enumerating pairs $(r,\{s_1,\ldots,s_m\})$,
where $r$ is a state that can be reached via a series of transitions
from $p$, and $\{s_1,\ldots,s_m\}$ are all the states that can be
reached via the same series of transitions from $q_1, \ldots, q_k$. A
counterexample is found when the algorithm encounters a pair
$(r,\{s_1,\ldots,s_m\})$ such that there exists a transition $r
\overset{a}{\rightarrow} ()$, but there is no transition $s_i
\overset{a}{\rightarrow} ()$ for any $i \in [m]$. 

\begin{example}\label{ex:down-incl}
    If A is the NFTA equivalent to the system from Example \ref{ex:ta-sys},
    to check $\mathcal{L}(A, p) \subseteq \mathcal{L}(A, q)$, we start with
    $(p, \{q\})$. A possible run of the algorithm is:
    \begin{center}
        \resizebox{\textwidth}{!}
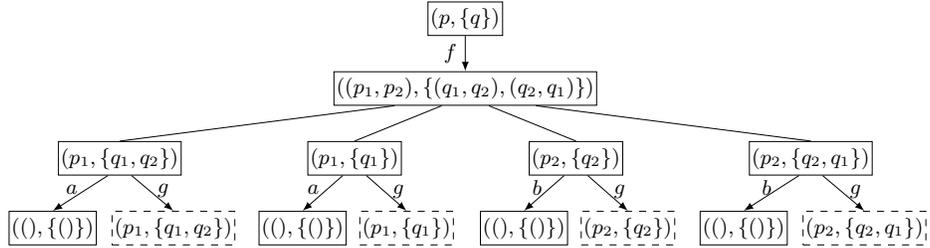
{%
            \begin{tikzpicture}[pstate/.style={minimum height=0.5cm, inner sep=1pt}, news/.style={draw=black}, olds/.style={draw=black, dashed}, ends/.style={draw=black}, lbl/.style={outer sep=0.15pt}]
            \node (pstate1) [pstate, news] {\small $(p, \{q\})$};
            \node (pstate2) [pstate, news, below=15pt of pstate1] {\small $((p_1, p_2), \{(q_1, q_2),(q_2,q_1)\})$};
            \node (pstate3) [pstate, news, below left=15pt and 5.15cm of pstate2.south, anchor=north] {\small $(p_1, \{q_1, q_2\})$};
            \node (pstate4) [pstate, news, below left=15pt and 1.65cm of pstate2.south, anchor=north] {\small $(p_1, \{q_1\})$};
            \node (pstate5) [pstate, news, below right=15pt and 1.65cm of pstate2.south, anchor=north] {\small $(p_2, \{q_2\})$};
            \node (pstate6) [pstate, news, below right=15pt and 5.15cm of pstate2.south, anchor=north] {\small $(p_2, \{q_2, q_1\})$};
            
            \node (pstate31) [pstate, ends, below left=15pt and 1cm of pstate3.south, anchor=north] {\small $((), \{()\})$};
            \node (pstate32) [pstate, olds, below right=15pt and 0.8cm of pstate3.south, anchor=north] {\small $(p_1, \{q_1, q_2\})$};
            
            \node (pstate41) [pstate, ends, below left=15pt and 0.775cm of pstate4.south, anchor=north] {\small $((), \{()\})$};
            \node (pstate42) [pstate, olds, below right=15pt and 0.775cm of pstate4.south, anchor=north] {\small $(p_1, \{q_1\})$};
            
            \node (pstate51) [pstate, ends, below left=15pt and 0.775cm of pstate5.south, anchor=north] {\small $((), \{()\})$};
            \node (pstate52) [pstate, olds, below right=15pt and 0.775cm of pstate5.south, anchor=north] {\small $(p_2, \{q_2\})$};
            
            \node (pstate61) [pstate, ends, below left=15pt and 1cm of pstate6.south, anchor=north] {\small $((), \{()\})$};
            \node (pstate62) [pstate, olds, below right=15pt and 0.8cm of pstate6.south, anchor=north] {\small $(p_2, \{q_2, q_1\})$};
            
            \draw [-latex] (pstate1.south) -- (pstate2.north) node [left,pos=0.5] {\small $f$};
            \draw ([xshift=-30pt]pstate2.south) -- (pstate3.north);
            \draw ([xshift=-10pt]pstate2.south) -- (pstate4.north);
            \draw ([xshift=10pt]pstate2.south) -- (pstate5.north);
            \draw ([xshift=30pt]pstate2.south) -- (pstate6.north);
            
            \draw [-latex] ([xshift=-5pt]pstate3.south) -- (pstate31.north) node [left,pos=0.4] {\small $a$};
            \draw [-latex] ([xshift=5pt]pstate3.south) -- (pstate32.north) node [right,pos=0.425] {\small $g$};
            
            \draw [-latex] ([xshift=-5pt]pstate4.south) -- (pstate41.north) node [left,pos=0.38] {\small $a$};
            \draw [-latex] ([xshift=5pt]pstate4.south) -- (pstate42.north) node [right,pos=0.45] {\small $g$};
            
            \draw [-latex] ([xshift=-5pt]pstate5.south) -- (pstate51.north) node [left,pos=0.35] {\small $b$};
            \draw [-latex] ([xshift=5pt]pstate5.south) -- (pstate52.north) node [right,pos=0.45] {\small $g$};
            
            \draw [-latex] ([xshift=-5pt]pstate6.south) -- (pstate61.north) node [left,pos=0.35] {\small $b$};
            \draw [-latex] ([xshift=5pt]pstate6.south) -- (pstate62.north) node [right,pos=0.45] {\small $g$};
            \end{tikzpicture}}
    \end{center}
    The algorithm performs two types of moves: transitions and split
    actions. The arrows labeled by function symbols $f,g,a$ and $b$ are
    transitions, for instance the arrow labeled by $f$ takes $p$ into
    the tuple $(p_1,p_2)$ by the transition rule $p \arrow{f}{}
    (p_1,p_2)$ and $\set{q}$ into the set of tuples
    $\set{(q_1,q_2),(q_2,q_1)}$, corresponding to the transition rules
    $q \arrow{f}{} (q_1,q_2)$ and $q \arrow{f}{} (q_2,q_1)$. However the
    pair of tuples $((p_1,p_2), \set{(q_1,q_2), (q_2,q_1)})$ is
    problematic because it asserts the following: \(\mathcal{L}(A,p_1)
    \times \mathcal{L}(A,p_2) \subseteq \mathcal{L}(A,q_1) \times
    \mathcal{L}(A,q_2) \cup \mathcal{L}(A,q_2) \times
    \mathcal{L}(A,q_1)\). Using several properties of the Cartesian
    product\footnote{See \cite[Theorem 1]{HolikLSV11}.} this proof
    obligation can be split into several conjunctive subgoals. The
    split move used above corresponds to considering simultaneously the
    pairs $(p_1, \{q_1, q_2\})$, $(p_1, \{q_1\})$, $(p_2, \{q_2\})$ and
    $(p_2, \{q_2, q_1\})$, asserting that $\mathcal{L}(A,p_1)
    \subseteq \mathcal{L}(A,q_1)$ and $\mathcal{L}(A,p_2) \subseteq
    \mathcal{L}(A,q_2)$. The other possibilities are: \begin{inparaenum}[(1)]
        \item \(\{(p_1, \{q_1, q_2\}), (p_1, \{q_1\}), (p_1, \{q_2\}), (p_2, \{q_2, q_1\})\}\),
        \item \(\{(p_1, \{q_1, q_2\}), (p_2, \{q_1\}), \\ (p_1, \{q_2\}), (p_2,\{q_2, q_1\})\}\) and 
        \item \(\{(p_1, \{q_1, q_2\}), (p_2, \{q_1\}), (p_2, \{q_2\}), (p_2,
        \{q_2, q_1\})\}.\) \end{inparaenum} The algorithm does not further
    expand nodes $((), S)$ with $() \in S$, for which inclusion holds
    trivially, or $(p,S)$ which has a predecessor $(p,S')$ such that $S'
    \subseteq S$ (enclosed in dashed boxes). In the latter case, any
    counterexample that can be found from $(p,S)$ could have been
    discovered from $(p,S')$. \hfill$\blacksquare$
\end{example}


\subsection{A Proof Search Semi-algorithm} \label{sec:proof-search}

In this section we give a set of inference rules for proving the
validity of entailments between predicates defined in an inductive
system. We describe the inference rules using a Gentzen-style sequent
calculus.

\begin{wrapfigure}{r}{0.45\textwidth}
    \vspace{-1.75\baselineskip}
    \hspace{-5pt} $\proofrulepivot{(\IR)}
    {\lseq_p \vdash \rseq_p} {\lseq \vdash \rseq} {\lseq_1 \vdash \rseq_1 ~\ldots~ \lseq_n \vdash \rseq_n} {\begin{array}{@{}ll@{}}\scriptscriptstyle{\text{side}}\vspace*{-1mm} \\ \scriptscriptstyle{\text{conditions}}\end{array}} {\scriptscriptstyle{\mathbf{C}}}$
    \vspace{-2\baselineskip}
\end{wrapfigure}

We denote \emph{sequents} as $\lseq \vdash \rseq$, where $\lseq$ and
$\rseq$ are sets of formulae. The commas in the sequents are read as
set union, thus contraction rules are not necessary. A singleton
$\set{p(\vec{x})}$ is denoted as $p(\vec{x})$, and a sequent of the
form $p(\vec{x}) \vdash q_1(\vec{x}), \ldots, q_n(\vec{x})$ is said to
be \emph{basic}. An inference rule $\ir$ has $\#(\ir)\geq0$
\emph{antecedents} $\lseq_i \vdash \rseq_i$, for $i = 1, \ldots,
\#(\ir)$ and a \emph{consequent} $\lseq \vdash \rseq$. We denote an
empty antecedent list of an inference rule by $\top$.

Additionally, certain inference rules with no antecedents may have a
\emph{pivot} $\lseq_p \vdash \rseq_p$, which is always a predecessor
of the consequent in the transitive closure of the
consequent-antecedent relation. The path between the pivot
and the consequent is subject to a \emph{pivot constraint}, defined
next. 

For conciseness of presentation, we refer to \emph{inference rule
    schemata} $\IR$, that are possibly infinite sets of inference rules
sharing the same pattern. Without entering formal details, we assume
that checking whether a given inference rule instance belongs to a
given schema is straightforward and that there are finitely many
instances of a certain schema for a given pivot and consequent.

\begin{definition}\label{def:derivation-tree}
    A \emph{derivation} is a possibly infinite tree $\mathcal{D} =
    (V,v_0,S,R,P)$, where $V$ is a set of vertices, $v_0 \in V$ is the
    root node, and for each vertex $v \in V$, $S(v)$ is a sequent, $R(v)$
    is a rule schema labeling $v$ and, if $v \neq v_0$, then $P(v)$ is the
    parent of $v$. In particular, for each $v \in V \setminus \set{v_0}$,
    $S(v)$ is an antecedent of an instance of $R(P(v))$, with consequent
    $P(v)$. A \emph{proof} is a finite derivation.
\end{definition}

Observe that, by this definition, the antecedent list of each terminal
(leaf) node in a proof is necessarily empty and, since the branching
degree of a proof is finite, each path in the proof must be finite, by
K\"onig's Lemma. Below we define (possibly infinite) traces, by
considering the continuation of a path via a backlink from the
consequent to the pivot of a rule.

\begin{definition}\label{def:traces}
    Given a derivation $\mathcal{D} = (V,v_0,S,R,P)$, a \emph{backlink}
    is a pair $(u, v)$ with $u, v \in V$ such that $v$ is the pivot of
    the instance of $R(u)$ applied at $u$. A \emph{trace} is a
    sequence of vertices $\tau=v_1,v_2,\ldots$ such that, for all
    $i\geq2$, either $v_{i-1} = P(v_i)$ or $(v_{i-1},v_i)$ is a backlink.
    A \emph{path} of $\tau$ is any finite subsequence
    $\pi = v_i,\ldots,v_j$ such that $i < j$ and $v_{k-1}=P(v_k)$, for
    all $k \in [i+1,j]$. If, moreover, $(v_j,v_i)$ is a backlink, then
    $\pi$ is a \emph{direct path}.
\end{definition}

Additionally, if $\mathcal{D} = (V,v_0,S,R,P)$ is a derivation and
$\tau = v_1, v_2,\ldots$ is a trace in $\mathcal{D}$, we say that
$\tau$ contains a backlink if there exists a position $i > 1$ such
that $(v_{i-1}, v_{i})$ is a backlink. Given a path or a finite
trace $\pi = v_1,\ldots,v_k$ with $k\geq2$, we
denote by $\typelabel(\pi)$ the sequence $R(v_1),\ldots, R(v_{k-1})$
of inference rule schemata that are applied on $\pi$.

The pivot constraint $\mathbf{C}$ of a rule schema $\IR$ is a set of
finite sequences of rule schemata, such that, for any instance $\ir$
of $\IR$, if $\pi$ is the direct path from the pivot of $\ir$ to its
consequent in any derivation, then $\typelabel(\pi) \in
\mathbf{C}$. 

\begin{proposition}\label{prop:direct-path}
    Given a proof $\mathcal{D}$, any infinite trace of $\mathcal{D}$
    contains infinitely many direct paths.
\end{proposition}

Given a system $\sys$, an interpretation $\I$ and a set of predicates
$p^{\sigma_1,\ldots,\sigma_n}$, $q_1^{\sigma_1,\ldots,\sigma_n}$, $\ldots$,
$q_k^{\sigma_1,\ldots,\sigma_n} \in \preds$, a set of inference rule
schemata $\mathcal{R}$ is: \begin{inparaenum}[(i)]
    \item \emph{sound} if, for any proof $\mathcal{D}=(V,v_0,S,R,P)$ with
    $S(v_0) = p(\vec{x}) \vdash q_1(\vec{x}), \ldots,
    q_n(\vec{x})$, we have $p \models_\sys^\I q_1, \ldots, q_n$, 
    \item \emph{complete} if $p \models_\sys^\I q_1, \ldots, q_n$ implies
    the existence of a proof $\mathcal{D}=(V,v_0,s,r,p)$, where $S(v_0)
    = p(\vec{x}) \vdash q_1(\vec{x}), \ldots, q_n(\vec{x})$.
\end{inparaenum}

A \emph{strategy} is a set of sequences $\mathbf{S}$ of inference rule
schemata. A sequence $s$ is a \emph{valid prefix} for $\mathbf{S}$ if
$s\cdot s' \in \mathbf{S}$ for another sequence $s'$. A derivation
$\mathcal{D}$ (proof) is an $\mathbf{S}$-derivation
($\mathbf{S}$-proof) if the sequence of inference rules along each
maximal path of $\mathcal{D}$ belongs to $\mathbf{S}$.

\begin{algorithm}[htb]
    {\begin{algorithmic}[0]
            \State {\bf data structure}:
            $\mathsf{Node}(\mathit{Seq},\mathit{CList},\mathit{Parent},\mathit{Rule})$,
            where:
            \begin{compactitem}
                \item $\mathit{Seq}$ is the sequent that labels the node,
                \item $\mathit{CList}$ is the list of children nodes, 
                \item $\mathit{Parent}$ is the link to the parent of the node,
                \item $\mathit{Rule}$ is the inference rule with consequent
                $\mathit{Seq}$.
            \end{compactitem}
            \State {\bf input}: \begin{compactitem}
                \item an inductive system $\sys$, 
                \item a sequent $p(\vec{x}) \vdash q_1(\vec{x}), \ldots,
                q_n(\vec{x})$,
                \item a set $\mathcal{R}$ of inference rule schemata and a
                strategy $\mathbf{S}$
            \end{compactitem}
            \State {\bf output}: a proof $\mathcal{D}=(V,v_0,S,R,P)$ such
            that $S(v_0) = p(\vec{x}) \vdash q_1(\vec{x}), \ldots,
            q_n(\vec{x})$ 
    \end{algorithmic}}
    {\begin{algorithmic}[1]
            \State $\mathsf{Root} \leftarrow \mathsf{Node}(p(\vec{x})  \vdash
            q_1(\vec{x}), \ldots, q_n(\vec{x}), [], \mathit{nil}, \mathit{nil})$
            
            \State $\mathsf{WorkList} \leftarrow \set{\mathsf{Root}}$ 
            
            \While{$\mathsf{WorkList} \neq \emptyset$}
            
            \State remove $N$ from $\mathsf{WorkList}$ and match it with 
            $\mathsf{Node}(\lseq\vdash\rseq,\mathit{CList},P,R)$
            
            \State let $\pi$ be the path between $\mathsf{Root}$ and $N$
            
            \State let $\mathcal{R}_N \subseteq \mathcal{R}$ be the inference rule schemata applicable on $N.\mathit{Seq}$ and $\pi$
            
            \State let $\mathcal{R}_N^0 \subseteq \mathcal{R}_N$ be the subset of $\mathcal{R}_N$ with empty antecedent lists
            
            \If{$\typelabel(\pi) \cdot \IR$ is a valid prefix of $\mathbf{S}$ for some $\IR \in \mathcal{R}_N^0$} \label{ln:close}
            
            \State $N.\mathit{Rule} \leftarrow \IR$
            
            \State mark $N$ as closed
            
            \EndIf
            
            \If{$N$ not closed and $\typelabel(\pi) \cdot \IR$ is a valid prefix of $\mathbf{S}$ for some $\IR \in \mathcal{R}_N$}
            
            \State let $\ir$ be an instance of $\IR$ such that
            $N.\mathit{Seq}$ is the consequent of $\ir$
            
            \For{each antecedent $\lseq' \vdash \rseq'$ of $\ir$ }
            \State $N' \leftarrow \mathsf{Node}(\lseq' \vdash \rseq', [], N, \mathit{nil})$
            
            \State Append $N'$ to $N.\textit{CList}$
            
            \State Append $N'$ to $\mathsf{WorkList}$
            \EndFor
            
            \If{$N.CList$ is empty}
            mark $N$ as closed
            \EndIf
            
            \EndIf
            
            \EndWhile
    \end{algorithmic}}
    \caption{Proof search semi-algorithm.}
    \label{alg:proof-search}
\end{algorithm}

Given an input sequent $p(\vec{x}) \vdash q_1(\vec{x}), \ldots, 
q_n(\vec{x})$, a set $\mathcal{R}$ of inference rules and a
strategy $\mathbf{S}$, the proof search semi-algorithm 
(\ref{alg:proof-search}) uses a worklist iteration to build a
derivation of $p(\vec{x}) \vdash q_1(\vec{x}), \ldots, 
q_n(\vec{x})$. When a sequent is
removed from the worklist, it chooses (non-deterministically) an
inference rule and an instance whose consequent matches the current
sequent, if one exists. In order to speed up termination, the nodes
matching a rule with zero antecedents are considered eagerly (line
\ref{ln:close}). It is manifest that, if a proof of the input
sequent exists, then there exists a finite execution of the
semi-algorithm (\ref{alg:proof-search}) leading to it.

\subsection{Inference Rules}\label{sec:inference-rules}

Figure \ref{fig:ta-rules} gives a set $\RInd$ of
inference rule schemata for the entailment problem. To shorten the
presentation, we write $\tuple{\lseq_i \vdash \rseq_i}_{i=1}^n$ for
$\lseq_1 \vdash \rseq_1, \ldots, \lseq_n \vdash \rseq_n$.

\begin{figure}[htb]
    \begin{adjustbox}{max width=\textwidth}
        \begin{tabular}{@{}l@{}}
            $\proofrule{(\LU)}{\lseq \vdash \rseq}{\tuple{R_i(\vec{x},\vec{y}_i),\lseq \setminus p(\vec{x}) \vdash \rseq}_{i=1}^n}{\begin{array}{@{}l@{}} \scriptstyle p(\vec{x}) \in \lseq \text{, } p(\vec{x}) \leftarrow_\sys R_1(\vec{x},\vec{y}_1) \mid \ldots \mid R_n(\vec{x},\vec{y}_n) \\[-2pt] \scriptstyle \vec{y}_1, \ldots, \vec{y}_n \text{ fresh variables } \end{array}}$ \\[20pt]
            
            $\proofrule{(\RU)}{\lseq \vdash \rseq}{\lseq \vdash \set{\exists \vec{y}_i \,.\, \bigwedge R_i(\vec{x},\vec{y}_i)}_{i=1}^n, \rseq \setminus p(\vec{x})}{\begin{array}{@{}l@{}} \scriptstyle p(\vec{x}) \in \rseq \text{, } p(\vec{x}) \leftarrow_\sys R_1(\vec{x},\vec{y}_1) \mid \ldots \mid R_n(\vec{x},\vec{y}_n) \\[-2pt] \scriptstyle \vec{y}_1, \ldots, \vec{y}_n \text{ fresh variables } \end{array}}$ \\[20pt]
            
            $\proofrule{(\RD)}{\phi(\vec{x},\vec{x}_1,\ldots,\vec{x}_n),p_1(\vec{x}_1),\ldots,p_n(\vec{x}_n) \vdash \{\exists \vec{y}_j \,.\, \psi_j(\vec{x},\vec{y}_j) \wedge \mathcal{Q}_j(\vec{y}_j)\}_{j=1}^k}{p_1(\vec{x}_1),\ldots,p_n(\vec{x}_n) \vdash \{\mathcal{Q}_j\theta \mid \theta \in S_j\}_{j=1}^i}{\begin{array}{@{}l@{}} \scriptstyle \phi \models^\I \bigwedge_{j=1}^i \exists \vec{y}_j . \psi_j \\[-1.5pt] \scriptstyle \phi \not\models^\I \bigvee_{j=i+1}^k \exists \vec{y}_j . \psi_j \\[-1.5pt] \scriptstyle S_j \subseteq \subst{\phi,\psi_j}, j \in [i] \end{array}}$ \\[20pt]
            
            \begin{tabular}{@{}l@{\;}@{}l@{\;}@{}l@{}}
                $\proofrule{(\RI)}{\lseq \vdash p(\vec{x}) \land q(\vec{x}) \land \mathcal{Q},\rseq}{\lseq \vdash p(\vec{x}) \land \mathcal{Q},\rseq \;\;\; \lseq \vdash q(\vec{x}) \land \mathcal{Q},\rseq}{}$ & $\proofrule{(\AX)}{\lseq \vdash \rseq}{\top}{\scriptstyle \bigwedge\lseq \models^\I \bigvee \rseq}$ & $\proofrulepivot{(\ID)}{\lseq \vdash \rseq}{\lseq\theta \vdash \rseq'\theta}{\top}{\begin{array}{@{}l@{}} \scriptstyle \theta \text{ flat injective} \\[-4pt] \scriptstyle \text{substitution} \\[-1pt]
                    \rseq \subseteq \rseq'\end{array}}{\scriptstyle \RInd^* \cdot \LU \cdot \RInd^*}$\end{tabular}\\[30pt]
            
            $\proofrule{(\SP)}{p_1(\vec{x}_1),\ldots,p_n(\vec{x}_n) \vdash \mathcal{Q}_1(\vec{x}_1,\ldots,\vec{x}_n), \ldots, \mathcal{Q}_k(\vec{x}_1,\ldots,\vec{x}_n)}{\langle p_{\bar{\imath}_j}(\vec{x}) \vdash \{q_{\bar{\imath}_j}^\ell(\vec{x}) \mid \ell \in [k],~ f_j(\overline{\mathcal{Q}}_\ell) = \bar{\imath}_j\} \rangle_{j=1}^{n^k}}{\begin{array}{@{}l@{}} \scriptstyle \forall i,j \in [n] \,.\, \vec{x}_i \cap \vec{x}_j = \emptyset, ~ \bar{\imath} \in {[n]}^{n^k} \\[-2pt] \scriptstyle \mathcal{Q}_i = \bigwedge_{j=1}^{n} q_j^i(\vec{x}_j), \overline{\mathcal{Q}}_i=\langle q_1^i,\ldots,q_n^i \rangle \\[-2pt] \scriptstyle \cf(\overline{\mathcal{Q}}_1,\ldots,\overline{\mathcal{Q}}_k) = \set{f_1,\ldots,f_{n^k}} \end{array}}$ 
        \end{tabular}
    \end{adjustbox}
    \caption{ The set $\RInd$ of inference rule schemata for inductive entailments. }
    \vspace*{-\baselineskip}
    \label{fig:ta-rules}
\end{figure}

The inference rules $(\LU)$ and $(\RU)$ correspond to the
unfolding of a predicate atom $p(\vec{x})$ occurring on the left- and
right-hand side of a sequent $\lseq \vdash \rseq$, respectively. By
unfolding, we mean the replacement of $p(\vec{x})$ with
the set of predicate rules in $p(\vec{x}) \leftarrow_\sys R_1 \mid \ldots \mid
R_n$. Observe that the left unfolding yields a set of sequents for
each $R_i$ that must be all proved, whereas the right unfolding
simply replaces $p(\vec{x})$, on the right-hand side of the sequent,
with a set of formulae in which the subgoal variables of
$R_1,\ldots,R_n$ are existentially quantified.

The inference rules $(\RD)$ eliminate the constraints from both the
left- and right-hand sides. The existentially quantified variables on
the right-hand side are eliminated using a finite set of substitutions
that witness the entailments between the left and right constraints.
Since $(\RD)$ can create conjunctions of predicates with the same
arguments, we assume that every application of $(\RD)$ is followed
by a cleanup of the right-hand side of its antecedent(s) using a
weakening rule $(\RI)$ until all such conjunctions of predicates
are eliminated.

In the case of tree automata language inclusion checking, the $(\LU)$,
$(\RU)$ and $(\RD)$ rules are taken all at once, by a transition move
of the language inclusion algorithm of \cite{HolikLSV11} (Example
\ref{ex:down-incl}). This is natural because the transition rules
(predicate rules) of tree automata are controlled uniquely by the
function symbols labeling the root of the current input term. Because
function symbols can only be compared via equality, the constraints
within the predicate rules of a tree automata match
unambiguously. For instance, we have $x \teq f(x_1,x_2)
\models^\tinyherb \exists y_1 \exists y_2 ~.~ x \teq g(y_1,y_2)$ if
and only if $f$ and $g$ are the same function symbol, in which case
the only substitution witnessing the validity of the entailment is
$\theta(x_i) = y_i$, for $i=1,2$.

However, when considering general constraints, matching amounts to
discovering non-trivial substitutions that prove an entailment between
existentially quantified formulae in the logic in which the
constraints are written. Moreover, the matching step implemented by
the $(\RD)$ rule is crucial for proving completeness of the set of
inference rules, by generalizing from the simple case of tree automata
constraints and discovering general properties of the set of
constraints that allow matching to be complete. Such properties are
detailed in \S\ref{sec:restrictions}. 

We introduce a set of universal predicate rules $\mathcal{S}_{\mathsf{univ}}
= \{\langle \{\top\}, p^k_{\mathsf{u}}(x_1, \ldots, x_k) \rangle \mid k
\geq 0 \}$ and assume that any system $\mathcal{S}$ contains it by default.
If, after the cleanup done by applying ($\RI$), there are conjunctions
of predicates $\mathcal{Q}$ on the right hand side such that there exist
sets of subgoals $\vec{x}_i$ for which $\exists p \in \sys \, . \, 
p(\vec{x}_i) \in \lseq$, but $\forall q \in \sys \, . \, q(\vec{x}_i)
\not\in \mathcal{Q}$, then we add $p^{\card{\vec{x}_i}}_{\mathsf{u}}
(\vec{x}_i)$ to $\mathcal{Q}$. While not changing the semantics of the
entailment, this makes sure that, after every application of ($\RD$),
there always are the same number of predicates on the left hand side
and in every set of predicate conjunctions on the right hand side, 
thus enabling the application of ($\SP$).

The $(\SP)$ rules break up a sequent without constraints, of the form
$p_1(\vec{x}_1),\ldots,$ $p_n(\vec{x}_n) \vdash \bigwedge_{j=1}^n
q_j^1(\vec{x}_j), \ldots, \bigwedge_{j=1}^n q_j^k(\vec{x}_j)$, into
$n$ basic sequents, with left-hand sides $p_1(\vec{x}_1), \ldots,
p_n(\vec{x}_n)$, respectively. Given a set of tuples
$\{\overline{\mathcal{Q}}_1,\ldots,\overline{\mathcal{Q}}_k\}
\subseteq \preds^n$, for some $n \geq 1$, a \emph{choice function} $f$
maps each tuple $\overline{\mathcal{Q}}_i$ into an index
$f(\overline{\mathcal{Q}}_i) \in [n]$ corresponding to a given
coordinate in the tuple. Let
$\cf(\overline{\mathcal{Q}}_1,\ldots,\overline{\mathcal{Q}}_k)$ be the
set of such choice functions. This set has cardinality $n^k \leq
n^{\card{\preds}^n}$, for any set of $n$-tuples of predicates. Observe
that $(\SP)$ is applied to each tuple $\bar\imath \in [n]^{n^k}$ of
choices, indexed by the set of choice functions. Intuitively, the
$(\SP)$ inference rules correspond to the split moves of the tree
automata language inclusion algorithm (Example \ref{ex:down-incl}).

The rules ($\AX$) close the current branch of the proof if the sequent
can be proved using a decision procedure for the underlying constraint
logic, by possibly treating predicate symbols as uninterpreted boolean
functions. In the case of tree automata, this is similar to the
matching of constant symbols that leads to pairs with empty left-hand
sides (Example \ref{ex:down-incl}).

Finally, let us explain the infinite descent rules $(\ID)$, which are
the only rules to introduce backlinks in a proof, from the consequent
vertex labeled with a sequent $\lseq\theta \vdash \rseq'\theta$ to a
predecessor (pivot) vertex labeled with $\lseq \vdash \rseq$. The
condition $\RInd^* \cdot \LU \cdot \RInd^*$ requires that a vertex
labeled with a rule $(\LU)$ must occur on the direct path between the
pivot and the consequent. The soundness of $(\ID)$ rules is based on
the following argument. Assuming that the consequent $\lseq\theta
\vdash \rseq'\theta$ denotes a non-valid entailment, there exists a
valuation $\nu \in \mu\sys^\I(\bigwedge\lseq\theta) \setminus
\mu\sys^\I(\bigvee\rseq'\theta)$ which contradicts the
entailment. Then, because $\theta$ is an injective substitution, the
restriction of $\theta$ to $\fv{\lseq \cup \rseq'}$ has an inverse
and, because moreover $\rseq' \subseteq \rseq$, we obtain that $\nu' =
\nu \circ \theta^{-1}$ is a counterexample for the pivot sequent,
i.e. $\nu' \in \mu\sys^\I(\bigwedge\lseq) \setminus
\mu\sys^\I(\bigvee\rseq)$. By the local soundness of the rules in
$\RInd \setminus \set{\ID}$ we deduce that there exists a path in the
proof, on which this counterexample can be propagated downwards. But
this path may not encounter a leaf labeled $(\AX)$, because this would
violate its side condition, and can be potentially extended to an
infinite trace. However, then we could also propagate the
counterexample along a trace with an infinite number of direct paths
(Proposition \ref{prop:direct-path}). Since $(\LU)$ must occur on each
direct path, by the pivot condition, we use an additional
\emph{ranking} assumption (Definition \ref{def:fol-ranked}) to show
that no infinite decreasing sequence of counterexamples exists, hence
there was no counterexample to begin with.

In analogy with tree automata, the language inclusion algorithm of
\cite{HolikLSV11} stops expanding a branch in the search tree whenever
it has discovered a pair $(p,S')$ that has a predecessor $(p,S)$, with
$S \subseteq S'$. Just as for the $(\ID)$ inference rules, backtracking
relies on the Infinite Descent principle \cite{Bussey18}, that forbids
infinitely descending sequences of counterexamples. 

\begin{example}\label{ex:ta}
%
Considering the inductive system from Example \ref{ex:ta-sys}, we can use $\RInd$ to build the following proof for the sequent $p(x)\vdash q(x)$:
\vspace*{-\baselineskip}
\begin{center}
\resizebox{1\textwidth}{!}{%
\begin{tikzpicture}[nd/.style={outer sep=0.15pt}, lbl/.style={outer sep=0.15pt}]
	\node (seq1) [] {$p_1(x) \vdash q_1(x), q_2(x)$};
	\node (seq2) [outer sep = -8pt, above=0pt of seq1]{
		\begin{tikzpicture}
		\node(seq3) [nd] {\small $x\approx a \vdash q_1(x), q_2(x)$};
		\node(seq4) [nd, above=0pt of seq3] {\small $x\approx a \vdash x \approx a, \exists y_1 \,.\, x\approx g(y_1) \land q_1(y_1), q_2(x)$};	
		\node(seq5) [nd, above=0pt of seq4] {\small $\top$};				
		
		\node(seq6) [nd, right=4cm of seq3.east] {\small $x\approx g(x_1), p_1(x_1) \vdash q_1(x), q_2(x)$};
		\node(seq7) [nd, above=0pt of seq6] {\small $x\approx g(x_1), p_1(x_1) \vdash x \approx a, \exists y_1 \,.\, x\approx g(y) \land q_1(y_1), q_2(x)$};
		\node(seq8) [nd, above=0pt of seq7] {\small $\begin{array}{@{}r@{\,}c@{\,}l@{}}x\approx g(x_1), p_1(x_1) & \vdash & x \approx a, \exists y_1 \,.\, x\approx g(y_1) \land q_1(y_1), \\ & & x \approx b, \exists y_1 \,.\,x\approx g(y_1) \land q_2(y_1) \end{array}$};
		\node(seq9) [nd, above=0pt of seq8] {\small $p_1(x) \vdash q_1(x), q_2(x)$};
		\node(seq10) [nd, above=0pt of seq9] {\small $\top$};		
		
		\draw (seq3.south west) -- (seq6.south east);		
		\draw (seq4.south west) -- (seq4.south east);
		\draw (seq4.north west) -- (seq4.north east);
		\draw (seq7.south west) -- (seq7.south east);
		\draw (seq8.south west) -- (seq8.south east);
		\draw (seq8.north west) -- (seq8.north east);
		\draw (seq9.north west) -- (seq9.north east);		

		\node(lbl1) [lbl, left=0pt of seq3.south west] {\scriptsize $\LU$};
		\node(lbl2) [lbl, left=0pt of seq4.south west] {\scriptsize $\RU$};
		\node(lbl3) [lbl, left=0pt of seq4.north west] {\scriptsize $\AX$};		
		\node(lbl4) [lbl, left=0pt of seq7.south west] {\scriptsize $\RU$};
		\node(lbl5) [lbl, left=0pt of seq8.south west] {\scriptsize $\RU$};
		\node(lbl6) [lbl, left=0pt of seq8.north west] {\scriptsize $\RD$};
		\node(lbl7) [lbl, left=0pt of seq9.north west] {\scriptsize $\ID$};				
								
		\end{tikzpicture}		
	};			
	
	\path[dashed, rounded corners=2mm, draw=black, -latex] let \p1=(seq9.east),\p2=(seq1.east) in ([xshift=-3.65cm,yshift=0.35cm]\x1, \y1) -- ([xshift=-1.25cm,yshift=0.35cm]\x1,\y1) -- ([xshift=-1.25cm]\x1,\y2) -- (\x2, \y2);
\end{tikzpicture}}

\vspace{3pt}

\resizebox{0.75\textwidth}{!}{%
\begin{tikzpicture}[nd/.style={outer sep=0.15pt}, lbl/.style={outer sep=0.15pt}]
	\node (seq1) [nd] {\small $p(x) \vdash q(x)$};
	\node (seq2) [nd, above=0pt of seq1] {\small $x\approx f(x_1, x_2), p_1(x_1), p_2(x_2) \vdash q(x)$};
	\node (seq3) [nd, above=0pt of seq2] {\small $\begin{array}{r@{\,}c@{\,}l} x\approx f(x_1, x_2), p_1(x_1), p_2(x_2) & \vdash & \exists y_1, y_2 \,.\, x\approx f(y_1, y_2) \land q_1(y_1) \land q_2(y_2), \\ & & \exists y_1, y_2 \,.\, x\approx f(y_1, y_2) \land q_2(y_1) \land q_1(y_2) \end{array}$};	
	\node (seq4) [nd, above=0pt of seq3] {\small $p_1(x_1), p_2(x_2) \vdash q_1(x_1) \land q_2(x_2), q_2(x_1) \land q_1(x_2)$};
	\node (seq5) [outer sep = -2pt, above=0pt of seq4] {
		\begin{tikzpicture}
		\node (seq51) [] {\small $p_1(x) \vdash q_1(x), q_2(x)$};
		\node (seq52) [right=12pt of seq51] {\small $p_1(x) \vdash q_1(x)$};
		\node (seq53) [right=12pt of seq52] {\small $p_2(x) \vdash q_2(x)$};
		\node (seq54) [right=12pt of seq53] {\small $p_2(x) \vdash q_2(x), q_1(x)$};
		\end{tikzpicture}
	};
	
	\draw (seq2.south west) -- (seq2.south east);
	\draw (seq3.south west) -- (seq3.south east);
	\draw (seq3.north west) -- (seq3.north east);	
	\draw (seq5.south west) -- (seq5.south east);
	
	\node(lbl1) [lbl, left=0pt of seq2.south west] {\textsc{lu}};
	\node(lbl2) [lbl, left=0pt of seq3.south west] {\textsc{ru}};
	\node(lbl3) [lbl, left=0pt of seq3.north west] {\textsc{rd}};
	\node(lbl4) [lbl, left=0pt of seq5.south west] {\textsc{sp}};	
\end{tikzpicture}}
\end{center}
The dashed arrow indicates the pivot of the $\ID$ rule. For space
reasons, some branches following the application of ($\SP$) are
omitted. The full proof is provided as additional material. \hfill\qed
\end{example}

\subsection{Restricting the Set of Constraints}
\label{sec:restrictions}

The following definitions introduce sufficient conditions that ensure
the soundness and completeness of the set of inference rules given in
\S\ref{sec:inference-rules}. These definitions are not
bound to a particular logic or interpretation, and will be extended to
logics other than multisorted first-order logic, such as Separation
Logic \cite{Reynolds02} (\S\ref{sec:sl}).

Effectively checking these restrictions
for a given system is subject to the existence of a decision procedure
for the $\exists^*\forall^*$ fragment of the logic in which the
constraints are written. Formally, this is the set of prenex normal
form sentences of the form $\exists x_1 \ldots \exists x_n \forall y_1
\ldots \forall y_m ~.~ \phi$, where $\phi$ is quantifier-free. For
First Order Logic with the canonical interpretation, this problem,
known as disunification, has been shown decidable in
\cite{ComonLescanne89}, with tighter complexity bounds given in
\cite{Pichler03}.

The first restriction requires that, given any models for the subgoals
of a predicate rule, it be possible to find an all-encompassing model
that also satisfies the constraint of the rule. This restriction is
necessary because lifting it leads, in general, to the undecidability
of the entailment problem, as it is the case for tree automata with
equality and disequality constraints\footnote{See e.g.\ \cite[Theorem
    4.2.10]{Tata05}.}.

\begin{definition}\label{def:fol-non-filtering}
    Given an interpretation $\I$, a first order inductive system $\sys$ is
    \emph{non-filtering} iff, for every $\langle\{\phi, q_1(\vec{x}_1),
    \ldots, q_n(\vec{x}_n)\}, p(\vec{x})\rangle \in\sys$, for all $i \in
    [n]$ and $\overline{v}_i \in \mu\sys^\I(q_i)$, there exists a
    valuation $\nu$ such that $\nu(\vec{x}_i) = \overline{v}_i$ and
    $\I, \nu \models \phi$. An inductive system is non-filtering if
    all predicate rules are non-filtering.
\end{definition}

\begin{example}\label{ex:fol-non-filtering}
    The system in Example \ref{ex:fol-sys} is non-filtering. If we added the
    rule $\tuple{\set{x \teq f(x_1, x_2)\wedge x_1 \teq x_2, 
            p_1(x_1), p_2(x_2)},p(x)}$, we would break this restriction, because it rejects all 
    subgoals models assigning different values to $x_1$ and $x_2$. \hfill\qed
\end{example}

\begin{lemma}\label{lemma:fol-non-filtering}
    The problem ``Given a first order inductive system $\sys$, is
    $\sys$ non-filtering?'' is undecidable in the Herbrand interpretation.
\end{lemma}

Due to this negative result, we adopt a stronger (sufficient)
condition, which requires that $\forall \vec{x}_1 \ldots \forall
\vec{x}_n \exists \vec{x} \,.\,
\phi(\vec{x},\vec{x}_1,\ldots,\vec{x}_n)$ holds, for each constraint
$\phi$ in the system. In the canonical Herbrand
interpretation, the latter problem becomes decidable, because each
constraint $\phi$ is a conjunction of equalities $s \teq t$ and
disequalities $\neg s \teq t$ between terms over the variables
$\vec{x} \cup \bigcup_{i=1}^n \vec{x}_i$.
Establishing the validity of $\forall \vec{x}_1 \ldots \forall
\vec{x}_n \exists \vec{x} \,.\, \phi(\vec{x},\vec{x}_1,\ldots,\vec{x}_n)$
reduces to checking that $\exists \vec{x}_1 \ldots \exists \vec{x}_n
\forall \vec{x} \,.\, \\ \neg\phi(\vec{x},\vec{x}_1,\ldots,
\vec{x}_n)$ is unsatisfiable. Because we assumed that
constraints do not contain disjunctions, $\neg\phi$ is a disjunction
of equalities and disequalities, thus it is trivially in conjunctive
normal form. Since the satisfiability of the formulae $\exists \vec{x}
\forall \vec{y} \,.\, \phi(\vec{x},\vec{y})$, with $\phi$ in conjunctive
normal form, is \np-complete\footnote{See \cite[Theorem
5.2]{Pichler03}.}, our validity problem is in co-\np.

The second restriction guarantees that the principle of Infinite
Descent \cite{Bussey18} can be applied to close a branch of the proof
tree. We fix an interpretation $\I$ and assume that
$(\sigma^\I,\preceq_{\I,\sigma})$ is a wfqo, for each sort $\sigma \in
\ssorts{\Sigma}$. Given a valuation $\nu$ and a set of variables
$\vec{x} \subseteq \vars$, we denote by $\sem{\vec{x}}_{\nu}$ the
multiset $\sem{ \nu(x) \mid x \in \vec{x} }$. For two multisets
$\sem{\vec{x}}_{\nu}$ and $\sem{\vec{y}}_{\mu}$, where $\nu$ and $\mu$
are two valuations, we write $\sem{\vec{x}}_{\nu}
\preceq^{\forall\exists}_\I \sem{\vec{y}}_{\mu}$ (respectively
$\sem{\vec{x}}_{\nu} \prec^{\forall\exists}_\I \sem{\vec{y}}_{\mu}$)
iff for each variable $x^\sigma \in \vec{x}$ there exists a variable
$y^\sigma \in \vec{y}$, of the same sort $\sigma$, such that $\nu(x)
\preceq_{\I,\sigma} \mu(y)$ (respectively $\nu(x) \prec_{\I,\sigma}
\mu(y)$). It is easy to see that, in every chain
$\sem{\vec{x}_1}_{\nu_1} \sim \sem{\vec{x}_2}_{\nu_2} \sim \ldots \sim
\sem{\vec{x}_k}_{\nu_k}$, where $\sim$ is either
$\succeq^{\forall\exists}_\I$ or $\succ^{\forall\exists}_\I$, we have
$\sem{\vec{x}_1}_{\nu_1} \succ^{\forall\exists}_\I
\sem{\vec{x}_k}_{\nu_k}$ iff $\sem{\vec{x}_i}_{\nu_i}
\succ^{\forall\exists}_\I \sem{\vec{x}_{i+1}}_{\nu_{i+1}}$ for at
least some $i \in [k-1]$.

\begin{proposition}\label{prop:forall-exists-wqo}
    Given a signature $\Sigma$ and an interpretation $\I$, let $S =
    \bigcup_{\sigma \in \ssorts{\Sigma}} \sigma^\I$ be the set of values
    in the interpretation of each sort in $\Sigma$. Then
    $(\mathcal{M}(S),\prec^{\forall\exists}_\I)$ is a wfqo provided that
    $(\sigma^\I,\preceq_{\I,\sigma})$ is a wfqo, for each $\sigma \in
    \ssorts{\Sigma}$.
\end{proposition}

\begin{definition}\label{def:fol-ranked}
    A first order inductive system $\sys$ is \emph{ranked in the
        interpretation $\I$} iff, for every constraint
    $\phi(\vec{x},\vec{x}_1,\ldots,\vec{x}_n)$ in $\sys$, with goal
    variables $\vec{x}$ and subgoal variables
    $\bigcup_{i=1}^n\vec{x}_i$, and every valuation $\nu$, such that
    $\I,\nu \models \phi$, we have $\sem{\bigcup_{i=1}^n\vec{x}_i}_{\nu}
    \prec^{\forall\exists}_\I \sem{\vec{x}}_{\nu}$.
\end{definition}

\begin{example}\label{ex:fol-ranked}
  The system from Example \ref{ex:fol-sys} is ranked, because
  the only constraints involving subgoal variables are\begin{inparaenum}[(i)]
  \item\label{it1:fo-ranked-ex} $x \teq f(x_1,x_2)$ and 
  \item\label{it2:fo-ranked-ex} $x \teq g(x_1)$ 
  \end{inparaenum} and for each valuation $\nu$ we have
  $\nu(x_1) \sqsubseteq \nu(x)$ and $\nu(x_2) \sqsubseteq \nu(x)$, if
  $\nu$ satisfies the constraint (\ref{it1:fo-ranked-ex}) and
  $\nu(x_1) \sqsubseteq \nu(x)$, if $\nu$ satisfies the constraint
  (\ref{it2:fo-ranked-ex}), where $\sqsubseteq$ is the subtree relation
  described in \S\ref{sec:herb}. \hfill\qed
\end{example}

For the Herbrand interpretation, we consider the natural subtree order
on ground terms $(\mathcal{T}_\Sigma, \subtreeq)$ and obtain the
following:

\begin{lemma}\label{lemma:fol-ranked}
    The problem ``Given a first order inductive system $\sys$, is $\sys$
    ranked in the Herbrand interpretation'' is in co-\np.
\end{lemma}

Considering again systems whose constraints are interpreted in the
canonical Herbrand interpretation, it is natural to ask whether a
given system is ranked in the subtree order
$(\T_\Sigma,\sqsubseteq)$. Since the satisfiability of the
quantifier-free fragment of the first order logic with a binary
relation symbol interpreted as the subterm relation is an \np-complete
problem \cite{Venkataraman87}, one can effectively decide if a given
system is ranked. For each constraint
$\phi(\vec{x},\vec{x}_1,\ldots,\vec{x}_n)$ we check if the formula
$\phi \wedge \bigvee_{y \in \vec{x}_1 \cup \ldots \cup \vec{x}_n}
\bigwedge_{x \in \vec{x}} \left(y \teq x \vee \neg y \sqsubseteq
x\right)$ is unsatisfiable, where $\sqsubseteq$ is interpreted as the
subtree order. Since the size of the latter formula is polynomially
bounded in the size of $\phi$, the problem of checking if a given
system is ranked is in co-\np.

The third restriction guarantees that all constraints can be
eliminated from a sequent, by instantiating the subgoal variables on
the right-hand side using finitely many substitutions mapping to
the subgoal variables from the left-hand side. Given a sequent
Given a sequent
$\phi(\vec{x}, \vec{x}_1, \ldots, \vec{x}_n),p_1(\vec{x}_1),\ldots, p_n(\vec{x}_n) \vdash \exists \vec{y}_1
\ldots \exists \vec{y}_m \,.\, \psi(\vec{x}, \vec{y}_1, \\ \ldots, \vec{y}_m) \wedge q_1(\vec{y}_1) \wedge \ldots
\wedge q_m(\vec{y}_m)$, if the entailment $\phi \models^\I \exists
\vec{y}_1 \ldots \exists \vec{y}_m \,.\, \psi$ is valid, then we can
replace it with $p_1(\vec{x}_1), \ldots, p_n(\vec{x}_n)
\vdash \set{q_1(\vec{y}_1\theta) \wedge \ldots \wedge q_m(\vec{y}_m\theta) \mid \theta \in
    S}$, where $\phi \models^\I \psi\theta$, for each $\theta \in
S$. This elimination of constraints from sequents is generally sound
but incomplete. The above entailment is valid iff
\(\phi(\vec{x},\vec{x}_1,\ldots,\vec{x}_n) \models^\I
\psi'(\vec{x},\vec{x}_1,\ldots,\vec{x}_m)\), where $\psi'$ is obtained
from $\psi$ by replacing each $y \in \vec{y}_1 \cup \ldots \cup
\vec{y}_m$ with a Skolem function symbol
$f_y(\vec{x},\vec{x}_1,\ldots,\vec{x}_n)$ not occurring in $\phi$ or
$\psi$\footnote{We assume w.l.o.g. that these function symbols belong
to the signature, i.e.\ $f_y \in \sfuns{\Sigma}$.}. A complete proof
rule based on this replacement has to consider all possible
interpretations of these Skolem witnesses. This is
impossible in general, as their definitions are not bound
to any particular form. For completeness, we require
that these functions are defined as flat substitutions ranging
over $\vec{x} \cup \bigcup_{i=1}^n\vec{x}_i$. This condition
ensures moreover that there are finitely many possible interpretations
of the Skolem witnesses.

\begin{definition}\label{def:fol-fvi}
    An inductive system $\sys$ has the
    \emph{finite instantiation} property iff for any two constraints
    $\phi(\vec{x},\vec{x}_1,\ldots,\vec{x}_n)$ and
    $\psi(\vec{x},\vec{y}_1,\ldots, \vec{y}_m)$ from $\sys$, with goal
    variables $\vec{x}$ and subgoal variables $\bigcup_{i=1}^n
    \vec{x}_i$ and $\bigcup_{j=1}^m \vec{y}_j$, respectively, the set
    $\subst{\phi,\psi} = \{\theta : \bigcup_{i=1}^m\vec{y}_i \rightarrow
    \T_\Sigma(\vec{x},\vec{x}_1,\ldots,\vec{x}_n) \mid \phi \models^\I
    \psi\theta\}$
    is finite. Moreover, $\sys$ has the \emph{finite
        variable instantiation (fvi)} property iff for all $i \in [m]$
    there exists $j \in [n]$ such that $\vec{y}_i\theta = \vec{x}_j$,
    for each $\theta \in \subst{\phi,\psi}$.
\end{definition}

\begin{example}\label{ex:fol-fvi}
    Consider the constraints $\phi \equiv x \teq f(x_1,x_2)$ and $\psi
    \equiv x \teq f(y_1,y_2)$. Then $\phi \models^\tinyherb \exists y_1
    \exists y_2 \,.\, \psi \Leftrightarrow \phi \models^\tinyherb \psi\theta$, where
    $\theta(x_1) = y_1$ and $\theta(x_2) = y_2$, i.e.\ $\subst{\phi,\psi}
    = \set{\theta}$. \hfill\qed
\end{example}

Observe that, whenever $\sys$ has the fvi property, a constraint with
no subgoal variables may not entail a constraint with more than one
subgoal variables. If $\sys$ has the fvi property, $\phi(\vec{x})
\models^\I \exists \vec{y}_1 \ldots \exists \vec{y}_m \,.\,
\psi(\vec{x},\vec{y}_1,\ldots,\vec{y}_m)$ and $m>0$ imply
$\subst{\phi,\psi} \neq \emptyset$. But then each flat substitution
$\theta \in \subst{\phi,\psi}$ would have an empty range, which is not
possible.

Below we give an upper bound for the complexity of the problem whether
a given inductive system has the fvi property, in the canonical Herbrand
interpretation of constraints. It is unclear, for now, whether the
bound below can be tightened, because the exact complexity of the
satisfiability of equational problems is still unknown, in
general\footnote{Converting a formula into CNF requires exponential
    time at most, thus \nexptime~ is an upper bound.}.

\begin{lemma}\label{lemma:fol-fvi}
    The problem ``Given a first order inductive system $\sys$,
    does $\sys$ have the fvi property?'' is in \nexptime~ in the Herbrand
    interpretation. If there exists a constant $K > 0$, independent of the
    input, such that for each constraint $\phi(\vec{x},\vec{x}_1,\ldots,\vec{x}_n)$,
    with goal variables $\vec{x}$ and subgoal variables $\bigcup_{i=1}^n \vec{x}_i$,
    respectively, we have $\card{\vec{x}_i} \leq K$, then the problem is in \np.
\end{lemma}

The last condition required for completeness is also related to the
elimination of constraints from sequents. Intuitively, we do not
allow two constraints to overlap, having at least one model in common,
without one entailing the other. 

\begin{definition}\label{def:non-overlapping}
    An inductive system $\sys$ is \emph{non-overlapping} iff,
    for any two constraints $\phi(\vec{x},\vec{x}_1,\ldots,\vec{x}_n)$ and
    $\psi(\vec{x},\vec{y}_1,\ldots,\vec{y}_m)$ in $\sys$, with goal
    variables $\vec{x}$ and subgoal variables $\bigcup_{i=1}^n \vec{x}_i$
    and $\bigcup_{j=1}^m \vec{y}_j$ respectively, $\phi \wedge \psi$ is
    satisfiable only if $\phi \models^\I \exists \vec{y}_1 \ldots \exists
    \vec{y}_m \,.\, \psi$ for a given $\I$.
\end{definition}

\begin{example}\label{ex:fol-non-overlapping}
    The system from Example \ref{ex:fol-sys} is non-overlapping,
    because, for instance, $x \teq f(x_1,x_2) \wedge x \teq f(y_1,y_2)$ is
    satisfiable and $x \teq f(x_1,x_2) \models^\tinyherb \exists y_1
    \exists y_2 \,.\, x \teq f(y_1,y_2)$, whereas $x \teq f(x_1,x_2) \wedge x
    \teq g(y_1)$ is unsatisfiable. \hfill\qed
\end{example}

\noindent For a non-overlapping system, if
$\phi(\vec{x},\vec{x}_1,\ldots,\vec{x}_n) \wedge
\psi(\vec{x},\vec{y}_1,\ldots,\vec{y}_m)$ is a satisfiable conjunction
of constraints, then the formulae $\exists \vec{x}_1 \ldots \exists
\vec{x}_n \,.\, \phi$ and $\exists \vec{y}_1 \ldots \vec{y}_m \,.\,
\psi$ are equivalent.

\begin{lemma}\label{lemma:fol-non-overlapping}
    The problem ``Given a first order inductive system $\sys$, is
    $\sys$ non-overlapping?'' is in \np~ in the Herbrand interpretation.
\end{lemma}

\subsection{Soundness and Completeness}
\label{sec:soundness-completeness}

In this section we address the problems of soundness and completeness
of the set $\RInd$ of inference rules (Definition
\ref{sec:soundness-completeness}) and show that $\RInd$ is sound for
entailments in a given ranked system $\sys$, whereas completeness is
guaranteed if, moreover, $\sys$ is non-filtering, non-overlapping and
has the fvi property. We point out that the soundness and completeness
are independent of any particular interpretation of First Order Logic
and depend only on the restrictions from \S\ref{sec:restrictions}.

\vspace{5pt}
\paragraph{Soundness}
Here we develop the argument for soundness for the $\RInd$ set of
inference rules. First, we prove that each rule from $\RInd \setminus
\set{\ID}$ is \emph{locally sound} meaning that, if the consequent
$\lseq \vdash \rseq$ denotes an invalid entailment,
i.e. $\mu\sys^\I(\bigwedge\lseq) \not\subseteq \mu\sys^\I(\bigvee\rseq)$
then at least one of its antecedents $\lseq_i \vdash \rseq_i$ denotes
an invalid entailment. Moreover, we can relate the counterexample of
the consequent to the corresponding counterexample of the antecedent
by a wfqo.

\begin{lemma}\label{lemma:local-soundness}
    Given an inductive system $\sys$ that is ranked in the interpretation
    $\I$, for each instance of a schema $\IR \in
    \RInd\setminus\set{\ID}$, with consequent $\lseq\vdash\rseq$ and
    antecedents $\set{\lseq_i\vdash\rseq_i}_{i=1}^n$ and each valuation
    $\nu \in \mu\sys^\I(\bigwedge\lseq) \setminus
    \mu\sys^\I(\bigvee\rseq)$, there exists a valuation $\nu_i \in
    \mu\sys^\I(\bigwedge\lseq_i) \setminus \mu\sys^\I(\bigvee\rseq_i)$
    such that $\sem{\fv{\lseq_i}}_{\nu_i} \preceq^{\forall\exists}_\I
    \sem{\fv{\lseq}}_{\nu}$, for some $i \in [n]$.
\end{lemma}

Local soundness allows to define a reachability relation between
countermodels and we write $\nu \nextval \nu'$ if $\nu$ is a
counterexample of the consequent and $\nu'$ is the counterexample of
the antecedent, as in the statement of Lemma
\ref{lemma:local-soundness}.

\begin{definition}\label{def:counterexample-path}
    A path $v_1,v_2,\ldots,v_k$ in a proof $\mathcal{D} = (V,v_0,S,R,P)$
    built with $\RInd$ is a \emph{counterexample path} if there exists a sequence of
    valuations $\nu_1,\nu_2,\ldots,\nu_k$ such that, for all $i\in[k]$ we
    have: \begin{inparaenum}[(i)]
        \item $\nu_i \in \mu\sys^\I(\bigwedge\lseq_i) \setminus
        \mu\sys^\I(\bigvee\rseq_i)$, where $S(v_i)=\lseq_i \vdash \rseq_i$, and
        \item if $i < k$ then $\nu_i \nextval \nu_{i+1}$. 
    \end{inparaenum}
\end{definition}

Next, we prove that any direct counterexample path in a proof causes a
strict descent of multisets in the multiset ordering, provided that
the system is ranked. 

\begin{lemma}\label{lemma:direct-counterexample-path}
    Given a system $\sys$ and an interpretation $\I$ such that $\sys$ is
    ranked in $\I$, let $\mathcal{D} = (V,v_0,S,R,P)$ be a proof and let
    $\pi = v_1,\ldots,v_k$ be a direct counterexample path in
    $\mathcal{D}$, with valuations $\nu_1,\ldots,\nu_k$, for a backlink
    $(v_k,v_1)$. Then $\sem{\fv{\lseq_1}}_{\nu_1}
    \succ^{\forall\exists}_\I \sem{\fv{\lseq_k}}_{\nu_k}$, where $S(v_i)
    = \lseq_i \vdash \rseq_i$, for all $i \in [k]$.
\end{lemma}

Next, we extend the reachability relation $\nextval$ to backlinks and
show that, in any infinite trace in a proof there exists an infinite
strictly descending sequence of multisets associated to countermodels,
which clearly contradicts the well-foundedness of the interpretation
domain on which the ranking assumption is based (Definition
\ref{def:fol-ranked}). This allows us to conclude that, in each proof
using the rules $\RInd$, there cannot be a counterexample for the
sequent labeling the root.

\begin{theorem}\label{thm:ta-soundness}
    Given a ranked inductive system $\sys$, if a sequent $\lseq \vdash
    \rseq$ has a proof $\mathcal{D}=(V,v_0,S,R,P)$ built with $\RInd$,
    and $S(v_0) = \lseq \vdash \rseq$, then the entailment
    $\bigwedge\lseq \models_\sys^\I \bigvee\rseq$ holds.
\end{theorem}

\paragraph{Completeness} We prove that the set of inference rules
$\RInd$ is complete for entailments between predicates in
inductive systems that are ranked, non-filtering, non-overlapping and
have the fvi property (\S\ref{sec:restrictions}). A derivation is said to be
\emph{maximal} if it cannot be extended by an application of an
inference rule, and \emph{irreducible} if it cannot be rewritten into
a smaller derivation of the same sequent by an $(\ID)$ application.
Note that the proof search semi-algorithm 
(\ref{alg:proof-search}) generates only irreducible derivations,
because $(\ID)$ is always applied before any other inference rules.

A derivation $\mathcal{D}$ is \emph{structured} if, on each path of
$\mathcal{D}$, between any two consecutive applications of $(\LU)$
there exists an application of $(\RD)$.
Intuitively, unstructured derivations constitute poor candidates
for proofs. For instance, a derivation consisting only of applications
of $(\LU)$ rules will only grow the size of the left-hand sides of the
sequents, without making progress towards $\top$ or a
counterexample. Observe that each subtree of a structured derivation
is also structured. We denote by $\DS(\lseq \vdash \rseq)$ the set of
irreducible, maximal and structured derivations rooted in $\lseq
\vdash \rseq$.

\begin{lemma}\label{lemma:max-irred-finite}
    If the inductive system $\sys$ has the fvi property, then the following
    properties of derivations built with $\RIndsl$ hold: \begin{compactenum}
        \item\label{it1:max-irred-finite} any irreducible and structured
        derivation is finite, and
        \item\label{it2:max-irred-finite} for any sequent $\lseq \vdash
        \rseq$, the set $\DS(\lseq \vdash \rseq)$ is finite.
    \end{compactenum}
\end{lemma}

\begin{definition}\label{def:tree-shaped}
A set $F = \set{\phi_1, \ldots,
  \phi_n,q_1(\vec{x}_1),\ldots,q_m(\vec{x}_m)}$ is \emph{tree-shaped}
if and only if $\phi_1, \ldots, \phi_n$ are constraints,
$q_1(\vec{x}_1), \ldots q_m(\vec{x}_m)$ are predicate atoms, and there
exist trees $t_1,\ldots,t_k$ such that:
\begin{compactitem}
\item each node labeled with a constraint
  $\phi_i(\vec{y},\vec{y}_1,\ldots,\vec{y}_n)$ in some tree $t_\ell$,
  $\ell\in[k]$ has exactly $n$ children and for all $j \in [n]$,
  the $j$-th child is labeled either \begin{inparaenum}[(i)]
  \item with a constraint whose goal variables are $\vec{y}_j$, or
  \item with a predicate atom $q_k(\vec{y}_j)$, and
  \end{inparaenum}
\item a predicate atom $q_i(\vec{x}_i)$ may occur only on the frontier
  of a tree $t_j$, for some $j\in[k]$.
\end{compactitem}
If $k=1$ we say that $F$ is \emph{singly-tree shaped}.
\end{definition}
Tree-shaped sets can be uniquely represented by trees labeled with
formulae, thus we use sets of trees instead of sets of formulae
interchangeably. We write $\lseq \vdash \rseq \leadsto \lseq' \vdash
\rseq'$ iff $\lseq' \vdash \rseq'$ occurs in a derivation from
$\DS(\lseq \vdash \rseq)$. Next, we prove an invariant on the shape of
the sequents occurring in a proof of a basic sequent $p(\vec{x})
\vdash q(\vec{x})$.

\begin{lemma}\label{lemma:tree-shaped}
  Given a system $\sys$ and predicates $p, q_1, \ldots, q_n$,
  in every sequent $\lseq \vdash \rseq$ such that
  $p(\vec{x}) \vdash q_1(\vec{x}), \ldots, q_n(\vec{x}) \leadsto \lseq \vdash \rseq$, $\lseq$
  is a tree-shaped set and $\rseq$ consists of finite conjunctions of
  tree-shaped sets, with all subgoal variables existentially
  quantified.
\end{lemma}

The following lemma characterizes the cases in which the root of a
derivation is an invalid entailment, from which a counterexample can
be extracted. This is crucial in establishing our completeness result
(Theorem \ref{thm:ta-completeness}). We write $\lseq \vdash \rseq
\leadsto \lseq' \vdash \rseq'$ iff $\lseq' \vdash \rseq'$ occurs in a
derivation from $\DS(\lseq \vdash \rseq)$.

\begin{lemma}\label{lemma:cex}
    Given an interpretation $\I$, a non-filtering and non-overlapping
    inductive system $\sys$ with the fvi property, the predicate atoms
    $p(\vec{x}), q_1(\vec{x}), \ldots, q_n(\vec{x})$, and a sequent $\lseq
    \vdash \rseq$ such that $p(\vec{x}) \vdash q_1(\vec{x}), \ldots,
    q_n(\vec{x}) \leadsto \lseq \vdash \rseq$, if every derivation
    $\mathcal{D} \in \DS(\lseq \vdash \rseq)$ contains a leaf $\lseq'
    \vdash \emptyset$ then there exists a valuation $\nu \in
    \mu\sys^\I(\bigwedge\lseq) \setminus \mu\sys^\I(\bigvee\rseq)$.
\end{lemma}

The following theorem proves that $\RInd$ is complete in
the canonical interpretation, and provides a proof search strategy.

\begin{theorem}\label{thm:ta-completeness}
    Given an interpretation $\I$ and a non-filtering, non-overlapping
    inductive system $\sys$ that is ranked in the interpretation $\I$
    and has the fvi property, let $p, q_1, \ldots,
    q_n$ be predicates occurring in $\sys$. Then the entailment $p
    \models_\sys^\I q_1, \ldots, q_n$ holds only if the sequent
    $p(\vec{x}) \vdash q_1(\vec{x}), \ldots, q_n(\vec{x})$ has an
    $\mathbf{S}$-proof with the set of inference rules $\RInd$, where
    $\mathbf{S}$ is defined by the regular expression $(\LU \cdot \RU^*
    \cdot \RD \cdot \RI^* \cdot \SP?)^* \cdot \LU? \cdot \RU^* \cdot
    (\AX \mid \ID)$.
\end{theorem}

The proof search semi-algorithm (\ref{alg:proof-search})
from \S \ref{sec:proof-search} only explores
irreducible derivations. If executed with the
strategy $\mathbf{S}$ from Theorem \ref{thm:ta-completeness}, 
these derivations are also structured. By Lemma
\ref{lemma:max-irred-finite} (\ref{it1:max-irred-finite}),
irreducible and structured derivations are finite, thus every execution
of the semi-algorithm is guaranteed to terminate. If, moreover, the
input inductive system $\sys$ is ranked, non-filtering,
non-overlapping and has the fvi property, then
$\RInd$ is complete and semi-algorithm (\ref{alg:proof-search})
becomes a decision procedure for this class of entailment problems.

\section{Inductive Predicate Entailments in Separation Logic}
\label{sec:sl}

In this section we apply the method described in \S
\ref{sec:canonical-proof-system} to deciding entailments between
predicates whose defining rules use constraints from a fragment of
Separation Logic \cite{Reynolds02}. These predicates are common for
specifications of recursive data structures implemented using
pointers, thus having complete sets of proof rules for these systems
is important for obtaining decision procedures that solve verification
conditions generated by program analysis tools. Using a similar
approach as for first order logic, we give a set of inference rules
and prove completeness under a number of (decidable) restrictions on
the set of constraints that occur in the system.

Throughout this section, we consider a signature $\Sigma$, such that
$\ssorts{\Sigma} = \set{\locs,\Bool}$ and $\sfuns{\Sigma} =
\emptyset$, i.e.\ the only sorts are the boolean and \emph{location}
sort, with no function symbols defined on it, other than
equality. Observe that, in this case $\T_\Sigma (\vec{x}) = \vec{x}$,
for any $\vec{x} \subseteq \vars$, i.e.\ the only terms occurring in a
formula are variables of sort $\locs$. In the rest of this section we
consider systems whose constraints are Separation Logic ($\seplog$)
formulae, generated by the following syntax:
\[\varphi \Coloneqq \bot \mid x \teq y \mid \emp \mid x \mapsto (y_1, \ldots, y_k) \mid 
\varphi_1 * \varphi_2 \mid \neg\varphi_1 \mid \varphi_1 \wedge
\varphi_2 \mid \exists x \,.\, \varphi_1\] where $k > 0$ is a fixed
constant. As usual, we consider that the constraints of an inductive
system are quantifier-free $\seplog$ formulae in the above
fragment. For a set of formulae $F =
\set{\varphi_1,\ldots,\varphi_n}$, we write $\Asterisk F$ for
$\varphi_1 * \ldots * \varphi_n$ if $F \neq \emptyset$, and
$\emp$ if $F = \emptyset$. The size of a formula is the
number of variables and connectives occurring in it. The size of a
system is the sum of the sizes of its constraints.

Most definitions of common recursive data structures employed by
programers (e.g.\ lists, trees, etc.) use a restricted fragment of
quantifier-free $\seplog$, consisting of formulae $\symhp \wedge
\symhs$, called \emph{symbolic heaps}, in the following syntax, for
\emph{pure} ($\symhp$) and \emph{spatial} ($\symhs$) formulae defined
as follows:
\[\begin{array}{rclcrcl}
\symhp & \Coloneqq & x \teq y \mid \neg x \teq y \mid \symhp_1 \land \symhp_2 & \hspace*{0.5cm} &
\symhs & \Coloneqq & \emp \mid x \mapsto (y_1, \ldots, y_k) \mid \symhs_1 * \symhs_2
\end{array}\]

In the rest of this section, we fix an interpretation $\I$ such that
$\I(\locs) = \locsi$ is a countably infinite set and omit to specify
$\I$ any further. A \emph{heap} is a finite partial mapping $h :
\locsi \rightharpoonup_{\mathsf{fin}} \locsi^{k}$ associating
locations with $k$-tuples of locations. We denote by $\dom(h)$ the set
of locations on which $h$ is defined, by $\img(h)$ the set of
locations occurring in the range of $h$, and by $\heaps$ the set of
heaps. Two heaps $h_1$ and $h_2$ are disjoint if $\dom(h_1) \cap
\dom(h_2) = \emptyset$. In this case $h_1 \uplus h_2$ denotes their
union, which is undefined if $h_1$ and $h_2$ are not disjoint. Given a
valuation $\nu : \vars \rightarrow \locsi$ and a heap $h$, the
semantics of $\seplog$ formulae is defined as:
\[\begin{array}{lcl}
\nu,h \models^{\tinyseplog} x \teq y & \iff & \nu(x) = \nu(y) \\
\nu,h \models^{\tinyseplog} \emp & \iff & h = \emptyset \\
\nu,h \models^{\tinyseplog} x \mapsto (y_1,\ldots,y_k) & \iff & h = \{\tuple{\nu(x),(\nu(y_1), \ldots, \nu(y_k))}\} \\
\nu,h \models^{\tinyseplog} \phi_1 * \phi_2 & \iff & \exists h_1,h_2 \in \heaps \,.\,
 h=h_1\uplus h_2 \text{ and } \nu,h_i \models^{\tinyseplog} \phi_i, i \in [2] \\
\nu,h \models^{\tinyseplog} \exists x \,.\, \varphi(x) & \iff & 
\nu[x \leftarrow \ell],h \models^{\tinyseplog} \varphi(x) \text{, for some } \ell \in \locsi
\end{array}\]
The semantics of boolean connectives is the usual one, omitted for
brevity.

An assignment $\X$ maps each predicate $p(x_1,\ldots,x_n)$ to a set
$\X(p) \subseteq \locsi^n \times \heaps$. For a set $F =
\{\phi,p_1(\vec{x}_1), \ldots, p_m(\vec{x}_m)\}$, where $\phi$ is a
$\seplog$ formula and $p_1(\vec{x}_1), \ldots, p_m(\vec{x}_m) \in
\preds$, we define $\X(\Asterisk F) = \{(\nu,h_0\uplus
\biguplus_{i=1}^m h_i) \mid \nu,h_0 \models^{\tinyseplog} \phi,~
(\nu(\vec{x}_j),h_j) \in \X(p_j),~ j \in [m]\}$. The least solution
$\mu\sys^\tinyseplog$ of a system $\sys$ is the least fixed point of the
function \(\mathbb{F}^\tinyseplog_\sys\), where
\(\mathbb{F}^\tinyseplog_\sys(\X)\) maps each predicate $p(\vec{x})
\in \preds$, such that $p(\vec{x}) \leftarrow_\sys R_{1} \mid \ldots
\mid R_{m}$, into the set $\bigcup_{i=1}^m \{(\nu(\vec{x}),h) \mid
(\nu,h) \in \X(\Asterisk R_i)\}$. Observe that the heaps from the
subgoal assignments are separately joined with a heap satisfying the
constraint of the rule to obtain a heap for the goal. In this section
we consider entailments between predicates $p \models^\tinyseplog q_1, \ldots, q_n$
if and only if $\mu\sys^\tinyseplog(p) \subseteq
\bigcup_{i=1}^n\mu\sys^\tinyseplog(q_i)$. As before, we extend this notation to
$\seplog$ formulae and write $\phi \models^\tinyseplog \psi$ for
$\mu\sys^\tinyseplog(\phi) \subseteq \mu\sys^\tinyseplog(\psi)$,
where, for an arbitrary $\seplog$ formula $\varphi$,
$\mu\sys^\tinyseplog(\varphi)$ is defined recursively on its
structure. 

\begin{example}\label{ex:sl-system}
Consider the following system, with symbolic heap constraints: 
\begin{center}
\(\begin{array}{@{\;}l@{\;}c@{\;}l@{\;}}
	ls^+(x, y) & \leftarrow_\sys & x \mapsto y \mid y \teq y' \land x \mapsto z, ls^+(z, y') \\[1pt]
	ls^e(x, y) & \leftarrow_\sys & x \teq y \land \emp \mid y \teq y' \land x \mapsto z, ls^o(z, y')\\[1pt]
	ls^o(x, y) & \leftarrow_\sys & x \mapsto y \mid y \teq y' \land x \mapsto z, ls^e(z, y') \\[1pt]
	\widehat{ls}^+(x, y) & \leftarrow_\sys & y \teq y' \land x \mapsto z * ls^e(z,y') \mid y \teq y' \land x \mapsto z *ls^o(z,y')
\end{array}\)
\end{center}
Intuitively, $ls^+(x,y)$ defines the set of finite list segments of at
least one element between $x$ and $y$, $ls^e$ and $ls^o$ are list
segments of even and odd length, respectively, and $\widehat{ls}^+(x,
y)$ is the definition of a list segment consisting of one element
followed by an even or an odd list segment. It is immediate to see
that both entailments $ls^+ \models^\tinyseplog \widehat{ls}^+$ and
$\widehat{ls}^+ \models^\tinyseplog ls^+$ hold. \hfill\qed
\end{example}

The following negative result \cite{Atva14,AGHKO14} justifies a number
of restrictions on the set of $\seplog$ constraints occurring in a
system\footnote{See e.g.\ \cite[Theorem 2]{Atva14} and \cite[Theorem
3]{AGHKO14}.} (\S\ref{sec:sl-restrictions}).

\begin{theorem}\label{thm:sl-entailment-undec}
  The entailment problem is undecidable for systems with symbolic heap
  constraints.
\end{theorem}

\subsection{Inference Rules}
\label{sec:sl-rules}

\begin{figure}[htb]
    \begin{adjustbox}{max width=\textwidth}
        \begin{tabular}{@{}l@{}}
            \begin{tabular}{@{}l@{\;}@{}l@{}} $\proofrule{(\AXsl)}{\lseq \vdash \rseq}{\top}{\scriptstyle \Asterisk \lseq \models^\tinyseplog \bigvee \rseq}$ & $\proofrule{(\RUsl)}{\lseq \vdash \rseq}{\lseq \vdash \set{\exists \vec{y}_i \,.\, \Asterisk R_i(\vec{x},\vec{y}_i)}_{i=1}^n, \rseq \setminus p(\vec{x})}{\begin{array}{@{}l@{}} \scriptstyle p(\vec{x}) \in \rseq \text{, } \vec{y}_1, \ldots, \vec{y}_n \text{ fresh } \\[-2pt] \scriptstyle p(\vec{x}) \leftarrow_\sys R_1(\vec{x},\vec{y}_1) \mid \ldots \mid R_n(\vec{x},\vec{y}_n)\end{array}}$\end{tabular} \\[20pt]
            
            $\proofrule{(\RDsl)}{\phi(\vec{x},\vec{x}_1,\ldots,\vec{x}_n),p_1(\vec{x}_1),\ldots,p_n(\vec{x}_n) \vdash \{\exists \vec{y}_j \,.\, \psi_j(\vec{x},\vec{y}_j) * \mathcal{Q}_j(\vec{y}_j)\}_{j=1}^k}{p_1(\vec{x}_1),\ldots,p_n(\vec{x}_n) \vdash \{\mathcal{Q}_j\theta \mid \theta \in S_j\}_{j=1}^i}{\begin{array}{@{}l@{}} \scriptstyle \phi \models^\tinyseplog \bigwedge_{j=1}^i \exists \vec{y}_j . \psi_j \\[-1.5pt] \scriptstyle \phi \not\models^\tinyseplog \bigvee_{j=i+1}^k \exists \vec{y}_j . \psi_j \\[-1.5pt] \scriptstyle S_j \subseteq \subst{\phi,\psi_j}, j \in [i] \end{array}}$ \\[20pt]
            
            $\proofrule{(\SPsl)}{p_1(\vec{x}_1),\ldots,p_n(\vec{x}_n) \vdash \mathcal{Q}_1(\vec{x}_1,\ldots,\vec{x}_n), \ldots, \mathcal{Q}_k(\vec{x}_1,\ldots,\vec{x}_n)}{\langle p_{\bar{\imath}_j}(\vec{x}) \vdash \{q_{\bar{\imath}_j}^\ell(\vec{x}) \mid \ell \in [k],~ f_j(\overline{\mathcal{Q}}_\ell) = \bar{\imath}_j\} \rangle_{j=1}^{n^k}}{\begin{array}{@{}l@{}} \scriptstyle \forall i,j \in [n] \,.\, \vec{x}_i \cap \vec{x}_j = \emptyset, ~ \bar{\imath} \in {[n]}^{n^k} \\[-2pt] \scriptstyle \mathcal{Q}_i = \Asterisk_{j=1}^{n} q_j^i(\vec{x}_j), \overline{\mathcal{Q}}_i=\langle q_1^i,\ldots,q_n^i \rangle \\[-2pt] \scriptstyle \cf(\overline{\mathcal{Q}}_1,\ldots,\overline{\mathcal{Q}}_k) = \set{f_1,\ldots,f_{n^k}} \end{array}}$ 
        \end{tabular}
    \end{adjustbox}
    \caption{The set $\RIndsl$ of inference rule schemata for $\seplog$ inductive entailments.}
    \label{fig:sl-rules}
\end{figure}

We modify the $\RInd$ proof system (\S\ref{sec:inference-rules})
by systematically replacing boolean with spatial conjunctions,
in order to match the semantics of the rules in a system, which
separately join the constraint and subgoal heaps into a goal heap.
We obtain the proof system $\RIndsl = \{\LU, \RUsl, \RDsl, \RI, 
\SPsl, \AXsl, \ID\}$, where the inference rule schemata $\RUsl$, 
$\RDsl$, $\SPsl$ and $\AXsl$ are depicted in Figure \ref{fig:sl-rules}.

\begin{example}\label{ex:sl-proof}
Below we give a proof for the sequent $ls^+(x, y) \models^\utree
\widehat{ls}^+(x,y)$, using the rules in Figure \ref{fig:sl-rules}:
\vspace*{-14pt}
\begin{center}
\begin{adjustbox}{max width=\textwidth}
\begin{tikzpicture}[nd/.style={outer sep=0.15pt}, lbl/.style={outer sep=0.15pt}]
	\node (seq1) [nd] {\small $ls^+(x,y) \vdash \widehat{ls}^+(x, y)$};
	\node (seq2) [nd, above=0pt of seq1] {\small $y \teq z_2 \land x \mapsto z_1, ls^+(z_1,z_2) \vdash \widehat{ls}^+(x, y)$};	
	\node (seq3) [nd, above=0pt of seq2] {\small $\begin{array}{@{}l@{\,}c@{\,}l@{}} y \teq z_2 \land x \mapsto z_1, ls^+(z_1,z_2) & \vdash & \exists u_1 \exists u_2 \, . \, y \teq u_2 \land x \mapsto u_1 * ls^e(u_1,u_2), \\ & & \exists u_1 \exists u_2 \, . \, y \teq u_2 \land x \mapsto u_1 *ls^o(u_1,u_2)	\end{array}$};
	\node (seq4) [nd, above=0pt of seq3] {\small $ls^+(z_1,y) \vdash ls^e(z_1, y), ls^o(z_1, y)$};
	\node (seq5) [nd, above=0pt of seq4] {\small $y \teq u_2 \land z_1 \mapsto u_1, ls^+(u_1,u_2) \vdash ls^e(z_1, y), ls^o(z_1, y)$};	
	\node (seq6) [nd, above=0pt of seq5] {\small $y \teq u_2 \land z_1 \mapsto u_1, ls^+(u_1,u_2) \vdash z_1 \teq y \land \emp, \exists v_1 \exists v_2 \, . \, y \teq v_2 \land z_1 \mapsto v_1 * ls^o(v_1, v_2), ls^o(z_1, y)$};
	\node (seq7) [nd, above=0pt of seq6] {\small $\begin{array}{@{}l@{\,}c@{\,}l@{}} y \teq u_2 \land z_1 \mapsto u_1, ls^+(u_1,u_2) & \vdash & z_1 \teq y \land \emp, \exists v_1 \exists v_2 \, . \, y \teq v_2 \land z_1 \mapsto v_1 * ls^o(v_1, v_2), \\ & & z_1 \mapsto y, \exists v_1 \exists v_2 \, . \, y \teq v_2 \land z_1 \mapsto v_1 * ls^e(v_1, v_2) \end{array}$};	
	\node (seq8) [nd, above=0pt of seq7] {\small $ls^+(u_1,y) \vdash ls^o(u_1, y), ls^e(u_1, y)$};
	\node (seq9) [nd, above=0pt of seq8] {\small $\top$};	
			
	\draw (seq2.south west) -- (seq2.south east);
	\draw (seq3.south west) -- (seq3.south east);
	\draw (seq3.north west) -- (seq3.north east);
	\draw (seq5.south west) -- (seq5.south east);
	\draw (seq6.south west) -- (seq6.south east);
	\draw (seq6.north west) -- (seq6.north east);
	\draw (seq7.north west) -- (seq7.north east);	
	\draw (seq8.north west) -- (seq8.north east);
	
	\node(lbl1) [lbl, left=0pt of seq2.south west] {\scriptsize $\LU$};
	\node(lbl2) [lbl, left=0pt of seq3.south west] {\scriptsize $\RU_\tinyseplog$};
	\node(lbl3) [lbl, left=0pt of seq3.north west] {\scriptsize $\RD_\tinyseplog$};
	\node(lbl4) [lbl, left=0pt of seq5.south west] {\scriptsize $\LU$};		
	\node(lbl5) [lbl, left=0pt of seq6.south west] {\scriptsize $\RU_\tinyseplog$};
	\node(lbl6) [lbl, left=0pt of seq6.north west] {\scriptsize $\RU_\tinyseplog$};
	\node(lbl7) [lbl, left=0pt of seq7.north west] {\scriptsize $\RD_\tinyseplog$};
	\node(lbl8) [lbl, left=0pt of seq8.north west] {\scriptsize $\ID$};			
	
	\path[dashed, rounded corners=2mm, draw=black, -latex] let \p1=(seq8.east),\p2=(seq4.east) in (\x1, \y1) -- ([xshift=4.25cm]\x1,\y1) -- ([xshift=4.25cm]\x1,\y2) -- (\x2, \y2);
\end{tikzpicture}
\end{adjustbox}
\end{center}
For space reasons, several simple branches of the proof are
omitted. \hfill\qed
\end{example}

\subsection{Restricting the Set of Constraints}
\label{sec:sl-restrictions}

In this section we state the counterparts of the semantic
restrictions introduced in \S\ref{sec:restrictions},
necessary for soundness and completeness. Moreover, we give complexity
bounds for the problem of deciding whether a certain system, with
quantifier-free $\seplog$ and symbolic heap constraints, respectively,
complies with these restrictions.

\begin{definition}\label{def:sl-non-filtering}
    An $\seplog$ inductive system $\sys$ is \emph{non-filtering} iff, for
    every $\langle\{\phi, q_1(\vec{x}_1), \\ \ldots, q_n(\vec{x}_n)\},
    p(\vec{x})\rangle \in \sys$, for all $i \in [n]$ and
    $(\overline{\ell}_i,h_i) \in \mu\sys^\tinyseplog(q_i)$, where $h_i$
    are pairwise disjoint heaps, there exists a valuation $\nu$ and a heap
    $h$, disjoint from $\biguplus_{i=1}^n h_i$, such that $\nu,h
    \models^\tinyseplog \phi$ and $\nu(\vec{x}_i) = \overline\ell_i$, for
    all $i\in[n]$.
\end{definition}

\begin{example}\label{ex:sl-non-filtering}
    The system from Example \ref{ex:sl-system} is non-filtering because
    there exists a model $\nu,h \models^\tinyseplog y \teq y' \wedge x
    \mapsto z$, such that $\nu(y') = \nu'(y')$, $\nu(z) = \nu'(z)$ and
    $\dom(h) \cap \dom(h') = \emptyset$, for each given pair
    $(\nu',h')$. Since the set $\locsi$ is infinite, it is always possible
    to  find a value $\nu(x) \not\in \dom(h')$. \hfill\qed
\end{example}

As opposed to the case of systems with first-order constraints, under
the Herbrand interpretation (Lemma \ref{lemma:fol-non-filtering}),
the non-filtering property is decidable for systems with $\seplog$
constraints. This is because one can build an over-approximation of
the least solution, that is both necessary and sufficient to
characterize the satisfiability of a quantifier-free $\seplog$ formula
using predicate atoms \cite{Brotherston14}. The lemma below establishes
the upper bound for the complexity of deciding whether a given inductive
system is non-filtering.

\begin{lemma}\label{lemma:sl-non-filtering}
    The problem ``Given an inductive system $\sys$ with $\wand$-free
    $\seplog$ constraints, is $\sys$ non-filtering?'' is in {\expspace}.
\end{lemma}

Next, we turn to the ranking condition, that ensures the soundness of
applying the principle of Infinite Descent to a system with $\seplog$
constraints. In the absence of a natural wfqo on the set of locations
$\locsi$ (since there are no relations other than equality defined on
it), we consider the following wfqo on heaps. For any $h_1,h_2 \in
\heaps$, we have $h_1 \unlhd h_2$ iff there exists $h \in \heaps$ such
that $h_2 = h_1 \uplus h$. We write $h_1 \lhd h_2$ if, moreover, $h
\neq \emptyset$.

\begin{definition}\label{def:sl-ranked}
    An $\seplog$ inductive system $\sys$ is \emph{ranked}
    iff for every predicate rule $\langle\{\phi, q_1(\vec{x}_1), \ldots, 
    q_n(\vec{x}_n)\}, p(\vec{x})\rangle \in \sys$ and every $(\overline\ell_i,h_i)
    \in \mu\sys(q_i)$, $i \in [n]$ there exists $(\overline\ell,h) \in
    \mu\sys(p)$ such that $\biguplus_{i=1}^n h_i \lhd h$.
\end{definition} 

\begin{example}\label{ex:sl-ranked}
    The system of Example \ref{ex:sl-system} is ranked because each rule
    with at least one subgoal has a constraint $y \teq y' \wedge x \mapsto
    z$, which does not admit an empty heap model. \hfill\qed
\end{example}

\begin{lemma}\label{lemma:sl-ranked}
    The problem ``Given an $\seplog$ inductive system $\sys$, is
    $\sys$ ranked in the subheap order $(\heaps, \unlhd)$?'' is in \pspace.
    When considering symbolic heap constraints, the problem is in {\p}.
\end{lemma}

We continue with the finite variable instantiation (fvi) property
(cf. Definition \ref{def:fol-fvi}) for quantifier-free $\seplog$
constraints. We show that this problem is decidable and provide
several upper bounds.

\begin{example}\label{ex:sl-fvi}
    The system from Example \ref{ex:sl-system} has the fvi property,
    because the entailment $y\teq y_1 \wedge x \mapsto z_1
    \models^\tinyseplog \exists y_2 \exists z_2 \,.\, y \teq y_2 \wedge x
    \mapsto z_2$ is witnessed by a single substitution $\theta(y_2) = y_1$
    and $\theta(z_2) = z_1$. \hfill\qed
\end{example}

\begin{lemma}\label{lemma:sl-fvi}
    The problem ``Given an $\seplog$ inductive system $\sys$, does $\sys$ have the fvi property?'' is in {\pspace} if $\sys$ has quantifier-free and $\wand$-free $\seplog$ constraints, and in {\sigmatwop} if $\sys$ has symbolic heap constraints.
\end{lemma}

A related question is whether an inductive system with quantfier-free
$\seplog$ constraints is non-overlapping (cf. Definition \ref{def:non-overlapping}). 

\begin{example}\label{ex:sl-non-overlapping}
    The system from Example \ref{ex:sl-system} is non-overlapping because
    the only constraints with a satisfiable conjunction are $x \mapsto y$
    and $y \teq y' \wedge x \mapsto z$ and both entailments $x \mapsto y
    \models^\tinyseplog \exists y' \exists z \,.\, y \teq y' \wedge x
    \mapsto z$ and $y \teq y' \wedge x \mapsto z \models^\tinyseplog x
    \mapsto y$ are valid. \hfill\qed
\end{example}

\begin{lemma}\label{lemma:sl-non-overlapping}
    The problem ``Given an $\seplog$ inductive system $\sys$, is $\sys$ non-overlapping?'' is in {\pspace} if $\sys$ has quantifier-free and $\wand$-free $\seplog$ constraints, and in {\sigmatwop} if $\sys$ has symbolic heap constraints.
\end{lemma}


\subsection{Soundness and Completeness}
\label{sec:sl-soundness-completeness}

\paragraph{Soundness} We develop the argument for the soundness of the
$\RIndsl$ set of inference rules in similar fashion as we did for $\RInd$.
The following lemma is the counter-part of Lemma \ref{lemma:local-soundness}
and gives the local soundness of $\RIndsl\setminus\set{\ID}$.

\begin{lemma}\label{lemma:sl-local-soundness}
    Given a ranked $\seplog$ inductive system $\sys$, for each instance of an inference rule schema in $\RIndsl\setminus\set{\ID}$, having the consequent $\lseq\vdash\rseq$ and antecedents $\lseq_i\vdash\rseq_i$ with $i \in [n]$, and each $(\nu, h) \in \mu\sys^\tinyseplog(\bigwedge\lseq) \setminus \mu\sys^\tinyseplog(\bigvee\rseq)$, there exists $(\nu_i, h_i) \in \mu\sys^\tinyseplog(\bigwedge\lseq_i) \setminus \mu\sys^\tinyseplog(\bigvee\rseq_i)$ for some $i \in [n]$ such that $h_i \unlhd h$.
\end{lemma}

We consider a reachability relation between counterexamples and write $(\nu, h) \nextval (\nu', h')$ when, given any instance of an inference rule in $\RIndsl \setminus {\ID}$, $(\nu, h)$ is a counterexample of the consequent and $(\nu', h')$ is a counterexample of one of its antecedents obtained from $(\nu, h)$, as shown in the proof of Lemma \ref{lemma:sl-local-soundness}. With this in mind, we revisit the definition of a counterexample path and adapt it to $\seplog$.

\begin{definition}\label{def:sl-counterexample-path}
    A path $v_1,v_2,\ldots,v_k$ in a proof $\mathcal{D} = (V,v_0,S,R,P)$
    built with $\RIndsl$ is a \emph{counterexample path} if there exists a sequence of
    pairs $(\nu_1, h_1), (\nu_2, h_2), \\ \ldots, (\nu_k, h_k)$ such that, for all $i\in[k]$ we
    have: \begin{inparaenum}[(i)]
        \item $(\nu_i, h_i) \in \mu\sys^\tinyseplog(\bigwedge\lseq_i) \setminus
        \mu\sys^\tinyseplog(\bigvee\rseq_i)$, where $S(v_i)=\lseq_i \vdash \rseq_i$, and
        \item if $i < k$ then $(\nu_i, h_i) \nextval (\nu_{i+1}, h_{i+1})$. 
    \end{inparaenum}
\end{definition}

The following lemma is the counterpart of Lemma 
\ref{lemma:direct-counterexample-path} and shows that any direct counterexample
path in a proof causes a strict heap descent provided that the system is ranked.

\begin{lemma}\label{lemma:sl-direct-counterexample-path}
    Given a ranked $\seplog$ system $\sys$, let $\mathcal{D} = (V,v_0,S,R,P)$
    be a proof built with $\RIndsl$ and let $\pi = v_1,\ldots,v_k$ be a direct
    counterexample path in $\mathcal{D}$, with pairs $(\nu_1, h_1), \ldots,
    (\nu_k, h_k)$, for a backlink $(v_k,v_1)$. Then $h_1 \rhd h_k$, where $S(v_i)
    = \lseq_i \vdash \rseq_i$, for all $i \in [k]$.
\end{lemma}

Finally, the following theorem extends the reachability relation $\nextval$ to backlinks and shows that any infinite trace in a proof leads to an infinite strictly decreasing sequence of heaps, contradicting the well-foundedness of the wfqo which makes the inductive system be ranked (Definition \ref{def:sl-ranked}). Then there cannot exist a counterexample for any sequent labelling the root of a proof built with $\RIndsl$ and the associated entailment must hold.

\begin{theorem}\label{thm:sl-soundness}
    Given a ranked $\seplog$ inductive system $\sys$, if a sequent
    $\lseq \vdash \rseq$ has a proof $\mathcal{D}=(V,v_0,S,R,P)$
    built with $\RIndsl$, and $S(v_0) = \lseq \vdash \rseq$,
    then the entailment $\Asterisk\lseq \models_\sys^\tinyseplog
    \bigvee\rseq$ holds.
\end{theorem}

\paragraph{Completeness} The $\RIndsl$ proof system is not
complete for $\seplog$ entailments, even for those systems which
comply with the conditions in \S\ref{sec:sl-restrictions}. This section
proves the completeness of the set $\RIndsl$ for a more
restricted class of entailment problems. The existence of a complete
set of inference rules for the general entailment problem of $\seplog$
is, to our knowledge, still open.

We consider assignments $\X$ mapping a predicate $p$ of arity $n$ into
a subset of $\locsi^n \times \heaps \times \trees$. For a singly-tree
shaped set (Definition \ref{def:tree-shaped}) represented as a tree
$T$, we define $\X(\Asterisk T)$ to be the set of tuples $(\nu,h,t)$,
where $\nu : \bigcup_{p \in \dom(T)} \fv{T(p)} \rightarrow \locsi$ is
a valuation, $h$ is a heap and $t$ is a coverage tree for $h$, where:
\begin{compactitem}
\item for each $p \in \dom(T) \setminus \fr(T)$, we have $\nu,t(p)
  \models^\tinyseplog T(p)$,
\item for each $p \in \fr(T)$ where $T(p) = q(\vec{x})$, there exists
  $(\nu(\vec{x}),t(p),\sub{t}{p}) \in \X(q)$.
\end{compactitem}
The above definition is extended to the tree-shaped sets
$T_1,\ldots,T_k$ such that $\X(\Asterisk\set{T_1,\ldots,T_k}) =
\{(\nu,h_1\uplus\ldots\uplus h_k,\{t_1,\ldots,t_k\}) \mid
(\nu,h_i,t_i) \in \X(\Asterisk T_i),~ i\in[k]\}$. With this
interpretation of predicates and formulae, the least solution
$\mu\sys^\utree$ of a system $\sys$ is the least fixed point of the function
\(\mathbb{F}^\utree_\sys(\X)\), mapping each predicate $p \in \preds$ into
$\bigcup_{i=1}^m \set{(\nu(\vec{x}),h,t) \mid (\nu,h,t) \in
  \X(\Asterisk R_i)}$, where $p(\vec{x}) \leftarrow_\sys R_{1} \mid
\ldots \mid R_{m}$.

For each $(\nu,h,t) \in \mu\sys^\utree(p)$, for some $p \in \preds$,
we say that $t$ is an \emph{unfolding tree} for the singly-tree shaped
set $T$. Then the entailment problem becomes $p \models_\sys^\utree q_1, \ldots, q_n$ iff $\mu\sys^\utree(p) \subseteq \bigcup_{i=1}^n \mu\sys^\utree(q_1)$, given predicates $p,q_1, \ldots, q_n \in \preds$. It is not difficult to prove that $p \models_\sys^\utree
q_1, \ldots, q_n$ implies $p(\vec{x}) \models_\sys^\tinyseplog q_1, \ldots, q_n$, but
not vice versa.

Akin to the derivations built with $\RInd$,
we denote by $\DSSL(\lseq \vdash \rseq)$ the set of irreducible,
maximal and structured\footnote{A derivation is structured if and only
  if there is an occurrence of ($\RD_\tinyseplog$) between any two
  consecutive applications of ($\LU$).} derivations of $\lseq \vdash
\rseq$ built with $\RIndsl$.  Also,
we write $\lseq \vdash \rseq \leadsto^\tinyseplog \lseq' \vdash
\rseq'$ if $\lseq' \vdash \rseq'$ occurs inside a derivation from
$\DSSL(\lseq \vdash \rseq)$.

The following lemmas are the $\seplog$ counterparts of the ones in Section \ref{sec:soundness-completeness}. Since $\RIndsl$ is structurally the same as $\RInd$, we need not redo the proofs, as they closely resemble the ones for Lemma \ref{lemma:max-irred-finite} and Lemma \ref{lemma:tree-shaped}.

\begin{lemma}\label{lemma:sl-max-irred-finite}
    If the $\seplog$ inductive system $\sys$ has the fvi property, then the following
    properties of derivations built with $\RIndsl$ hold: \begin{compactenum}
        \item\label{it1:sl-max-irred-finite} any irreducible and structured
        derivation is finite, and
        \item\label{it2:sl-max-irred-finite} for any sequent $\lseq \vdash
        \rseq$, the set $\DSSL(\lseq \vdash \rseq)$ is finite.
    \end{compactenum}
\end{lemma}

\begin{lemma}\label{lemma:sl-tree-shaped}
    Given an $\seplog$ inductive system $\sys$ and predicates $p, q_1, \ldots, q_n$,
    in every sequent $\lseq \vdash \rseq$ such that
    $p(\vec{x}) \vdash q_1(\vec{x}), \ldots, q_n(\vec{x}) \leadsto^\tinyseplog \lseq \vdash \rseq$, $\lseq$ is a tree-shaped set and $\rseq$ consists of finite separating conjunctions of     tree-shaped sets, with all subgoal variables existentially quantified.
\end{lemma}

The following lemma proves an invariant that relates tree-shaped sets
with their corresponding unfolding trees, and is the counterpart of
Lemma \ref{lemma:cex}, needed to prove the completeness of $\RIndsl$,
for systems with symbolic heap constraints, with the above definition
of entailments.

\begin{lemma}\label{lemma:sl-u-cex}
    Given a non-filtering and non-overlapping $\seplog$ system $\sys$ with
    quantifier-free $\seplog$ constraints, having the fvi property,
     the predicate atoms
    $p(\vec{x}), q_1(\vec{x}), \ldots, q_n(\vec{x})$, and a sequent $\lseq
    \vdash \rseq$ such that $p(\vec{x}) \vdash q_1(\vec{x}), \ldots,
    q_n(\vec{x}) \leadsto^\tinyseplog \lseq \vdash \rseq$, if every derivation $D \in
    \DSSL(\lseq \vdash \rseq)$ contains a leaf $\lseq' \vdash \emptyset$
    then there exists a valuation $\nu$, a heap $h$ and a set of
    unfolding trees $U$ such that $(\nu,h,U) \in
    \mu\sys^\utree(\Asterisk\lseq) \setminus \mu\sys^\utree(\bigvee\rseq)$.
\end{lemma}

The completeness proof for $\RIndsl$ mirrors closely the proof of Theorem
\ref{thm:ta-completeness} (\S\ref{sec:soundness-completeness}). 

\begin{theorem}\label{thm:sl-completeness}
    Given a ranked, non-filtering, non-overlapping $\seplog$
    inductive system $\sys$ with the fvi property, let $p, q_1, \ldots,
    q_n$ be predicates occurring in $\sys$. Then the entailment $p
    \models_\sys^\utree q_1, \ldots, q_n$ holds only if the sequent
    $p(\vec{x}) \vdash q_1(\vec{x}), \ldots, q_n(\vec{x})$ has an
    $\mathbf{S}$-proof with the set of inference rules $\RIndsl$, where
    $\mathbf{S}$ is defined by the regular expression $(\LU \cdot \RUsl^*
    \cdot \RDsl \cdot \RI^* \cdot \SPsl?)^* \cdot \LU? \cdot \RUsl^* \cdot
    (\AXsl \mid \ID)$.
\end{theorem}

Because each $\mathbf{S}$-derivation is structured, the proof search
semi-algorithm \ref{alg:proof-search} terminates on all inputs, when
given $\mathcal{S}$ as strategy. A direct consequence of Theorem
\ref{thm:sl-completeness} is that algorithm \ref{alg:proof-search} is
a decision procedure for entailments $p \models^\utree q_1, \ldots, q_n$,
when $p, q_1, \ldots, q_n$ are defined by an inductive system $\sys$
with quantifier-free $\seplog$
constraints, that is ranked, non-filtering, non-overlapping and has
the fvi property. As discussed in \S\ref{sec:sl-restrictions}, the
problem whether a given system enjoys these properties is decidable.

\section{Conclusions}

We present a cyclic proof system for entailments between inductively
defined predicates written using (multisorted) First Order Logic,
based on Fermat's principle of Infinite Descent. The advantage of this
principle over classical induction is that the inductive invariants
are produced during proof search, whereas induction requires them to
be provided. The soundness of this principle is coined by a semantic
restriction on the constraints of the inductive system, that asks that
models generated by unfoldings decrease in a well-founded domain. On
the other hand, completeness relies on an argument inspired by the
theory of tree automata, that is applicable under three semantic
restrictions on the set of constraints. In general all these
restrictions are decidable, with computational complexities that
depend on the logical fragment in which the constraints of the
inductive system are written.

Moreover, we extend the proof system for First Order Logic to
Separation Logic and analyze its proof-theoretic properties. While
soundness is maintained by a similar ranking property as in First
Order Logic, completeness is lost, in general. We recover completeness
partially by restricting the semantics of entailments with a notion of
(matching) unfolding trees. Extending the proof system to handle
limited cases of entailments between divergent predicates, whose
unfolding trees do not match, but are related by reversal and rotation
relations is possible. The completeness and algorithmic properties of
such extensions, such as the decidability of the entailment problem,
are considered for future work.

\bibliographystyle{alpha}
\bibliography{refs}

\appendix

\section{Additional Material}

\subsection{Proof of Lemma \ref{lemma:finite-sets-wqo}}

\proof{Let $\mathcal{D}^*$ be the set of finite sequences of elements
    from $\mathcal{D}$, where $u_i$ denotes the $i$-th element of $u \in
    \mathcal{D}^*$ and $\len{u}$ is the length of $u$. The \emph{subword
        order} $\sw$ on $\mathcal{D}^*$ is defined as $u \sw v$ iff there
    exists a strictly increasing mapping $f : [\len{u}] \rightarrow
    [\len{v}]$ such that $u_i = v_{f(i)}$ for all $i \in [\len{u}]$.
    A qo $\preceq$ on $\mathcal{D}$ induces the
    following order on the set $\mathcal{D}^*$: for all $u,v \in
    \mathcal{D}^*$, $u \preceq^* v$ if there exists $v' \sw v$ such that
    $\len{u} = \len{v'}$ and $u_i \preceq v'_i$, for all $i =
    1,\ldots,\len{u}$. Because $\mathcal{D}$ is countable, there is an
    indexing of its elements. Then each finite set $S \in
    \finpow{\mathcal{D}}$ is uniquely represented as a finite word and
    the result follows from Higman's Lemma \cite{Higman52}, which states
    that $(\mathcal{D},\preceq)$ is a wqo only if
    $(\mathcal{D}^*,\preceq^*)$ is a wqo. \qed}

\subsection{Proof of Theorem \ref{thm:entailment-undec}}

\proof{The proof is by reduction from the
    inclusion problem for context-free languages, a known undecidable
    problem \cite[Theorem 5.10]{Sipser97}. Let $G =
    \tuple{\Xi,\Sigma,\Delta}$ be a context-free grammar, where $\Xi$ is
    the set of nonterminals, $\Sigma$ is the alphabet of terminals, and
    $\Delta$ is a set of productions $(X,w) \in \Xi \times (\Xi \cup
    \Sigma)^*$. For a nonterminal $X \in \Xi$, we denote by $L_X(G)
    \subseteq \Sigma^*$ the language produced by $G$ starting with $X$
    as axiom. The problem "given $X,Y \in \Xi$, does $L_X(G) \subseteq
    L_Y(G)$?" is undecidable. Given a context-free grammar $G =
    \tuple{\Xi,\Sigma,\Delta}$, we define a system $\sys_G$ as
    follows: \begin{compactitem}
        \item each nonterminal $X \in \Xi$ corresponds to a predicate
        $X(x^\sigma,y^\sigma)$, where $\sigma$ is the only sort used in the
        reduction,
        \item each alphabet symbol $a \in \Sigma$ corresponds to a function
        symbol $\overline{a}^{\sigma,\sigma}$, and a word $w = a_1\ldots a_n
        \in \Sigma^*$ is encoded by the context (i.e.\ the term with a hole)
        $\overline{w} = \overline{a_1}( \ldots \overline{a_n}(.) )$,
        \item each grammar rule $(X,u_1X_1 \ldots u_nX_nu_{n+1}) \in \Delta$
        corresponds to a rule:
        \[\tuple{\set{\phi(x,y,x_1,y_1,\ldots,x_{n},y_n), 
                X_1(x_1,y_1),\ldots,X_n(x_n,y_n)},X(x,y)}\]
        of $\sys_G$, where $\phi \equiv x \teq \overline{u_1}(x_1) ~\wedge~
        \bigwedge_{i=1}^{n-1} y_i \teq \overline{u_{i+1}}(x_{i+1}) ~\wedge~
        y_n \teq \overline{u_{n+1}}(y)$ . In particular, a rule
        $(\epsilon,X) \in \Delta$ maps into a rule $\tuple{\set{x \teq y},
            X(x,y)}$ of $\sys_G$.
    \end{compactitem} 
    We must check that, for any nonterminals $X,Y \in \Xi$, we have
    $L_X(G) \subseteq L_Y(G)$ if and only if $X \models_{\sys_G}^\tinyherb Y$. This
    is proved using the following invariant: 
    \[\forall w \in \Sigma^* ~.~ \left(\forall t \in \sigma^\tinyherb ~.~ 
    [x \leftarrow \overline{w}(t), y \leftarrow t] \in \mu\sys^\tinyherb_G(X)\right) \Leftrightarrow w \in L_X(G)\]
    where $[x \leftarrow t, y \leftarrow u]$ denotes the valuation mapping
    $x$ to $t$ and $y$ to $u$. \qed}

\subsection{Proof of Proposition \ref{prop:direct-path}}

\proof{Let $\tau$ be an infinite trace in $\mathcal{D} =
    (V,v_0,S,R,P)$. Since $V$ is finite, $\tau$ must contain infinitely
    many backlinks. Moreover, $V \times V$ is also finite and there can
    only be a finite number of backlink, thus there must exist a
    backlink $(v_{i-1},v_i)$ that repeats infinitely often in $\tau$.
    
    Next, we prove that for every finite trace $\rho=v_i, \ldots,
    v_{i-1}$, where $(v_{i-1},v_i)$ is a backlink, there exists a direct
    path in $\rho$, by induction on the number $N$ of backlinks in $\rho$.
    If $N=0$ the direct path is trivially $\rho$. For the induction step
    $N>0$, we suppose that the property holds for any $N'<N$. Let
    $\rho = v_i, \ldots, v_{j-1},v_j, \ldots, v_{i-1}$, where
    $(v_{j-1},v_j)$ is the last backlink on $\rho$. Then the suffix
    $v_j \ldots v_{i-1}$ of $\rho$ is a path. Since $(v_{i-1},v_i)$ is a
    backlink, then $v_i$ is a predecessor of $v_{i-1}$ in $\mathcal{D}$,
    thus $v_i,v_j$ and $v_{i-1}$ are on the same branch on $\mathcal{D}$.
    We distinguish two cases: \begin{compactitem}
        \item If $v_j$ is $v_i$ or a predecessor of $v_i$ then there exists
        a direct path from $v_i$ to $v_{i-1}$ in $\rho$ and we are done,
        because $\rho = v_i, \ldots, v_{j-1},v_j, \ldots, v_i, \ldots,
        v_{i-1}$ and there are no more backlinks between $v_j$ and
        $v_{i-1}$. 
        \item Else, if $v_i$ is a strict predecessor of $v_j$, we show that
        there must exist another occurrence of $v_j$ in the prefix $v_i,
        \ldots, v_{j-1}$ of $\rho$. Note that $v_i, v_j$ and $v_{j-1}$ occur
        on the same branch in $\mathcal{D}$. Suppose, by contradiction, that
        $v_k\neq v_j$, for all $k \in [i,j-1]$. To reach $v_{j-1}$ from
        $v_i$ there must exist a strict predecessor $u \in V$ of $v_j$ and
        a subsequence of $\rho$ from $u$ to some strict successor $u' \in V$
        of $v_j$ that goes through one or more backlinks (see the figure
        below). However this is not possible, since each backlink leads
        from a leaf of $\mathcal{D}$ to one of its predecessors, thus
        any trace starting at $v_i$, passing through $u$ and then
        following a different branch than the one on which $v_i, v_j$
        and $v_{j-1}$ reside, can only return to this branch at $u$ or
        predecessors of $u$, whereas $u'$ is a strict successor of $u$.
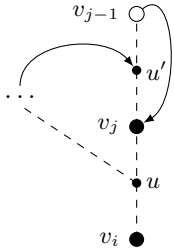
        \begin{center}
            \begin{tikzpicture}[nd/.style={circle,outer sep=0pt, inner sep=2pt}]
            \draw [dashed] (2,0) -- (2,1.5);
            
            \draw [dashed] (2,0.75) -- (0.5,1.75);
            \draw [dashed] (2,1.5) -- (2,3);
            
            \node (root) at (2,0) [nd, fill=black] {};
            \node (pivot) at (2,1.5) [nd, fill=black] {};
            \node (branch) at (2,0.75) [nd, fill=black, inner sep=1.25pt] {};
            \node (return) at (2,2.25) [nd, fill=black, inner sep=1.25pt] {};
            \node (id) at (2,3) [nd, draw=black, fill=white] {};
            
            \node (dots) at (0.45,1.9) {$\ldots$};
            
            \node [left=0pt of root.west, anchor=east] {\footnotesize $v_i$};
            \node [left=0pt of pivot.west, anchor=east] {\footnotesize $v_j$};
            \node [left=0pt of id.west, anchor=east] {\footnotesize $v_{j-1}$};
            \node [right=-1.75pt of branch.east] {\footnotesize $u$};
            \node [right=-1.75pt of return.east] {\footnotesize $u'$};
            
            \path [draw=black, fill=black, -latex] (id) edge[out=40, in=30] (pivot);
            \path [draw=black, fill=black, -latex] (dots) edge[out=90, in=130] (return);
            \end{tikzpicture}
        \end{center}  
        We have shown that there exists $k \in [i,j-1]$ such that $v_k =
        v_j$, thus we have a subsequence $\rho' = v_k,\ldots,v_{j-1}$ of
        $\rho$, where $(v_{j-1},v_k)$ is a backlink, containing $N'<N$ backlinks.
        By the induction hypothesis, $\rho'$ contains a direct
        path, which concludes the induction proof.
    \end{compactitem}
    Therefore, each finite subtrace $v_{i-1},\ldots,v_i$ contains a
    direct path. Because the backlink $(v_{i-1}, v_i)$ occurs infinitely
    often in $\tau$, there are infinitely many such subtraces in
    $\tau$, we can conclude that $\tau$ contains infinitely many direct
    paths.\qed}

\subsection{Proof of Lemma \ref{lemma:fol-non-filtering}}

\proof{ By reduction from the following undecidable problem: given a
    context-free grammar $G=\tuple{\Xi,\Sigma,\Delta}$, where $\Xi$ is
    the set of nonterminals, $\Sigma$ is the alphabet of terminals, and
    $\Delta$ is a set of productions $(X,w) \in \Xi \times (\Xi \cup
    \Sigma)^*$, and two nonterminals $X,Y \in \Xi$, is it the case that
    $L_X(G) \cap L_Y(G) \neq \emptyset$, where $L_X(G) \subseteq
    \Sigma^*$ denotes the language produced by $G$ starting with $X$ as
    axiom. We encode $G$ as a system, in the same way as done in the
    proof of Theorem \ref{thm:entailment-undec}, each nonterminal $Z \in
    \Xi$ corresponding to a predicate $Z(x,y)$. Then we encode the
    problem $L_X(G) \cap L_Y(G) \neq \emptyset$ using an additional rule
    $\tuple{\set{x_1 \teq x_2 \wedge y_1 \teq y_2, X(x_1,y_1),
            Y(x_2,y_2)}, P()}$. It is easy to check that the system is
    non-filtering iff $L_X(G) \cap L_Y(G) \neq \emptyset$. \qed}

\subsection{Proof of Proposition \ref{prop:forall-exists-wqo}}
\proof{ If $\sem{\vec{x}}_\nu \succ^{\forall\exists}_\I
    \sem{\vec{y}}_\mu$, since trivially $\sem{\vec{y}}_\mu =
    (\sem{\vec{x}}_\nu \setminus \sem{\vec{x}}_\nu) \cup
    \sem{\vec{y}}_\mu$, we have $\sem{\vec{x}}_\nu \succ^\dagger
    \sem{\vec{y}}_\mu$, where $\succ^\dagger$ is the Manna-Dershowitz
    multiset wfqo \cite{MannaDershowitz79}. Then an infinite strictly
    decreasing sequence in $\succ^{\forall\exists}_\I$ would imply the
    existence of an infinite strictly decreasing sequence in
    $\succ^\dagger$, contradicting \cite[Theorem
    1]{MannaDershowitz79}. \qed}

\subsection{Proof of Lemma \ref{lemma:fol-ranked}}

\proof{Since the satisfiability of the
    quantifier-free fragment of the first order logic with a binary
    relation symbol interpreted as the subterm relation is an \np-complete
    problem \cite{Venkataraman87}, one can effectively decide if a given
    system is ranked. For each constraint
    $\phi(\vec{x},\vec{x}_1,\ldots,\vec{x}_n)$ we check if the formula
    \[\phi \wedge \bigvee_{y \in \vec{x}_1 \cup \ldots \cup \vec{x}_n}
    \bigwedge_{x \in \vec{x}} \left(y \teq x \vee \neg y \sqsubseteq
    x\right)\] is unsatisfiable, where $\sqsubseteq$ is interpreted as the
    subtree order. Since the size of the latter formula is polynomially
    bounded in the size of $\phi$, the problem of checking if a given
    system is ranked is in co-\np. \hfill\qed}

\subsection{Proof of Lemma \ref{lemma:fol-fvi}}

\proof{ Let $\sys$ be a system and
    $\phi(\vec{x},\vec{x}_1,\ldots,\vec{x}_n)$,
    $\psi(\vec{x},\vec{y}_1,\ldots,\vec{y}_m)$ be two arbitrary
    constraints, with goal variables $\vec{x}$ and subgoal variables
    $\bigcup_{i=1}^n \vec{x}_i$ and $\bigcup_{j=1}^m \vec{y}_j$,
    respectively. Then $\sys$ has the fvi property if and only if the following
    entailment does not hold:
    \[\phi(\vec{x},\vec{x}_1,\ldots,\vec{x}_n) \models^\tinyherb 
    \exists \vec{y}_1 \ldots \exists \vec{y}_m ~.~
    \psi(\vec{x},\vec{y}_1,\ldots,\vec{y}_m) \wedge \bigvee_{j=1}^m
    \bigwedge_{i=1}^n \neg(\vec{y}_j \cong \vec{x}_i)\] where
    $\vec{y}_j \cong \vec{x}_i$ is a shorthand for $\left(\bigwedge_{y
        \in \vec{y}_j}\bigvee_{x \in \vec{x}_i} y \teq x\right) \wedge
    \left(\bigwedge_{x \in \vec{x}_i}\bigvee_{y \in \vec{y}_j} y \teq
    x\right)$. In other words, $\sys$ has the fvi property if and only
    if the following equational problem has a solution: 
    \[\exists \vec{x} \exists \vec{x}_1 \ldots \exists \vec{x}_n 
    \forall \vec{y}_1 \ldots \forall \vec{y}_m ~.~
    \phi(\vec{x},\vec{x}_1,\ldots,\vec{x}_n) \wedge
    \bigwedge_{j=1}^m\neg\psi(\vec{x},\vec{y}_1,\ldots,\vec{y}_m) \vee
    \bigvee_{i=1}^n \vec{y}_j \cong \vec{x}_i\] The last formula is not
    in CNF and expanding the formulae $\vec{y}_j \cong \vec{x}_i$ to
    obtain a CNF form causes a simply exponential blowup. Since checking
    the satisfiability of an equational problem in CNF is \np-complete,
    the above check can be performed in \nexptime. If the size of each
    set of subgoal variables is bound to a constant, not part of the
    input, the size of each clause in the CNF expansion of the above
    formula is constants, thus there are at most polynomially many such
    constants and we apply \cite[Theorem 5.2]{Pichler03} to obtain the
    \np~ upper bound. \qed}

\subsection{Proof of Lemma \ref{lemma:fol-non-overlapping}}
\proof{
    Given a first order inductive system $\sys$, in order to determine if $\sys$ has the non-overlapping property, it suffices to check that, for any two constraints $\phi, \psi$ of $\sys$, where $\vec{y}_1, \ldots, \vec{y}_m$ are the subgoal variables of $\psi$: \begin{enumerate*}[label=(\roman*)] \item \label{item:lm-nov-fol-item1} $\phi \land \psi$ is satisfiable and \item \label{item:lm-nov-fol-item2} $\phi \models^\I \exists \vec{y}_1 \ldots \exists \vec{y}_m \,.\, \psi$ is valid \end{enumerate*}. Checking the validity of the entailment in \ref{item:lm-nov-fol-item2} can be reduced to checking that the formula $\forall \vec{y}_1 \ldots \forall \vec{y}_m \,.\, \phi \land \lnot \psi$ is unsatisfiable.
    
    Under the Herbrand interpretation, since $\phi$ and $\psi$ are both conjunctions of literals, $\phi
    \land \psi$ and $\phi \land \lnot\psi$ are in conjunctive normal form. Then both problems \ref{item:lm-nov-fol-item1} and \ref{item:lm-nov-fol-item2} are in \np~ \cite{Pichler03}, and, thus the non-overlapping problem under the Herbrand interpretation is in \np~.\hfill\qed}

\subsection{Soundness of the $\SP$ inference rule in $\RInd$}

\begin{lemma}\label{lemma:sp}
    Given a system $\sys$, with predicates $p_1, \ldots, p_n$ and
    tuples of predicates $\overline{\mathcal{Q}}_i = \tuple{q^i_1,
        \ldots, q_n^i}$ in $\sys$, for all $i \in [k]$. Then
    \[\mu\sys^\I(p_1) \times \ldots \times \mu\sys^\I(p_n) \subseteq 
    \bigcup_{i=1}^k \mu\sys^\I(q_1^i) \times \ldots \times \mu\sys^\I(q_{n}^i)\]
    if and only if there exists a tuple $\bar{\imath} \in [n]^{n^k}$, such that:
    \[\mu\sys^\I(p_{\bar{\imath}_j}) \subseteq \bigcup
    \{\mu\sys^\I(q_{\bar{\imath}_j}^\ell) \mid \ell\in[k],
    f_j(\overline{\mathcal{Q}}_\ell) = \bar{\imath}_j\}\] for all
    $j \in [n^k]$, where
    $\cf(\overline{\mathcal{Q}}_1,\ldots,\overline{\mathcal{Q}}_k) =
    \set{f_1,\ldots,f_{n^k}}$.
\end{lemma}

\proof{ By \cite[Theorem 1]{HolikLSV11}, we have: \[\begin{array}{lc}
    \mu\sys^\I(p_1) \times \ldots \times \mu\sys^\I(p_n) \subseteq
    \bigcup_{i=1}^k \mu\sys^\I(q_1^i) \times \ldots \times \mu\sys^\I(q_{n}^i)
    & \Leftrightarrow \\ 
    \bigwedge_{j=1}^{n^k}\bigvee_{i=1}^n \left(\mu\sys^\I(p_i) 
    \subseteq \bigcup\{\mu\sys^\I(q_i^\ell) \mid \ell \in [k],~
    f_j(\overline{\mathcal{Q}}_\ell)=i\}\right)
    & \Leftrightarrow \\
    \bigvee_{\bar{\imath} \in [n]^{n^k}} \bigwedge_{j=1}^{n^k} \left(\mu\sys^\I(p_{\bar{\imath}_j}) 
    \subseteq \bigcup\{\mu\sys^\I(q_{\bar{\imath}_j}^\ell) \mid \ell \in [k],~
    f_j(\overline{\mathcal{Q}}_\ell)=\bar{\imath}_j\}\right)
    \end{array}\] 
    The last step is the expansion of the formula in
    disjunctive normal form. \qed}

\subsection{Proof of Lemma \ref{lemma:local-soundness}}
\proof{In the case of $(\AX)$, the lemma is trivially true because the list of antecedents is empty.
    For $(\LU)$, $(\RU)$, $(\RI)$, $(\RD)$ and $(\SP)$ we do the following case split.
    
    \vspace{10pt}
    \paragraph{Case $(\LU)$.} ~ Let $p(\vec{x}) \in \lseq$ be a predicate atom, where $p(\vec{x}) \leftarrow_\sys R_1(\vec{x}, \vec{x}_1) \mid \ldots \mid R_n(\vec{x}, \vec{x}_n)$ and $\vec{x}_i \subseteq \vars \setminus \fv{\lseq \cup \rseq}$ for each $i \in [n]$. The antecedents of $\lseq \vdash \rseq$ are $\lseq_i \vdash \rseq_i = R_i(\vec{x}, \vec{x}_i),\lseq \setminus p(\vec{x}) \vdash \rseq$, with $i \in [n]$. The least solution of $\lseq$ is
    \[\begin{array}{@{}lcl@{}}
    \mu\sys^\I(\bigwedge \lseq) & = & \mu\sys^\I(p(\vec{x})) \cap 
    \mu\sys^\I\left(\bigwedge \left(\lseq \setminus p(\vec{x}) \right)\right) \\[2pt]
    & = & \left(\bigcup_{i=1}^n \mu\sys^\I\left(\bigwedge R_i(\vec{x}) \right)\right) \cap 
    \mu\sys^\I\left(\bigwedge \left(\lseq \setminus p(\vec{x})\right)\right) \\[2pt]
    & = & \bigcup_{i=1}^n \mu\sys^\I \left(\bigwedge R_i(\vec{x}) \wedge \bigwedge \left(\lseq \setminus p(\vec{x})\right)\right)
    \end{array}\]
    If there exists $\nu \in \mu\sys^\I(\bigwedge\lseq) \setminus \mu\sys^\I(\bigwedge\rseq)$, then also $\nu \in \mu\sys^\I(\bigwedge R_i(\vec{x}, \vec{y}_i) \wedge \bigwedge (\lseq \setminus p(\vec{x}))) \subseteq \mu\sys^\I(\bigwedge \lseq)$ for some $i \in [n]$ and some $\vec{y}_i \subseteq \vars \setminus \fv{\lseq \cup \rseq}$. Then there also exists $\nu_i$ such that for every $x \in \fv{\lseq}$ we have $\nu_i(x) = \nu(x)$ and also $\nu_i(\vec{x}_i) = \nu(\vec{y}_i)$. Furthermore, because $\sys$ is ranked, $\sem{\vec{x}_i}_{\nu_i} \prec^{\forall\exists}_\I \sem{\vec{x}}_\nu$. Since $\fv{\lseq_i} = \fv{\lseq} \cup \vec{x}_i$ and $\vec{x} \subseteq \fv{\lseq}$, it follows that $\sem{\fv{\lseq_i}}_{\nu_i} \preceq^{\forall\exists}_\I \sem{\fv{\lseq}}_\nu$.
    
    \vspace{10pt}
    \paragraph{Case $(\RU)$.} Let $p(\vec{x}) \in \rseq$ be a predicate atom, where $p(\vec{x}) \leftarrow_\sys R_1(\vec{x}, \vec{x}_1) \mid \ldots \mid R_n(\vec{x}, \vec{x}_n)$ and $\vec{x}_i \subseteq \vars \setminus \fv{\lseq \cup \rseq}$ for each $i \in [n]$. Then $\lseq \vdash \rseq$ has only one antecedent $\lseq_1 \vdash \rseq_1 = \lseq \vdash \exists \vec{x}_1 . \bigwedge R_1(\vec{x},\vec{x}_1), \ldots, \exists \vec{x}_n. \bigwedge R_n(\vec{x},\vec{x}_n), \rseq \setminus p(\vec{x})$. Note that $\fv{\lseq_1} = \fv{\lseq}$. In this case, the least solution of $\rseq$ is
    \[\begin{array}{@{}lcl@{}}
    \mu\sys^\I(\bigvee \rseq) & =& \mu\sys^\I\left(p(\vec{x}) \lor \bigvee \left(\rseq \setminus p(\vec{x}) \right)\right) = \mu\sys^\I(p(\vec{x})) \cup \mu\sys^\I\left(\bigvee \left(\rseq \setminus p(\vec{x}) \right)\right) \\[2pt]
    & =& \left(\bigcup_{i=1}^n \mu\sys^\I\left(\bigwedge R_i(\vec{x}) \right)\right) \cup \mu\sys^\I\left(\bigvee \left(\rseq \setminus p(\vec{x})\right)\right) \\[2pt]
    & = & \left(\bigcup_{i=1}^n\mu\sys^\I\left(\exists \vec{x}_i . \bigwedge R_i(\vec{x}, \vec{x}_i) \right)\right) \cup \mu\sys^\I\left(\bigvee \left(\rseq \setminus p(\vec{x})\right)\right) \\[2pt]
    & =& \mu\sys^\I\left(\bigvee_{i=1}^n\exists \vec{x}_i . \bigwedge R_i(\vec{x}, \vec{x}_i) \right) \cup \mu\sys^\I\left(\bigvee \left(\rseq \setminus p(\vec{x})\right)\right) \\[2pt]
    & =& \mu\sys^\I\left(\bigvee_{i=1}^n\exists \vec{x}_i . \bigwedge R_i(\vec{x}, \vec{x}_i) \lor \bigvee \left(\rseq \setminus p(\vec{x})\right)\right) = \mu\sys^\I(\bigvee \rseq_1)
    \end{array}\]
    If there exists a valuation $\nu \in \mu\sys^\I(\bigwedge \lseq) \setminus \mu\sys^\I\left(\bigvee\rseq\right)$, then it is also the case that $\nu \in \mu\sys^\I(\bigwedge \lseq_1) \setminus \mu\sys^\I\left(\bigvee\rseq_1\right)$. Therefore, the counterexample for the antecedent is $\nu_1 = \nu$ and $\sem{\fv{\lseq_1}}_{\nu_1} \preceq^{\forall\exists}_\I \sem{\fv{\lseq}}_\nu$ holds trivially.
    
    \vspace{10pt}
    \paragraph{Case $(\RD)$.} Then the sequent $\lseq \vdash \rseq = \phi(\vec{x}, \vec{x}_1, \ldots, \vec{x}_n), p_1(\vec{x}_1),\ldots,p_n(\vec{x}_n) \vdash \{\exists \vec{y}_j . \psi_j (\vec{x}, \vec{y}_j) \land \mathcal{Q}_j(\vec{y}_j)\}_{j=1}^k$ has only one antecedent $\lseq_1 \vdash \rseq_1 = p_1(\vec{x}_1),\ldots,p_n(\vec{x}_n) \vdash \{\mathcal{Q}_j\theta \mid \theta \in S_j\}_{j=1}^i$. By the side condition of $(\RD)$, $\phi \models^\I \bigwedge_{j=1}^i \exists \vec{y}_j . \psi_j$.  Also, by Definition \ref{def:fol-fvi}, for each $\theta \in \subst{\phi,\psi_j}$, we have $\mu\sys^\I(\phi) \subseteq \mu\sys^\I(\psi_j\theta)$ for all $j \in [i]$. In this case, the least solution of $\rseq$ is
    \[\begin{array}{@{}lcl@{}}
    \mu\sys^\I\left(\bigvee\rseq\right) & =& \mu\sys^\I\left(\bigvee_{j=1}^k\exists \vec{y}_j . \psi_j \land \mathcal{Q}_j\right)\\[2pt]
    &\supseteq& \bigcup_{j=1}^k \mu\sys^\I\left(\exists \vec{y}_j . \psi_j \land \mathcal{Q}_j\right) \supseteq \bigcup_{j=1}^i \mu\sys^\I\left(\exists \vec{y}_j . \psi_j \land \mathcal{Q}_j\right) \\[2pt]
    & = &\bigcup_{j=1}^i \bigcup_{\theta \in \subst{\phi,\psi_j}} \mu\sys^\I\left((\psi_j \land \mathcal{Q}_j)\theta\right) \\[2pt]
    &=& \bigcup_{j=1}^i \bigcup_{\theta \in \subst{\phi,\psi_j}} \mu\sys^\I\left(\psi_j\theta \land \mathcal{Q}_j\theta\right) \\[2pt]
    & \supseteq& \bigcup_{j=1}^i \bigcup_{\theta \in \subst{\phi,\psi_j}} \left( \mu\sys^\I\left(\psi_j\theta\right) \cap \mu\sys^\I(\mathcal{Q}_j\theta)\right) \\[2pt]
    & = & \bigcup_{j=1}^i \bigcup_{\theta \in \subst{\phi,\psi_j}} \mu\sys^\I\left(\psi_j\theta\right) \cap \bigcup_{j=1}^i \bigcup_{\theta \in \subst{\phi,\psi_j}}\mu\sys^\I(\mathcal{Q}_j\theta)
    \end{array}\]
    Note that also \(\mu\sys^\I(\bigwedge \lseq)= \mu\sys^\I\left(\phi\right) \cap \mu\sys^\I\left(p_1(\vec{x}_1) \land \ldots \land p_n(\vec{x}_n)\right)\).
    If there exists a valuation $\nu \in \mu\sys^\I(\bigwedge \lseq) \setminus \mu\sys^\I\left(\bigvee\rseq\right)$, then
    \[\begin{array}{@{}l@{\;}@{}l@{}}
    \nu &\in \mu\sys^\I(\bigwedge \lseq) \setminus (\bigcup_{j=1}^i \bigcup_{\theta \in \subst{\phi,\psi_j}} \mu\sys^\I\left(\psi_j\theta\right) \cap \bigcup_{j=1}^i \bigcup_{\theta \in \subst{\phi,\psi_j}}\mu\sys^\I(\mathcal{Q}_j\theta)) \\
    & = \mu\sys^\I(\bigwedge \lseq) \setminus (\bigcup_{j=1}^i \bigcup_{\theta \in \subst{\phi,\psi_j}} \mu\sys^\I\left(\psi_j\theta\right)) \; \cup \\
    & \quad \mu\sys^\I(\bigwedge \lseq) \setminus (\bigcup_{j=1}^i \bigcup_{\theta \in \subst{\phi,\psi_j}}\mu\sys^\I(\mathcal{Q}_j\theta)) \\
    & = \emptyset \cup \mu\sys^\I(\bigwedge \lseq) \setminus (\bigcup_{j=1}^i \bigcup_{\theta \in \subst{\phi,\psi_j}}\mu\sys^\I(\mathcal{Q}_j\theta)) \\
    & \subseteq \mu\sys^\I\left(p_1(\vec{x}_1) \land \ldots \land p_n(\vec{x}_n)\right) \setminus (\bigcup_{j=1}^i \bigcup_{\theta \in \subst{\phi,\psi_j}}\mu\sys^\I(\mathcal{Q}_j\theta)) \\
    &\subseteq \mu\sys^\I\left(p_1(\vec{x}_1) \land \ldots \land p_n(\vec{x}_n)\right) \setminus (\bigcup_{j=1}^i \bigcup_{\theta \in S_j} \mu\sys^\I(\mathcal{Q}_j\theta)) \\
    & = \mu\sys^\I(\bigwedge \lseq_1) \setminus \mu\sys^\I(\bigvee \rseq_1)
    \end{array}\]
    Therefore, the counterexample for the antecedent is $\nu_1 = \nu$. Because $\phi$ is introduced to the left-hand side by left unfolding and $\sys$ is ranked, we have $\sem{\vec{x}_1 \cup \ldots \cup \vec{x}_n}_{\nu} \prec^{\forall\exists}_\I \sem{\vec{x}}_\nu$. Then $\sem{\fv{\lseq_1}}_{\nu_1} \prec^{\forall\exists}_\I \sem{\fv{\lseq}}_\nu$.
    
    \vspace{10pt}
    \paragraph{Case $(\RI)$.} Let $\rseq = \{p(\vec{x}) \land q(\vec{x}) \land \mathcal{Q}\} \cup \rseq'$. Then the antecedents of $\lseq \vdash \rseq$ are $\lseq_1 \vdash \rseq_1 = \lseq \vdash p(\vec{x}) \land \mathcal{Q},\rseq'$ and $\lseq_2 \vdash \rseq_2 = \lseq \vdash q(\vec{x}) \land \mathcal{Q},\rseq'$.
    In this case, the least solution of $\rseq$ is
    \[\begin{array}{@{}lcl@{}}
    \mu\sys^\I(\bigvee\rseq) & =& \mu\sys^\I\left(p(\vec{x}) \land q(\vec{x}) \land \mathcal{Q} \lor \bigvee \rseq'\right)\\[2pt]
    & =& \mu\sys^\I\left(p(\vec{x}) \land \mathcal{Q} \land q(\vec{x}) \land \mathcal{Q} \lor \bigvee \rseq'\right)\\[2pt]
    & =& \mu\sys^\I\left(p(\vec{x}) \land \mathcal{Q}\right) \cap \mu\sys^\I\left(q(\vec{x}) \land \mathcal{Q}\right) \cup \mu\sys^\I\left(\bigvee\rseq'\right) \\[2pt]
    & =& \left(\mu\sys^\I(p(\vec{x}) \land \mathcal{Q}) \cup \mu\sys^\I\left(\bigvee\rseq'\right)\right) \cap \left(\mu\sys^\I(q(\vec{x}) \land \mathcal{Q}) \cup \mu\sys^\I\left(\bigvee\rseq'\right)\right) \\[2pt]
    & = & \mu\sys^\I\left(p(\vec{x}) \land \mathcal{Q} \lor \bigvee\rseq'\right) \cap \mu\sys^\I\left(q(\vec{x}) \land \mathcal{Q} \lor \bigvee\rseq'\right) \\[2pt]
    & = & \mu\sys^\I(\bigvee \rseq_1) \cap \mu\sys^\I(\bigvee \rseq_2)
    \end{array}\]
    If there exists a valuation $\nu \in \mu\sys^\I(\bigwedge \lseq) \setminus \mu\sys^\I\left(\bigvee\rseq\right)$, then also $\nu \in \mu\sys^\I(\bigwedge \lseq) \setminus (\mu\sys^\I(\bigvee \rseq_1) \cap \mu\sys^\I(\bigvee \rseq_2)) = (\mu\sys^\I(\bigwedge \lseq) \setminus \mu\sys^\I(\bigvee \rseq_1)) \cup (\mu\sys^\I(\bigwedge \lseq) \setminus \mu\sys^\I(\bigvee \rseq_2))$. Therefore, $\nu \in \mu\sys^\I(\bigwedge \lseq_i) \setminus \mu\sys^\I(\bigvee \rseq_i)$ for some $i \in [2]$ and the counterexample for $\lseq_{i} \vdash \rseq_{i}$ is $\nu_i = \nu$. Because $\lseq = \lseq_{1} = \lseq_{2}$, we have $\sem{\fv{\lseq_i}}_{\nu_i} \preceq^{\forall\exists}_\I \sem{\fv{\lseq}}_\nu$.
    
    \vspace{10pt}
    \paragraph{Case $(\SP)$.} Then $\lseq \vdash \rseq = p_1(\vec{x}_1), \ldots, p_n(\vec{x}_n) \vdash \{\bigwedge_{i=1}^n q_i^j(\vec{x}_i) \}_{j=1}^k$. For each $\bar{\imath} \in [n]^{n^k}$, the antecedents of $\lseq \vdash \rseq$ are $\lseq^{\bar{\imath}}_{j} \vdash \rseq^{\bar{\imath}}_{j} =  p_{\bar{\imath}_j}(\vec{x}_{\bar{\imath}_j}) \vdash \{q_{\bar{\imath}_j}^\ell(\vec{x}_{\bar{\imath}_j}) \mid \ell \in [k],~ f_j(\overline{\mathcal{Q}}_\ell) = \bar{\imath}_j\}, j \in [n^k]$.
    
    If there exists a valuation $\nu \in \mu\sys^\I\left(\bigwedge_{i=1}^n p_i(\vec{x}_i)\right) \setminus \mu\sys^\I\left(\bigvee_{j=1}^k \bigwedge_{i=1}^n q_i^j(\vec{x}_i)\right)$, then as shown in the proof of Lemma \ref{lemma:sp}, there exists $j \in [n^k]$ and counterexamples $\nu_1, \ldots, \nu_n$ such that $\nu_i \in \mu\sys^\I(p_i(\vec{x}_i)) \setminus \bigcup\{\mu\sys^\I(q_i^\ell) \mid \ell \in [k],~ f_j(\overline{\mathcal{Q}}_\ell(\vec{x}_i)) )=i\}$ and $\nu(\vec{x}_i) = \nu_i(\vec{x}_i)$ for all $i \in [n]$. In other words, for all tuples $\bar{\imath} \in [n]^{n^k}$ we have \(\nu_{\bar{\imath}_j} \in \mu\sys^\I(p_{\bar{\imath}_j}(\vec{x}_{\bar{\imath}_j})) \setminus \bigcup \{\mu\sys^\I(q_{\bar{\imath}_j}^\ell(\vec{x}_{\bar{\imath}_j})) \mid \ell\in[k], f_j(\overline{\mathcal{Q}}_\ell) = \bar{\imath}_j\} = \mu\sys^\I(\bigwedge \lseq^{\bar{\imath}}_j) \setminus \mu\sys^\I(\bigvee \rseq^{\bar{\imath}}_j)\). Therefore, given such $j \in [n^k]$, the counterexample for each antecedent $\lseq^{\bar{\imath}}_{j} \vdash \rseq^{\bar{\imath}}_{j}$ is $\nu_{\bar{\imath}_j}$. Since $\fv{\lseq^{\bar{\imath}}_{j}} = \vec{x}_{\bar{\imath}_j} \subseteq \fv{\lseq}$ and $\nu(\vec{x}_{\bar{\imath}_j}) = \nu_i(\vec{x}_{\bar{\imath}_j})$, we have that $\sem{\fv{\lseq^{\bar{\imath}}_{j}}}_{\nu_i} \preceq^{\forall\exists}_\I \sem{\fv{\lseq}}_\nu$, for each $\bar{\imath} \in [n]^{n^k}$. \qed}

\subsection{Proof of Lemma \ref{lemma:direct-counterexample-path}}
\proof{Since $\pi$ is a direct path, it follows that $(\ID)$ does not occur in $\typelabel(\pi)$. Therefore, as shown in the proof of Lemma \ref{lemma:local-soundness}, which ensures the local soundness of $\RInd \setminus \{\ID\}$, we obtain a sequence of valuations $\nu_1 \nextval \ldots \nextval \nu_k$ such that $\sem{\fv{\lseq_1}}_{\nu_1} \succeq^{\forall\exists}_\I \ldots \succeq^{\forall\exists}_\I \sem{\fv{\lseq_k}}_{\nu_k}$.
    
    We know that $S(v_1) = \lseq_1 \vdash \rseq_{1}$,  $S(v_k) = \lseq_k \vdash \rseq_{k} = \lseq_1\theta \vdash \rseq'_{1}\theta$ with $\rseq_1 \subseteq \rseq'_{1}$ and $(\LU)$ occurs in $\typelabel(\pi)$, as required by the side condition of the $(\ID)$ instance applied at $v_k$. We show that $(\RD)$ is also required to occur in $\typelabel(\pi)$ by the following case analysis on the form of $\lseq_1$:
    \begin{enumerate}[label={(\roman*)}]
        \item If $\lseq_1 = \{p(\vec{x})\}$, then $(\LU)$ is the only inference rule applicable on $\lseq_1 \vdash \rseq$ which changes $\lseq$. This required application introduces a constraint on the left-hand side. In order to reach any sequent that has a constraint-free left-hand side, an application of $\RD$ is also required. To specifically reach $\lseq_1\theta\vdash \rseq'_1\theta$ with $\lseq_1\theta = \{p(\vec{x}\theta)\}$, $\typelabel(\pi)$ may contain additional occurrences of $(\LU)$ and $(\RD)$, but at least one of each is required; \label{item:rd-required-1}
        \item If $\lseq_1 = \{p_1(\vec{x}_1), \ldots, p_n(\vec{x}_n)\}$ with $n > 1$, then $(\LU)$ is immediately applicable, but it would introduce a constraint that cannot be reduced (this requires all arguments of predicate atoms to be subgoals in the constraint) thus making it impossible to reach any sequent with only predicate atoms on the left-hand side. Therefore, $(\SP)$ should be applied before the required $(\LU)$ and, later on, $(\RD)$ is needed to remove the constraint introduced by $(\LU)$. In order to specifically reach $\lseq_1\theta\vdash \rseq'_1\theta$ with $\lseq_1\theta = \{p_1(\vec{x}_1\theta), \ldots, p_n(\vec{x}_n\theta)\}$, $\typelabel(\pi)$ may contain additional occurrences of $(\SP)$, $(\LU)$ and $(\RD)$, but at least one $(\RD)$ is required; \label{item:rd-required-2}
        \item If $\lseq_1 = \{\phi, p_1(\vec{x}_1), \ldots, p_n(\vec{x}_n)\}$ with $n \geq 1$, then, just as in case \ref{item:rd-required-2}, applying $(\LU)$ before another inference rule that modifies the left-hand side introduces a second constraint impossible to reduce (this requires a single constraint on the left-hand side), thus preventing the reaching of any sequent with only one constraint on the left-hand side. Therefore, $(\RD)$ is necessary to remove $\phi$ before applying $(\LU)$. In order to specifically reach $\lseq_1\theta\vdash \rseq'_1\theta$ with $\lseq_1\theta = \{\phi\theta, p_1(\vec{x}_1\theta), \ldots, p_n(\vec{x}_n\theta)\}$, $\typelabel(\pi)$ may contain additional occurrences of $(\RD)$, $(\SP)$ and $(\LU)$, but at least one $(\RD)$ is required; \label{item:rd-required-4}
        \item If $\lseq_1 = \{\phi_1, \ldots, \phi_m\}$ with $m \geq 1$, then $(\LU)$ is not applicable on any path starting with $\lseq_1 \vdash \rseq_1$ because we cannot create a derivation with $\RInd$ that introduces predicate atoms on the left-hand side; \label{item:rd-required-3}
        \item If $\lseq_1 = \{\phi_1, \ldots, \phi_m, p_1(\vec{x}_1), \ldots, p_n(\vec{x}_n)\}$ with $m, n > 1$, then $(\LU)$ is the only inference rule applicable on $\lseq_1 \vdash \rseq_1$, and on all of the sequents derived from it, that modifies the left-hand side. Therefore, it is impossible to reach any sequent having $m$ constraints on the left-hand side, which includes $\lseq_1\theta\vdash \rseq'_1\theta$ with $\lseq_1\theta = \{\phi_1\theta, \ldots, \phi_m\theta, p_1(\vec{x}_1\theta), \ldots, p_n(\vec{x}_n\theta)\}$, because $(\LU)$ would introduce extra constraints in any derived sequent, which cannot be removed because $(\RD)$ is not applicable. \label{item:rd-required-5}
    \end{enumerate}
    It follows from cases \ref{item:rd-required-1}, \ref{item:rd-required-2}, \ref{item:rd-required-4}, i.e. the ones allowing the existence of the path $\pi$ with the necessary $(\LU)$ occurrence in $\typelabel(\pi)$, that $(\RD)$ is required to occur in $\typelabel(\pi)$. Then $\sem{\fv{\lseq_i}}_{\nu_i} \succ^{\forall\exists}_\I \sem{\fv{\lseq_{i+1}}}_{\nu_{i+1}}$ for some  $i \in [k-1]$, which leads to $\sem{\fv{\lseq_1}}_{\nu_1} \succ^{\forall\exists}_\I \sem{\fv{\lseq_{k}}}_{\nu_{k}}$. \qed}

\subsection{Proof of Theorem \ref{thm:ta-soundness}}
\proof{Suppose, by contradiction, that $\bigwedge\lseq
    \models_\sys^\I \bigvee\rseq$ does not hold, i.e.\ there exists a
    valuation $\nu_0 \in \mu\sys^\I(\bigwedge\lseq) \setminus
    \mu\sys^\I(\bigvee\rseq)$. Since $v_0$ is the root of the proof, it
    is also the consequent of an instance of $R(v_0)$ and, by Lemma
    \ref{lemma:local-soundness}, an antecedent of this inference rule
    has a counterexample $\nu_1$, such that $\nu_0 \nextval
    \nu_1$. Applying this argument iteratively, we build a path from
    $v_0$ to a leaf $v_k \in V$ and a sequence of valuations $\nu_0
    \nextval \nu_1 \nextval \ldots \nextval \nu_k$. 
    
    Now $R(v_k) \neq \AX$, because $(\AX)$ rules, by their side
    condition, cannot be applied on sequents that accept
    counterexamples. Since $v_k$ is a leaf, then
    $R(v_k) = \ID$ and let $v_{k+1}$ be the pivot of the instance of
    $(\ID)$ that applies at $v_k$. Then $S(v_k) = \lseq_k \vdash
    \rseq_k$ and necessarily $S(v_{k+1}) = \lseq_{k+1} \vdash
    \rseq_{k+1}$, where $\lseq_k = \lseq_{k+1}\theta$, 
    $\rseq_{k} = \rseq'_{k+1} \theta$ and $\rseq_{k+1} \subseteq
    \rseq'_{k+1}$, for some injective substitution $\theta :
    \fv{\lseq_{k+1} \cup \rseq_{k+1}} \rightarrow \fv{\lseq_k\cup
        \rseq_k}$. We can assume w.l.o.g. that $\theta$ is surjective,
    by defining $\theta(x) = x$ for each $x \in \fv{\lseq_k
        \cup \rseq_k} \setminus \theta(\fv{\lseq_{k+1}
        \cup \rseq_{k+1}})$. Since it is also injective, by the side condition
    of $\ID$, its inverse exists and $\nu_k\circ \theta^{-1}$ is a
    counterexample for $\lseq_{k+1} \vdash \rseq_{k+1}$. Thus we can
    extend the relation $\nextval$ with the pair
    $(\nu_k,\nu_k \circ \theta^{-1})$.
    
    This argument can be continued ad infinitum and we obtain an
    infinite trace $\tau = v_0,v_1, \ldots$ in $\mathcal{D}$ together
    with an infinite sequence of valuations $\nu_0 \nextval \nu_1
    \nextval \ldots$. If $S(v_i) = \lseq_i \vdash \rseq_i$, for each
    $i\geq0$, by Lemma \ref{lemma:local-soundness}, we have
    $\sem{\fv{\lseq_i}}_{\nu_i} \succeq_\I^{\forall\exists}
    \sem{\fv{\lseq_{i+1}}}_{\nu_{i+1}}$, for all $i\geq0$. By
    Proposition \ref{prop:direct-path}, $\tau$ contains infinitely many
    direct paths $\pi_j = v_{k_j}, \ldots, v_{\ell_j}$, where
    $\set{k_j}_{j\geq0}$ and $\set{\ell_j}_{j\geq0}$ are infinite
    strictly increasing sequences of integers such that $k_j < \ell_j
    \leq k_{j+1} < \ell_{j+1}$, for all $j\geq0$. By Lemma
    \ref{lemma:direct-counterexample-path}, we obtain that
    $\sem{\fv{\lseq_{k_j}}}_{\nu_{k_j}} \succ^{\forall\exists}_\I
    \sem{\fv{\lseq_{\ell_j}}}_{\nu_{\ell_j}}$, for all $j\geq0$. Since
    $\sem{\fv{\lseq_{k_j}}}_{\nu_{k_j}} \succeq^{\forall\exists}_\I
    \sem{\fv{\lseq_{k_{j+1}}}}_{\nu_{k_{j+1}}}$, for all $j\geq0$, we
    obtain a strictly decreasing sequence
    $\sem{\fv{\lseq_{k_0}}}_{\nu_{k_0}} \succ^{\forall\exists}_\I
    \sem{\fv{\lseq_{k_1}}}_{\nu_{k_1}} \succ^{\forall\exists}_\I
    \ldots$, which contradicts that $\succ^{\forall\exists}_\I$ is a
    wfqo, by Proposition \ref{prop:forall-exists-wqo}. We can thus
    conclude that there is no counterexample $\nu_0$ to start with, thus
    the entailment $\bigwedge\lseq \models_\sys^\I \bigvee\rseq$ holds.
    \qed}

\subsection{Proof of Lemma \ref{lemma:max-irred-finite}}

\proof{Let $\predno$ be the number of predicates in the inductive system
  $\sys$, $\ruleno$ the number of rules in $\sys$ and $\subgno$ the maximum
  number of subgoals occurring in the rules of $\sys$. Consider a structured
  derivation starting from a basic sequent $p(\vec{x}) \vdash q_1(\vec{x}),
  \ldots, q_n(\vec{x})$ and a path $\pi$ in this derivation that leads to another
  basic sequent $r(\vec{x}) \vdash s_1(\vec{x}), \ldots, s_m(\vec{x})$ without
  containing any other basic sequents. Clearly, only $\LU$, $\RU$, $\RI$,
  $\RD$ and $\SP$ can be applied on $\pi$, otherwise $\AX$ and $\ID$ would not
  allow us to reach the second basic sequent. The left-hand side of the
  sequents along the path $\pi$ allow us to apply $\LU$ at most once (on
  $p(\vec{x})$), $\RD$ at most once (after $\LU$, on the rule that replaces
  $p(\vec{x})$) and $\SP$ at most once (after $\RD$, if we are left with
  multiple predicates on the left-hand side). The right-hand side of the
  sequents along $\pi$ allows for $n$ applications of $\RU$, where $n$ is
  at most $\predno$, and $\ruleno * (\subgno - 1)$ applications of
  $\RI$ (there can be at most $\subgno - 1$ for every rule resulted from
  $\RU$ and reduced by $\RD$, and there can be at most $\ruleno$ rules).
  Thus, on any path between two consecutive basic sequents is of length
  at most $\baseno = 3 + \predno + \ruleno * (\subgno - 1)$, which
  is a constant determined by the system $\sys$.
	
  (\ref{it1:max-irred-finite}) Let $D$ be an irreducible
  structured derivation and suppose, by contradiction, that $\pi$ is
  an infinite path in $D$. Let $\rho$ is any subsequence of $\pi$ on
  which no ($\LU$) rule has been applied. The case in which $\rho$
  reaches its maximum possible length is when it starts right after the
  application of $\LU$ on a basic sequent and it extends until the next
  possible application of $\LU$, while encountering the next basic
  sequent and containing the results of applying all possible rules
  before $\LU$. The only rule that can be applied on a basic sequent
  before $\LU$ is $\RU$ and it can occur a maximum of $\predno$ times.
  Thus, $\rho$ must be finite, with a maximum length of $\baseno - 1 + 
  \predno$. Then ($\LU$) is applied infinitely often on
  $\pi$, and since $D$ is structured, also ($\RD$) must be applied
  infinitely often. But, since the antecedent of each application of
  an ($\RD$) rule contains no constraints, and moreover, the
  consequents thereof are of the form $\phi(\vec{x},
  \vec{x}_1,\ldots, \vec{x}_n), p_1(\vec{x}_1), \ldots,
  p_n(\vec{x}_n) \vdash \rseq$, the left hand side of such sequents
  must have been produced by a ($\LU$) rule with consequent of the
  form $p(\vec{x}) \vdash \rseq$, because ($\LU$) are the only rules
  introducing constraints on the left hand side of a sequent. But such
  sequents can only be the antecedents of ($\SP$) or ($\RD$) rules,
  with $n=1$ in the latter case. In the case of ($\SP$) rules, the
  right hand side $\rseq$ is a set consisting of predicates only, thus
  $p(\vec{x}) \vdash \rseq$ is a basic sequent. But this must be the
  case also for ($\RD$) rules, because $\sys$ has the fvi property and
  the assumption that ($\RI$) is used eagerly after an application of
  ($\RD$) to rule out conjunctions of predicates with the same
  argument list. Since there are finitely many predicates in $\sys$,
  the number of basic sequents is bounded thus some basic sequent must
  occur twice on $\pi$ and ($\ID$) is applicable, which contradicts
  the assumption that $D$ is irreducible. Then $\pi$ must be finite,
  and since it was chosen arbitrarily, we obtain that $D$ is finite,
  by an application of K\"onig's Lemma.

  (\ref{it2:max-irred-finite}) Suppose, by contradiction, that
  $\DS(\lseq \vdash \rseq)$ is infinite and let $D_1,D_2, \ldots$ be
  an infinite sequence of finite, maximal derivations of $\lseq \vdash
  \rseq$. By point (\ref{it1:max-irred-finite}), each derivation in
  $\DS(\lseq \vdash \rseq)$ is finite.  W.l.o.g. we assume that all
  $D_i$ are obtained by applying more than one rule --- in the opposite
  case, one can extract an infinite subsequence that satisfies this
  condition. The number of all possible basic sequents is $\predno *
  (2^{\predno} - 1)$, as there are $\predno$ predicates that can be on
  the left-hand side and $2^{\predno} - 1$ possible non-empty subsets
  of predicates on the right-hand side. Because we have a finite number
  of base sequents and the length of a path between two consecutive 
  base sequents is of finite length at most $\baseno$, then there must
  exist a derivation $D_i$ in the infinite sequence chosen above that
  contains a path in which the same basic sequent appears at least
  twice. But then this means that $D_i$ is reducible and, thus, that
  $D_i \not\in \DS(\lseq \vdash \rseq)$, which contrdicts our initial
  assumption. \qed}

\subsection{Proof of Lemma \ref{lemma:tree-shaped}}
\proof{By induction on the length of the path $\pi = p(\vec{x}) \vdash
    q_1(\vec{x}), \ldots, q_n(\vec{x}) = \lseq_1 \vdash \rseq_1,
    \ldots, \lseq_N \vdash \rseq_N
    = \lseq \vdash \rseq$ from the derivation in which $\lseq \vdash
    \rseq$ occurs. The case $N=1$ is trivial. Assuming that
    $\lseq_{N-1}$ and $\rseq_{N-1}$ are of the required form, we prove
    that $\lseq_N$ is tree-shaped and $\rseq_N$ consists of finite
    conjunctions of tree-shaped sets, in which all subgoal variables
    occur existentially quantified. We make a case split, based on the
    last inference rule on the path: \begin{compactitem}
        \item $(\LU)$ in this case $\lseq_{N-1}$ is tree-shaped and there
        exists a tree $t$ associated with $\lseq_{N-1}$ such that $t(\alpha) =
        r(\vec{y})$, for some frontier position $\alpha \in \fr(t)$ and
        $r(\vec{y}) \in \preds$. Then there exists a rule $R =
        \langle\{\phi(\vec{y},\vec{y}_1,\ldots,\vec{y}_m),r_1(\vec{x}_1),$
        $\ldots,r_h(\vec{y}_m)\},r(\vec{y})\rangle \in \sys$ such that
        $\lseq_N = R \cup \lseq_{N-1} \setminus \{r(\vec{y})\}$ and $\context{t}{q}
        \circ \tau_h(\phi(\vec{y},\vec{y}_1,\\ \ldots,\vec{y}_m),r_1(\vec{y}_1),
        \ldots, r_h(\vec{y}_m))$ replaces $t$ in the set of trees that
        represents $\lseq_N$.
        \item $(\RU)$ in this case there exists $r(\vec{y}) \in \rseq_{N-1}$ and a
        tree consisting of a single node labeled with $r(\vec{y})$. This
        tree is replaced in $\rseq_N$ by trees $t_1, \ldots, t_k$ corresponding to the
        tree-shaped sets $R_1(\vec{y},\vec{y}^1_1,\ldots,\vec{y}^1_{m_1}),
        \ldots, R_1(\vec{y},\vec{y}^m_1,\ldots,\vec{y}^k_{m_k})$, where
        $r(\vec{y}) \leftarrow_\sys R_1 \mid \ldots \mid R_m$ are the
        rules from the definition of $r(\vec{y})$. 
        \item the cases $(\RD)$ and $(\SP)$ are trivial, because the
        consequents of these rules consist of sequents of the form
        \(r_1(\vec{y}_1), \ldots, r_m(\vec{y}_m) \vdash
        \mathcal{Q}_1(\vec{y}_1,\ldots,\vec{y}_m), \ldots,
        \\ \mathcal{Q}_k(\vec{y}_1,\ldots,\vec{y}_m)\), where each
        $\mathcal{Q}_i$ is a conjunction of predicate atoms. \qed
\end{compactitem}}

\subsection{Proof of Lemma \ref{lemma:cex}}

\proof{ Let $\DS^*(\lseq \vdash \rseq)$ be the set of all derivations for
    $\lseq \vdash \rseq$ built with $\RInd$, together with all of their
    subtrees. Because, by Lemma \ref{lemma:max-irred-finite}, any
    derivation in $\DS(\lseq \vdash\rseq)$ is finite and $\DS(\lseq
    \vdash\rseq)$ itself is finite, then $\DS^*(\lseq \vdash \rseq)$ is
    also finite. Since $(\DS^*(\lseq \vdash \rseq),\sqsubseteq)$ is a wqo,
    it follows by Lemma \ref{lemma:finite-sets-wqo} that
    $(\finpow{\DS^*(\lseq \vdash \rseq)},\sqsubseteq^{\forall\exists})$ is
    also a wqo. Therefore, we can prove this lemma by induction on
    $(\finpow{\DS^*(\lseq \vdash \rseq)},\sqsubseteq^{\forall\exists})$.
    
    We have $\DS(\lseq \vdash \rseq) = \bigcup_{R} \DS_R(\lseq \vdash
    \rseq)$, where $\DS_R(\lseq \vdash \rseq)$ denotes the subset of
    $\DS(\lseq \vdash \rseq)$ consisting of derivations starting with an
    inference rule $R$. Observe that $R$ cannot be ($\AX$)
    because then no derivation in $\DS_R(\lseq \vdash \rseq)$ may contain
    a leaf $\lseq' \vdash \emptyset$, and $R$ cannot be $(\ID)$,
    because no derivation may start with an application of $(\ID)$. We
    distinguish the following remaining cases for $R$: \begin{compactitem}
        \item$(\LU)$ Let $r(\vec{y}) \in \lseq$ be the predicate atom chosen
        for replacement and $R_1,\lseq \setminus r(\vec{y}) \vdash \rseq,
        \ldots, R_m,\lseq \setminus r(\vec{y}) \vdash \rseq$ be the
        antecedents of $R$, where $r(\vec{y}) \leftarrow_\sys R_1 \mid
        \ldots \mid R_m$. It is sufficient to prove that there exists 
        $i \in [m]$ such that every derivation in $\DS(R_i,\lseq \setminus
        r(\vec{y}) \vdash \rseq)$ contains a leaf $\lseq' \vdash \emptyset$
        with no subgoals, and conclude by an application of the induction
        hypothesis. Suppose, by contradiction, that for each $i \in [n]$,
        there exists $D_i \in \DS(R_i,\lseq \setminus r(\vec{y}) \vdash
        \rseq)$ not containing such a leaf. Then there exists a derivation
        for $\lseq \vdash \rseq$ with the same property, which contradicts
        the hypothesis of the lemma. Thus, there must exist $i \in [m]$ such
        that every derivation $D \in \DS(R_i,\lseq \setminus r(\vec{y})
        \vdash \rseq)$ contains a leaf $\lseq' \vdash \emptyset$. By the
        induction hypothesis, there exists $\nu \in \mu\sys^\I(\bigwedge R_i
        \land \bigwedge (\lseq \setminus r(\vec{y}))) \setminus \mu\sys^\I(
        \bigvee \rseq) \subseteq \mu\sys^\I(\bigwedge\lseq) \setminus
        \mu\sys^\I(\bigvee\rseq)$, because $\mu\sys^\I(\bigwedge R_i \land
        \bigwedge(\lseq \setminus r(\vec{y}))) \subseteq \mu\sys^\I(
        r(\vec{y}) \land \bigwedge (\lseq \setminus r(\vec{y}))) =
        \mu\sys^\I(\bigwedge\lseq)$.
        \item$(\RU)$ Let $r(\vec{y}) \in \rseq$ be the predicate atom chosen
        for replacement, defined by $r(\vec{y}) \leftarrow_\sys
        R_1(\vec{y},\vec{z}_1) \mid \ldots \mid R_m(\vec{y},\vec{z}_m)$ and
        $\lseq \vdash \exists \vec{z}_1 ~.~ \bigwedge
        R_1(\vec{y},\vec{z}_1), \ldots, \exists \vec{z}_m \,.\, \\ \bigwedge
        R_m(\vec{y},\vec{z}_n)$ be the antecedent of $R$. Every $D \in
        \DS(\lseq \vdash \exists \vec{y}_1 \,.\, \bigwedge
        R_1(\vec{y},\vec{z}_1), \ldots, \\ \exists \vec{z}_n \,.\, \bigwedge
        R_m(\vec{y},\vec{z}_n),\rseq\setminus r(\vec{z}))$ contains a leaf
        $\lseq' \vdash \emptyset$, thus, by the induction hypothesis, there
        exists
        \[\begin{array}{rcl}
        \nu & \in & \mu\sys^\I(\bigwedge\lseq) \setminus\mu\sys^\I(\bigvee_{i=1}^n \exists \vec{y}_i \,.\, 
        \bigwedge R_i(\vec{x},\vec{y}_i) \vee \bigvee (\rseq\setminus r(\vec{y}))) \\ 
        & = & \mu\sys^\I(\bigwedge\lseq) \setminus\mu\sys^\I(\bigvee \rseq)
        \end{array}\]
        because
        \[\begin{array}{l}
        \mu\sys^\I(\bigvee_{i=1}^n \exists \vec{y}_i \,.\, \bigwedge
        R_i(\vec{x},\vec{y}_i) \vee \bigvee (\rseq\setminus r(\vec{y}))) = \\ 
        \mu\sys^\I(\bigvee_{i=1}^n \exists \vec{y}_i \,.\, \bigwedge R_i
        (\vec{x}, \vec{y}_i)) \cup \mu\sys^\I (\bigvee (\rseq\setminus r(\vec{y}))) = \\ 
        \bigcup_{i=1}^n\mu\sys^\I(\exists \vec{y}_i \,.\, \bigwedge R_i(\vec{x},\vec{y}_i)) 
        \cup (\mu\sys^\I(\bigvee \rseq) \setminus \mu\sys^\I(r(\vec{y}))) = \\ 
        \mu\sys^\I(r(\vec{y})) \cup (\mu\sys^\I(\bigvee \rseq) \setminus \mu\sys^\I(r(\vec{y}))) = \mu\sys^\I(\bigvee \rseq)
        \end{array}\]
        \item$(\RD)$ Let $\lseq =
        \set{\phi(\vec{y},\vec{y}_1,\ldots,\vec{y}_m),r_1(\vec{y}_1),\ldots,
            r_m(\vec{y}_m)}$ and $\rseq = \{\exists \vec{z}_1 \,.\,
        \psi_1(\vec{y},\\ \vec{z}_1) \wedge \mathcal{Q}_1(\vec{z}_1),
        \ldots, \exists \vec{z}_k \,.\, \psi_k(\vec{y},\vec{z}_k) \wedge
        \mathcal{Q}_k(\vec{z}_k)\}$, where $\phi, \psi_1, \ldots, \psi_k$
        are constraints, $r_1,\ldots,r_m$ are predicates, and
        $\mathcal{Q}_1, \ldots, \mathcal{Q}_k$ are conjunctions of
        predicates. W.l.o.g. we assume that $S_j = \subst{\phi,\psi_j}$,
        for all $j \in [k]$. Since $\sys$ has the fvi property, each $S_j$
        is finite. We distinguish the following cases: \begin{compactitem}
            \item $m=0$, i.e.\ $\lseq$ contains no predicate atoms. Since
            $p(\vec{x}) \vdash q_1(\vec{x}), \ldots, q_n(\vec{x}) \leadsto \lseq \vdash \rseq$, by
            Lemma \ref{lemma:tree-shaped}, $\lseq$ is tree-shaped, meaning
            that $\lseq = \phi(\vec{y})$, and $\phi$ has no subgoal
            variables. Again, we distinguish two cases: \begin{compactitem}
                \item if $k=0$ then $\rseq = \emptyset$ and, since $\phi$ is a
                constraint from $\sys$, it must be satisfiable. Thus any model
                of $\phi$ contradicts the entailment $\phi(\vec{y}) \models^\I
                \bot$, and moreover, such a model exists, because $\sys$ is
                non-filtering.
                \item else, if $k>0$ we have that $\phi(\vec{y}) \not\models^\I
                \bigvee_{j=1}^k \exists \vec{z}_j \,.\,
                \psi_j(\vec{y},\vec{z}_j)$ for each $j \in [k]$, because
                $\sys$ has the fvi property. In this case we can trivially
                find a counterexample for the entailment $\phi(\vec{y})
                \models^\I \bigvee\rseq$.
            \end{compactitem}
            \item $m>0$, then the antecedent of the rule $(\RD)$ is
            $r_1(\vec{y}_1),\ldots,r_m(\vec{y}_m) \vdash
            \{\mathcal{Q}_j\theta \mid \theta \in
            \subst{\phi,\psi_j}\}_{j=1}^i$. Moreover, by the side condition
            of the rule, we have $\phi \models^\I \bigwedge_{j=1}^i \exists
            \vec{z}_j \,.\, \psi_j$ and $\phi \not\models^\I \bigvee_{j=i+1}^k
            \exists \vec{z}_j \,.\, \psi_j$, via a possible reordering of
            $\rseq$.  Since every derivation $D \in
            \DS(r_1(\vec{y}_1),\ldots,r_m(\vec{y}_m) \vdash
            \{\mathcal{Q}_j\theta \mid \theta \in
            \subst{\phi,\psi_j}\}_{j=1}^i)$ must contain a leaf $\lseq'
            \vdash \emptyset$, by the induction hypothesis there must exist
            a counterexample $\nu \in\mu\sys^\I(\bigwedge_{\ell=1}^m
            p_\ell(\vec{y}_\ell)) \setminus
            \mu\sys^\I(\mathcal{Q}_j\theta)$, for all
            $j \in [i]$ and all $\theta \in \subst{\phi,\psi_j}$. Because
            we assumed that $\sys$ is non-filtering, there exists $\nu'$ such
            that $\I, \nu' \models \phi$ and $\nu$ and $\nu'$ agree on $\vec{y}_1,
            \ldots, \vec{y}_m$. Furthermore, because $\sys$ is assumed to be 
            non-overlapping and $\phi \not\models^\I \exists \vec{z}_j \,.\, \psi_j$, 
            for all $j \in [i+1,k]$, we obtain that $\phi \wedge \exists \vec{z}_j 
            \,.\, \psi_j$ is unsatisfiable, hence $\nu'$ is also a counterexample
            for the entailment $\phi \models^\I \exists \vec{z}_j \,.\, \psi_j$, 
            for each $j \in [i+1,k]$ and thus for the entailment $\phi 
            \models^\I \bigvee_{j=i+1}^k \exists \vec{z}_j \,.\, \psi_j \wedge 
            \mathcal{Q}_j$. 
            Suppose now, by contradiction, that $\nu'$ is a model of $\exists
            \vec{z}_j \,.\, \psi_j(\vec{y},\vec{z}_j) \wedge
            \mathcal{Q}_j(\vec{z}_j)$, for some $j \in [i]$. Then $\nu'$ is a
            model of $\exists \vec{z}_j \,.\, \psi_j(\vec{y}, \vec{z}_j)$
            also.  Since $\phi \models^\I \exists \vec{z}_j \,.\, \psi_j$ and
            $\sys$ has the fvi property, it must be the case that $\phi
            \models^\I \psi_j\theta$ for all $\theta \in \subst{\phi,\psi_j}$
            and, moreover, there is no other Skolem function that witnesses
            this entailment, besides the ones in $\subst{\phi,\psi_j}$.  But
            then it must be that $\nu'$ is a model of $\psi_j\theta$, for all
            $\theta \in \subst{\phi,\psi_j}$, and only for those
            substitutions. Since the range of each $\theta \in
            \subst{\phi,\psi_j}$ is $\vec{y}_1 \cup \ldots \cup\vec{y}_m$, we
            have that $\nu'\in \mu\sys^\I(\mathcal{Q}_j\theta)$ for some
            $\theta \in \subst{\phi,\psi_j}$, which contradicts the assumption
            that $\nu'$ is a counterexample of the antecedent. Then $\nu'$
            cannot be a model of $\exists \vec{z}_j \,.\,
            \psi_j(\vec{y},\vec{z}_j) \wedge\mathcal{Q}_j(\vec{z}_j)$, for
            some $j \in [i]$, and since it cannot be a model of the right-hand
            side for $j \in [i+1,k]$ either, it is a counterexample for the
            entailment $\bigwedge\lseq \models^\I \bigvee\rseq$, as required.
        \end{compactitem}
        
        \item $(\SP)$ Every derivation for $\lseq \vdash \rseq$ starts with
        the following inference rule, for some tuple of indices $(i_1,
        \ldots, i_{m^k}) \in [m]^{m^k}$:
        \[\infer{r_1(\vec{y}_1),\ldots,r_m(\vec{y}_m) \vdash 
            \mathcal{Q}_1(\vec{y}_1,\ldots,\vec{y}_m), \ldots,
            \mathcal{Q}_k(\vec{y}_1,\ldots,\vec{y}_m)}{\langle
            r_{i_1}(\vec{y}) \vdash \{q_{i_1}^\ell(\vec{y}) \mid \ell \in
            [k],~ f_1(\overline{\mathcal{Q}}_\ell) =
            i_1\}\rangle_{j=1}^{m^k}}\] By our assumption, each derivation $D
        \in \DS(\lseq \vdash \rseq)$ contains a leaf $\lseq' \vdash
        \emptyset$. Suppose, by contradiction, that there exists a tuple
        $(i_1,\ldots,i_{m^k}) \in [m]^{m^k}$ such that for all $j \in [m^k]$
        there exists $D \in \DS(r_{i_j}(\vec{y}) \vdash
        \{q_{i_j}^\ell(\vec{y}) \mid \ell \in [k],~
        f_j(\overline{\mathcal{Q}}_\ell) = i_j\})$ does not have a leaf
        $\lseq' \vdash \emptyset$. Then we can build a derivation for $\lseq
        \vdash \rseq$ that does not contain any such leaves, which
        contradicts our assumption. Thus it must be the case that for all
        tuples $(i_1,\ldots,i_{m^k}) \in [m]^{m^k}$ there exists $j \in
        [m^k]$ such that for all derivations $D \in \DS(r_{i_j}(\vec{y})
        \vdash \{q_{i_j}^\ell(\vec{y}) \mid \ell \in [k],~
        f_j(\overline{\mathcal{Q}}_\ell) = i_j\})$ contains a leaf $\lseq'
        \vdash \emptyset$. By the inductive hypothesis, for all tuples
        $(i_1,\ldots,i_{m^k}) \in [m]^{m^k}$ there exists $j \in [m^k]$ such
        that $\mu\sys (r_{i_j}(\vec{y})) \not\subseteq
        \mu\sys^\I(\{q_{i_j}^\ell(\vec{y}) \mid \ell \in [k],~
        f_j(\overline{\mathcal{Q}}_\ell) = i_j\})$. Then, by Lemma
        \ref{lemma:sp}, we have $\mu\sys^\I(r_1) \times \ldots \times
        \mu\sys^\I(r_m) \not\subseteq \bigcup_{i=1}^k \mu\sys^\I(q_1^i)
        \times \ldots \times \mu\sys^\I(q_{m}^i)$, proving the claim. \qed
    \end{compactitem}
}

\subsection{Proof of Theorem \ref{thm:ta-completeness}} \label{app:ta-completeness}

\proof{Since $\sys$ is non-filtering and non-overlapping, and
    moreover, it has the fvi property and also $p\models_\sys^\I q_1, \ldots, q_n$, by
    Lemma \ref{lemma:cex}, there exists a finite maximal, structured and
    irreducible derivation $D \in \DS(p(\vec{x}) \vdash q_1(\vec{x}), \ldots, q_n(\vec{x}))$
    which does not contain a leaf $\lseq \vdash \emptyset$. But then, no
    node in $D$ is of the form $\lseq \vdash \emptyset$, because all
    descendants of such a node must have empty right-hand sides as well.
    
    We first show that this derivation is actually a proof (i.e. all its
    leaves are $\top$). Suppose there exists a leaf that is not $\top$.
    This means that the leaf is a sequent $\lseq \vdash \rseq$, where
    $\rseq \neq \emptyset$. Let $\pi$ be the path in $D$ leading to this
    leaf. Since $D$ is a maximal derivation, $\pi$ cannot be extended
    any further by the application of an inference rule. Assume that the
    last inference rule applied on $\pi$ is $\IR$ and has the
    consequent $\lseq' \vdash \rseq'$. Since $p(\vec{x}) \vdash
    q_1(\vec{x}), \ldots, q_n(\vec{x}) \leadsto \lseq' \vdash \rseq'$, by Lemma
    \ref{lemma:tree-shaped}, $\lseq'$ is a tree-shaped set and $\rseq'$
    consists of existentially quantified finite conjunctions of
    tree-shaped sets. Also, $\IR$ cannot be $(\AX)$ or $(\ID)$, because then
    the leaf would be $\top$. We do a case split based on $R$: \begin{compactenum}
        \item $(\LU)$ Then $\lseq \vdash \rseq$ is of the form
        $R,\lseq' \setminus r(\vec{y}) \vdash \rseq'$, where
        $\tuple{R,r(\vec{y})} \in \sys$. If $\rseq'$ contains
        at least one predicate, we can still apply $(\RU)$, which
        contradicts the fact that $D$ is maximal. Otherwise, because
        $\rseq' \neq \emptyset$ consists of existentially quantified
        finite conjunctions of tree-shaped sets and it does not contain
        any predicates, then it must be the case that $\rseq'$ contains
        only existentially quantified conjunctions over rules from $\sys$,
        obtained from previous applications of $(\RU)$, or predicate
        conjunctions that are not singleton, obtained from previous
        applications of $(\RD)$. We distinguish the following cases:
        \begin{compactitem}
            \item $\lseq' \setminus r(\vec{y}) = \emptyset$. Then we can
            apply $(\RD)$ to the sequent $R \vdash \rseq'$ and
            extend $D$, which results in a contradiction.
            \item $\lseq' \setminus r(\vec{y}) \neq \emptyset$. If $(\IR)$
            is the first occurrence of a rule $(\LU)$ then it must be
            the case that $\lseq' \setminus r(\vec{y}) = \emptyset$,
            which contradicts our assumption. Then there must have been
            a previous application of $(\LU)$ on $\pi$, and, because $D$
            is structured, $(\RD)$ must have been applied between
            them. Therefore, since $\lseq'$ is tree-shaped, $\lseq'
            \setminus r(\vec{x})$ can only contain predicates, because
            the constraints introduced by $(\LU)$ are always eliminated
            by $(\RD)$. Then we can apply $(\LU)$ and extend $D$, which
            leads to a contradiction.
        \end{compactitem}
        \item $(\RU)$ Then $\lseq \vdash \rseq$ is of the form $\lseq'
        \vdash \{\exists \vec{z}_i ~. \bigwedge
        R_i(\vec{y},\vec{z}_i)\}_{i=1}^m, \rseq' \setminus
        r(\vec{y})$, where $r(\vec{y})$ is a predicate atom and
        $r(\vec{y}) \leftarrow_\sys R_1 \mid \ldots \mid R_m$. If $\rseq'
        \setminus r(\vec{y})$ contains at least one predicate, we can
        apply $(\RU)$ and extend $D$, contradiction. Otherwise, because
        $\rseq' \neq \emptyset$ consists of existentially quantified
        finite conjunctions of tree-shaped sets and it does not contain
        any predicates, then it must be the case that $\rseq'$ contains
        only existentially quantified conjunctions over rules from $\sys$,
        obtained from previous applications of $(\RU)$, or predicate
        conjunctions that are not singleton, obtained from previous
        applications of $(\RD)$. We distinguish the following cases:
        \begin{compactitem}
            \item $\lseq'$ contains a predicate atom. Then we can apply $(\LU)$
            and extend $D$, contradiction. 
            \item $\lseq'$ does not contain predicate atoms. Because
            $\lseq'$ is tree-shaped and $D$ is structured, $\lseq'$ can
            only contain a constraint with no subgoal variables. Then we
            can apply $(\RD)$ and extend $D$, contradiction.
        \end{compactitem}  
        \item $(\RD)$ Then $\lseq \vdash \rseq$ is of the form
        $r_1(\vec{y}_1),\ldots,r_m(\vec{y}_m) \vdash
        \mathcal{Q}_1(\vec{y}_1, \ldots, \vec{y}_m), \ldots,$
        $\mathcal{Q}_k(\vec{y}_1,\ldots, \vec{y}_m)$ and we can apply
        $(\LU)$ -- or even $(\RI)$, $(\RU)$ or $(\SP)$ if possible -- to
        $\lseq \vdash \rseq$, which means that we can still extend $\pi$,
        leading to a contradiction.
        \item $(\RI)$ $\lseq \vdash \rseq$ is of the form $\lseq \vdash r(\vec{y})
        \land \mathcal{Q}, \rseq''$. Since we only apply $(\RI)$ as cleanup after $(\RD)$, 
        $\lseq$ only contains predicate atoms and $\mathcal{Q}$ is a conjunction of
        predicate atoms. Then we can continue to apply $(\RI)$ if $r(\vec{y}) \land \mathcal{Q}$ or any member of $\rseq''$ contains some conjunction $s_1(\vec{z}) \land s_2(\vec{z})$, or apply $(\LU)$, $(\RU)$, or $(\SP)$,
        leading to a contradiction.
        
        \item $(\SP)$ Then $\lseq \vdash \rseq$ is of the form $r(\vec{y})
        \vdash s_1(\vec{y}), \ldots, s_m(\vec{y})$ and we can apply
        $(\LU)$ or $(\RU)$ to $\lseq \vdash \rseq$, which means that we
        can still extend $\pi$ and leads to a contradiction.
    \end{compactenum}  
    
    We will now show that the sequence of inference rules fired on each
    maximal path in $D$ is captured by the strategy $\mathbf{S}$. Let $\pi$ be
    an arbitrary maximal path in $D$. Since $D$ is a maximal derivation, $\pi$
    cannot be extended any further by the application of an inference
    rule. W.l.o.g. we assume that the first application of $(\LU)$ is
    not immediately preceded by an application of $(\RU)$ --- otherwise,
    one can obtain the same sequent by first applying $(\LU)$ before any
    $(\RU)$. The proof goes by induction on
    the number $N\geq1$ of basic sequents that occur on $\pi$.
    
    \vspace*{\baselineskip}\noindent If $N=1$, then the only basic
    sequent $p(\vec{x}) \vdash q_1(\vec{x}), \ldots, q_n(\vec{x})$ occurs on the first position
    of $\pi$. In this case $(\SP)$ is never applied on $\pi$, because
    its antecedent is a basic sequent, and thus $N>1$, contradiction.
    We distinguish two cases: \begin{compactenum}
        \item If $(\LU)$ is not applied on $\pi$, the only possibility is to
        directly apply $\AX \in \mathbf{S}$ to $p(\vec{x}) \vdash
        q_1(\vec{x}), \ldots, q_n(\vec{x})$, thus ending the path. Otherwise, $(\LU)$ is enabled,
        which contradicts the maximality of $\pi$.
        \item Else, if $(\LU)$ is applied on $\pi$, then it must be applied
        in the beginning, because only $(\LU)$ and $(\RU)$ are applicable
        on $p(\vec{x}) \vdash q_1(\vec{x}), \ldots, q_n(\vec{x})$ and we assumed that no instance of $(\RU)$ immediately precedes $(\LU)$. Assume that the first rule
        application on $\pi$ is:
        \[\infname{\LU} \infer[]{p(\vec{x}) \vdash q_1(\vec{x}), \ldots, q_n(\vec{x})}{\lseq' \vdash \rseq'}\]    
        Then $\tuple{\lseq',p(\vec{x})} \in \sys$ and $(\LU)$ cannot be
        applied again without applying $(\RD)$ first, due to the
        assumption that $D$ is structured. Since $\rseq' =
        \{q_1(\vec{x}), \ldots, q_n(\vec{x})\}$ after the first application of $(\LU)$, we can
        now apply $(\RU)$. Because $(\SP)$ is never applied on $\pi$,
        either ($\AX$) or $(\RD)$ can be applied next. In the first case,
        we obtain $\typelabel(\pi) \in \LU \cdot \RU \cdot \AX \in
        \mathbf{S}$. In the second case, if $n=1$ in the antecedent of
        $(\RD)$, since $\sys$ has the fvi property, we obtain that the
        antecedent of $(\RD)$ is a basic sequent, contradicting our
        assumption. Then it must be the case that $n>1$, and now
        $(\RD)$ is not applicable any longer, because the number of
        predicate atoms will always be bigger than the number of subgoal
        variables in the constraint, on the left-hand side. The only
        possibilities for continuation are then ($\AX$), $(\ID)$, $(\RI)$,
        $(\LU)$ and $(\RU)$. However, $(\RI)$ is applicable only a finite 
        number of times, equal to the number of predicate conjunctions with 
        the same arguments on the right and side, $(\RU)$ can also only be 
        applied a finite number of times, equal to the number of singleton predicate atoms
        on the right-hand side, and $(\LU)$ is applicable at most once, 
        because $(\RD)$ is no longer applicable. In both cases, $\pi$ is 
        not maximal, because $(\LU)$ is enabled. Then the only possibility
        is to end the path by ($\AX$) or $(\ID)$, obtaining $\LU \cdot \RU 
        \cdot \RD \cdot \RI^* \cdot \LU? \cdot \RU^* \cdot (\AX \mid \ID)
        \subseteq \mathbf{S}$.
    \end{compactenum}
    
    \noindent If $N>1$, let $\pi = \tau \cdot \rho$, where $\rho$ starts
    with the second occurrence of a basic sequent in $\pi$. As before,
    the first occurrence is the initial sequent $p(\vec{x}) \vdash
    q_1(\vec{x}), \ldots, q_n(\vec{x})$. Then the inference rule applied on the
    last vertex of $\tau$ is either $(\SP)$ or $(\RD)$. In the latter case,
    the sequent has a single predicate atom on the left-hand side. In the
    former case, there is an application of $(\RD)$ optionally followed by
    several applications of $(\RI)$ preceding the last vertex of $\tau$, and consider now the consequent of the last application of $(\RD)$
    on a vertex in $\tau$. As argued before, this consequent is a predicate rule of
    $\sys$ with goal $r(\vec{y})$, introduced by a previous
    application of $(\LU)$. Since, between the antecedent of this $(\LU)$ instance and
    the consequent of $(\RD)$, the left-hand side of the sequents is
    unchanged, the only possibility is that $(\RU)$ has been used between them, thus $\typelabel(\pi) \in \LU\cdot\RU^*\cdot \RD \cdot \RI^* \cdot
    \SP? \cdot \typelabel(\rho)$. By the inductive hypothesis, the sequence of rules on
    $\typelabel(\rho) \in \mathbf{S}$, thus $\typelabel(\pi) \in
    (\LU\cdot\RU^*\cdot \RD \cdot \RI^* \cdot \SP?) \cdot \mathbf{S} \subseteq
    \mathbf{S}$. \qed}

\subsection{Proof of Lemma \ref{lemma:sl-non-filtering}}

\proof{The abstraction we need is defined as the least fixed point of an
    operator $\mathbb{F}^\sharp_\sys$, denoted $\mu\sys^\sharp$. An
    \emph{abstract assignment} $\Y$ is a mapping of predicates
    $p^{\sigma_1 \ldots \sigma_n}$ into sets of pairs $(A,E)$, where $A
    \in \pow{[n]}$ is a set of allocated arguments and $E \subseteq [n]
    \times [n]$ is a set of equality constraints such that, for each model
    $(\tuple{\ell_1,\ldots,\ell_n},h) \in \mu\sys^\tinyseplog(p)$ there
    exists a pair $(A,E) \in \mu\sys^\sharp(p)$ such that $A = \set{i \in
        [n] \mid \ell_i \in \dom(h)}$ and $(i,j) \in E$ if and only if
    $\ell_i = \ell_j$. Given a quantifier-free $\seplog$ formula
    $\varphi(\vec{x})$, we define the following sets:
    \[\begin{array}{rcl}
    \mayalloc(\varphi) & = & \set{x \in \fv{\varphi} \mid \varphi \wedge \exists z_1 \ldots \exists z_k \,.\, x \mapsto (z_1,\ldots,z_k) * \top \text{ is satisfiable}} \\
    \mustalloc(\varphi) & = & \set{x \in \fv{\varphi} \mid \varphi \models^\tinyseplog \exists z_1 \ldots \exists z_k \,.\, x \mapsto (z_1,\ldots,z_k) * \top} \\
    \eqset(\varphi) & = & \set{(x,y) \in \fv{\varphi} \times \fv{\varphi} \mid \varphi \models^\tinyseplog x \teq y} 
    \end{array}\]
    Computing the above sets can be done in polynomial space, in general,
    using the decision procedures for quantifier-free \cite{CYOH01} and
    Bernays-Schoenfinkel-Ramsey $\seplog$ formulae \cite{Vmcai17}, and in
    polynomial time for symbolic heaps, respectively. Dually, we consider
    the formulae $\mathbf{A}(X) = \Asterisk_{x\in X} \exists z_1 \ldots
    \exists z_k \,.\, x \mapsto (z_1,\ldots,z_k)$ and $\mathbf{E}(R) =
    \bigwedge_{(x,y) \in R} x \teq y \wedge \bigwedge_{(x,y) \not\in R}
    \neg x\teq y$, for any set $X \subseteq \vars$ and relation $R
    \subseteq \vars \times \vars$ on variables.
    
    Given $R = \langle \{\phi(\vec{x}, \vec{y}_1,\ldots,\vec{y}_m),
    q_1(\vec{y}_1), \ldots, q_m(\vec{y}_m) \}, p(\vec{x}) \rangle \in
    \sys$, let $\vec{y}_i = \langle y^i_1, \ldots, y^i_{n_i} \rangle$ for
    all $i \in [m]$. For a tuple $\vec{P} = \langle (A_1,E_1), \ldots,
    (A_m,E_m) \rangle \in \mu\sys^\sharp(q_1) \times \ldots \times
    \mu\sys^\sharp(q_m)$ and a relation $C$ on the free variables of
    $\phi$, we define:
    \[\begin{array}{rcl}
    \omega_R(\vec{P},C) & \equiv & \Asterisk_{i \in [m]} \mathbf{A}(\{x_j
    \mid j \in A_i\}) \wedge \mathbf{E}(\{(y^i_r,y^i_s) \mid (r,s) \in
    E_i,~ i \in [m]\}) \wedge \mathbf{E}(C) \\ \eta_R(\vec{P},C) & \equiv
    & \exists \vec{y}_1 \ldots \exists \vec{y}_m \,.\, \phi(\vec{x},
    \vec{y}_1,\ldots,\vec{y}_m) * \omega_R(\vec{P},C)
    \end{array}\]
    Then an abstract assignment $\mathbb{F}_\sys^\sharp(\Y)$ maps each
    predicate $p$ into the set $\bigcup_{i=1}^m \Y(R_i)$, where
    $p(x_1,\ldots,x_n) \leftarrow_\sys R_1 \mid \ldots \mid R_m$ and
    $\Y(R)$ is the set of pairs $(A,E)$ for which there exists a tuple of
    pairs $\vec{P} \in \mu\sys^\sharp(q_1) \times \ldots \times
    \mu\sys^\sharp(q_m)$ and a relation $C \subseteq \fv{\phi} \times
    \fv{\phi}$ such that:
    \[\begin{array}{rcl}
    A & = & \{i \in [n] \mid x_i \in \mathcal{A},~\mustalloc(\eta_R(\vec{P},C)) 
    \subseteq \mathcal{A} \subseteq \mayalloc(\eta_R(\vec{P},C)) \} \\ 
    E & = &  \{(i,j) \in [n] \times [n] \mid (x_i,x_j) \in
    \eqset(\eta_R(\vec{P},C))\}
    \end{array}\]
    
    If $\eta_R(\vec{P},C)$ is unsatisfiable,
    \(\mustalloc(\eta_R(\vec{P},C)) = \vec{x}\) and
    \(\mayalloc(\eta_R(\vec{P},C)) = \emptyset\), and there is no choice
    for the set $\mathcal{A}$, thus no corresponding pair
    $(A,E)$.

  If $p(\vec{x}) \leftarrow_\sys R_1 \mid \ldots \mid R_m$ are all the predicate rules for $p$ in $\sys$, then the abstract assignment $\mathbb{F}_\sys^\sharp(\Y)$ maps the predicate $p$ to the set $\mathbb{F}_\sys^\sharp(\Y)(p) = \bigcup_{i=1}^m \Y(R_i)$. Similarly to $\mathbb{F}^\tinyseplog_\sys$ and $\mu\sys^\tinyseplog$, the operator $\mathbb{F}_\sys^\sharp$ is monotone and continuous, with $\mu\sys^\sharp$ as its least fixed point.
  
  Considering again the arbitrary predicate rule $R \in \sys$ described above, let $\vec{P} = \langle (A_1,E_1), \\ \ldots, (A_m,E_m) \rangle \in \mu\sys^\sharp(q_1) \times \ldots \times \mu\sys^\sharp(q_m)$ be a tuple of pairs and $Rel \subseteq \fv{\phi} \times \fv{\phi}$ a relation on variables, such that $\omega_R(\vec{P},Rel)$ is satisfiable. We claim that, if the formula $\phi * \omega_R(\vec{P},Rel)$ is satisfiable, then for each tuple of models $\langle (\overline{\ell}_1, h_1), \ldots, (\overline{\ell}_m,h_m) \rangle \in \mu\sys^\tinyseplog(q_1) \times \ldots \times \mu\sys^\tinyseplog(q_m)$, there exist a valuation $\nu$ and a heap $h$ such that $\nu,h \models^\tinyseplog \phi$ and $\nu,\biguplus_{i=1}^m h_i \models^\tinyseplog \omega_R(\vec{P},Rel)$, where $\nu(\vec{y}_i) = \overline{\ell}_i$, $\forall i \in [m]$, and $\dom(h) \cap (\bigcup_{i=1}^m \dom(h_i)) = \emptyset$.
  
  The proof idea for this claim is that, because $\omega_R(\vec{P},Rel)$ specifies exactly those variables which are allocated, as well as those which are not, and the pairs of variables which are equal, along with the ones which are not, the truth value of $\phi * \omega_R(\vec{P},Rel)$ is invariant under the renaming of the values of $\vec{x} \cup \bigcup_{i=1}^m \vec{y}_i$, as long as the allocations and equalities are preserved. Moreover, each tuple of models $\langle (\overline{\ell}_1, h_1), \ldots, (\overline{\ell}_m,h_m) \rangle \in \mu\sys^\tinyseplog(q_1) \times \ldots \times \mu\sys^\tinyseplog(q_m)$ is a model of $\omega_R(\vec{P},Rel)$, for some $\vec{P} = \langle (A_1,E_1), \ldots, (A_m,E_m) \rangle \in \mu\sys^\sharp(q_1) \times \ldots \times \mu\sys^\sharp(q_m)$ and $Rel \subseteq \fv{\phi} \times \fv{\phi}$. Then, for each predicate rule $R \in \sys$ we need to check the satisfiability of $\phi * \omega_R(\vec{P},Rel)$, for each $\vec{P}$ and $Rel$, such that $\omega_R(\vec{P},Rel)$ is satisfiable.
  
  Since there are finitely many variables in $\sys$, for a predicate $p^{\sigma_1 \ldots \sigma_n}$ of $\sys$ the set of pairs $(A,E)$ is finite, of cardinality at most $2^{n+n^2}$. Then $\mu\sys^\sharp(p)$ can be computed in an exponential number of steps\footnote{See \cite[Lemma 4.6]{Brotherston14} for an analogous construction for inductive systems with symbolic heap constraints.}, each step requiring polynomial space. These pairs can be stored in a table that requires $2^{\mathcal{O}(n^2)}$ space, indexed by $\mathcal{O}(n^2)$ bits, where each pair occupies $\mathcal{O}(n)$ bits. Checking the satisfiability of a formula $\phi * \omega_R(\vec{P},Rel)$ is possible in polynomial space.
  
  In conclusion, we can check if a predicate rule in $\sys$ is non-filtering, by checking the satisfiability of an exponential number of $\seplog$ formulae, where each satisfiability check can be done in polynomial space. Thus, the overall complexity of checking if an $\seplog$ inductive system is non-filtering is {\expspace}.\qed}

\subsection{Proof of Lemma \ref{lemma:sl-ranked}}
\proof{Deciding whether a given system is ranked is possible in polynomial
space when all constraints are quantifier-free $\seplog$ formulae, by
checking the validity of $\phi \models^\tinyseplog \neg\emp$,
i.e.\ the satisfiability of $\phi \wedge \emp$ for each constraint
$\phi$, which is in \pspace \cite{CYOH01}. This bound drops to
polynomial time for systems with symbolic heap constraints, because
each model of a symbolic heap $\symhp \wedge \symhs$ is empty iff
$\symhs$ does not contain atoms of the form $x \mapsto
(y_1,\ldots,y_k)$.}

\subsection{Proof of Lemma \ref{lemma:sl-fvi}}
\proof{
Similarly to the proof for Lemma \ref{lemma:fol-fvi}, an $\seplog$ inductive system $\sys$ has the fvi property if, for any two constraints $\phi(\vec{x},\vec{x}_1,\ldots,\vec{x}_n)$ and $\psi(\vec{x},\vec{y}_1,\ldots,\vec{y}_m)$ from $\sys$, with goal variables $\vec{x}$ and subgoal variables $\bigcup_{i=1}^n \vec{x}_i$ and $\bigcup_{j=1}^m \vec{y}_j$, respectively, the following entailment is not valid:
\begin{align*}
\phi(\vec{x},\vec{x}_1,\ldots,\vec{x}_n)\models^\tinyseplog \exists \vec{y}_1 \ldots \exists \vec{y}_m \,.\, \bigg(\psi(\vec{x}, \vec{y}_1,\ldots,\vec{y}_m) \land \bigvee_{j=1}^m \bigwedge_{i=1}^n \lnot(\vec{x}_i \cong \vec{y}_j)\bigg)
\end{align*}
where $\vec{x}_i \cong \vec{y}_j$ is shorthand for $\left(\bigwedge_{y \in \vec{y}_j}\bigvee_{x \in \vec{x}_i} x \teq y\right) \land \left(\bigwedge_{x \in \vec{x}_i}\bigvee_{y \in \vec{y}_j} x \teq y\right)$. Just as in the proof for Lemma \ref{lemma:fol-fvi}, this entailment is valid only if the following formula is unsatisfiable:
\begin{align*}
\exists \vec{x} \, \exists \vec{x}_1 \ldots \exists \vec{x}_n \, \forall \vec{y}_1 \ldots \forall \vec{y}_m \,.\, \bigg(\phi(\vec{x},\vec{x}_1,\ldots,\vec{x}_n) \land \bigwedge_{j=1}^m \bigg(\lnot \psi(\vec{x}, \vec{y}_1,\ldots,\vec{y}_m) \lor \bigvee_{i=1}^n \vec{x}_i \cong \vec{y}_j \bigg)\bigg)
\end{align*}
We know that checking the satisfiability for the above formula is in {\pspace} when $\phi$ and $\psi$ are quantifier-free and $\wand$-free $\seplog$ formulae \cite{Vmcai17} and, thus, the fvi problem is in {\pspace}. If, however, $\phi$ and $\psi$ are symbolic heaps, the initial entailment problem is in {\pitwop} \cite[Theorem 6]{AGHKO14}, thus the fvi problem is in {\sigmatwop}.\qed}

\subsection{Proof of Lemma \ref{lemma:sl-non-overlapping}}
\proof{Similarly to the first order case, as shown in the proof of Lemma \ref{lemma:fol-non-overlapping}, given an $\seplog$ inductive system $\sys$, in order to determine if $\sys$ has the non-overlapping property, it suffices to check that, for any two constraints $\phi, \psi$ of $\sys$, where $\vec{y}_1, \ldots, \vec{y}_m$ are the subgoal variables of $\psi$: \begin{enumerate*}[label=(\roman*)] \item \label{item:lm-nov-sl-item1} $\phi \land \psi$ is satisfiable and \item \label{item:lm-nov-sl-item2} $\phi \models^\tinyseplog  \\ \exists \vec{y}_1 \ldots \exists \vec{y}_m \,.\, \psi$ is valid (or, conversely, that $\forall \vec{y}_1 \ldots \forall \vec{y}_m \,.\, \phi \land \lnot \psi$ is unsatisfiable) \end{enumerate*}.

When $\phi$ and $\psi$ are quantifier-free and $\wand$-free $\seplog$ formulae, then \ref{item:lm-nov-sl-item2} is in {\pspace} \cite{Vmcai17} and, thus, the non-overlapping problem is in {\pspace}. If, however, $\phi$ and $\psi$ are symbolic heaps, then the satisfiability problem \ref{item:lm-nov-sl-item1} for symbolic heaps is in {\np} \cite{CYOH01} and the entailment \ref{item:lm-nov-sl-item2} between existentially quantified symbolic heaps is {\pitwop}-complete \cite{AGHKO14}, and, thus, the non-overlapping problem is in {\pitwop}.
\qed}

\subsection{Soundness of the $\SPsl$ rule type in $\mathcal{R}_\mathsf{Indsl}$}
\begin{lemma}\label{lemma:spsl}
    Given a system $\sys$, with predicates $p_1(\vec{x}_1), \ldots,
    p_n(\vec{x}_n)$, such that $\vec{x}_i \cap \vec{x}_j = \emptyset$,
    for all $1 \leq i < j \leq n$, and let $\overline{\mathcal{Q}}_i =
    \langle q^i_1(\vec{x}_1),\ldots, q_n^i(\vec{x}_n)\rangle$ in
    $\sys$, for all $i \in [k]$, be tuples of
    predicates. Then \[\mu\sys(p_1(\vec{x}_1) * \ldots *
    p_n(\vec{x}_n)) \subseteq \bigcup_{i=1}^k \mu\sys(q^i_1(\vec{x}_1)
    * \ldots * q^i_n(\vec{x}_n))\] \vspace*{-0.5\baselineskip} if and
    only if there exists a tuple $\bar{\imath} \in [n]^{n^k}$, such
    that:
    \[\mu\sys(p_{\bar{\imath}_j}) \subseteq \bigcup 
    \{\mu\sys(q_{\bar{\imath}_j}^\ell) \mid \ell\in[k],
    f_j(\overline{\mathcal{Q}}_\ell) = \bar{\imath}_j\}\] for all
    $j \in [n^k]$, where $\cf(\overline{\mathcal{Q}}_1,\ldots,
    \overline{\mathcal{Q}}_k) = \set{f_1,\ldots,f_{n^k}}$.
\end{lemma}

\proof{Let $\bigotimes_{i=1}^n \mu\sys(r_i) \overset{\text{def}}{=} 
    \{(\tuple{\vec{u}_1, \ldots, \vec{u}_n}, h_1 \uplus \ldots \uplus 
    h_n) \mid (\vec{u}_i, h_i) \in \mu\sys(r_i), i \in [n]\}$, for 
    some predicates $r_1(\vec{x}_1), \ldots, r_n(\vec{x}_n)$ in $\sys$, and
    $\mathcal{U}_k = \locsi^k \times \heaps$. Then the following property holds:    
    \[\begin{array}{l}\bigotimes_{i=1}^n \mu\sys(r_i) = \bigcap_{i=1}^n 
    \left(\bigotimes_{j=1}^{i-1} \mathcal{U}_{\card{\vec{x}_j}} \otimes
    \mu\sys(r_i) \otimes \bigotimes_{j=i+1}^{n} \mathcal{U}_{\card{\vec{x}_j}}
    \right) \end{array}\]
    
    Using this property, the inclusion we need to prove can be rewritten as
    \[\begin{array}{lc}\mu\sys(p_1(\vec{x}_1) * \ldots * p_n(\vec{x}_n)) 
    \subseteq \bigcup_{i=1}^k \mu\sys(q_1^i(\vec{x}_1) * \ldots * q^i_n(\vec{x}_n)) 
    & \Leftrightarrow \\[5pt]
    
    \bigotimes_{i=1}^n \mu\sys(p_i) \subseteq \bigcup_{i=1}^k \bigotimes_{j=1}^n
    \mu\sys(q_j^i) & \Leftrightarrow \\[5pt]
    
    \bigotimes_{i=1}^n \mu\sys(p_i) \subseteq \bigcup_{i=1}^k \bigcap_{j=1}^n
    \left(\bigotimes_{\ell=1}^{j-1} \mathcal{U}_{\card{\vec{x}_\ell}} \otimes 
    \mu\sys(q^i_j) \otimes \bigotimes_{\ell=j+1}^{n} \mathcal{U}_{\card{\vec{x}_\ell}}
    \right) & \end{array}\]
    
    As in the proof of \cite[Theorem 1]{HolikLSV11}, because the power set 
    lattice $(2^V, \subseteq)$ of any set $V$ is a completely distributive
    lattice, for any doubly indexed family $\{S_{j,k} \in 2^V \mid j \in J, k 
    \in K_j\}$ it holds that $\bigcup_{j \in J} \bigcap_{k \in K_j} S_{j, k}
    = \bigcap_{f \in F} \bigcup_{j \in J} S_{j, f(j)}$, where $F$ is the set
    of choice functions $f$ choosing for each index $j \in J$ some index
    $f(j) \in K_j$. In our case, let $F =\cf(\overline{\mathcal{Q}}_1,\ldots,
    \overline{\mathcal{Q}}_k)$. Then,
    \[\begin{array}{lc}
    \bigotimes_{i=1}^n \mu\sys(p_i) \subseteq \bigcup_{i=1}^k \bigcap_{j=1}^n
    \left(\bigotimes_{\ell=1}^{j-1} \mathcal{U}_{\card{\vec{x}_\ell}} \otimes 
    \mu\sys(q^i_j) \otimes \bigotimes_{\ell=j+1}^{n} \mathcal{U}_{\card{\vec{x}_\ell}}
    \right) & \Leftrightarrow \\[5pt]
    
    \bigotimes_{i=1}^n \mu\sys(p_i) \subseteq \bigcap_{f\in F}
    \bigcup_{i=1}^k \left(\bigotimes_{\ell=1}^{f(\overline{\mathcal{Q}}_i)-1} 
    \mathcal{U}_{\card{\vec{x}_\ell}} \otimes \mu\sys(q^i_{f(\overline{\mathcal{Q}}_i)})
    \otimes \bigotimes_{\ell=f(\overline{\mathcal{Q}}_i)+1}^{n} 
    \mathcal{U}_{\card{\vec{x}_\ell}} \right) & \Leftrightarrow \\[5pt]
    
    \forall f\in F \,. \, \bigotimes_{i=1}^n \mu\sys(p_i) \subseteq
    \bigcup_{i=1}^k \left(\bigotimes_{\ell=1}^{f(\overline{\mathcal{Q}}_i)-1} 
    \mathcal{U}_{\card{\vec{x}_\ell}} \otimes \mu\sys(q^i_{f(\overline{\mathcal{Q}}_i)})
    \otimes \bigotimes_{\ell=f(\overline{\mathcal{Q}}_i)+1}^{n} \mathcal{U}_{\card{\vec{x}_\ell}}
    \right) &  (*) \end{array}\]
    
    For a fixed $f$, we can rewrite the right hand-side of the inclusion as
    
    \[\begin{array}{lc}
    \bigcup_{i=1}^k \left(\bigotimes_{\ell=1}^{f(\overline{\mathcal{Q}}_i)-1} 
    \mathcal{U}_{\card{\vec{x}_\ell}} \otimes \mu\sys(q^i_{f(\overline{\mathcal{Q}}_i)})
    \otimes \bigotimes_{\ell=f(\overline{\mathcal{Q}}_i)+1}^{n} 
    \mathcal{U}_{\card{\vec{x}_\ell}}\right) & = \\[5pt]
    
    \bigcup_{j=1}^n \bigcup_{i \in [k], f(\overline{\mathcal{Q}}_i) = j} 
    \left(\bigotimes_{\ell=1}^{j-1}\mathcal{U}_{\card{\vec{x}_\ell}} \otimes \mu\sys(q^i_j)
    \otimes \bigotimes_{\ell=j+1}^{n} \mathcal{U}_{\card{\vec{x}_j}}\right) & = \\[5pt]
    
    \bigcup_{j=1}^n \left(\bigotimes_{\ell=1}^{j-1}\mathcal{U}_{\card{\vec{x}_\ell}} \otimes
    \left(\bigcup_{i \in [k], f(\overline{\mathcal{Q}}_i) = j}\mu\sys(q^i_j) \right)
    \otimes \bigotimes_{\ell=j+1}^{n} \mathcal{U}_{\card{\vec{x}_j}} \right) &
    \end{array}\]
    
    Then the inclusion query $(*)$ becomes
    \[\begin{array}{lc}
    \forall f\in F \,. \, \bigotimes_{i=1}^n \mu\sys(p_i) \subseteq \bigcup_{j=1}^n
    \left(\bigotimes_{\ell=1}^{j-1}\mathcal{U}_{\card{\vec{x}_\ell}} \otimes \left(
    \bigcup_{i \in [k], f(\overline{\mathcal{Q}}_i) = j} \mu\sys(q^i_j) \right)
    \otimes \bigotimes_{\ell=j+1}^{n} \mathcal{U}_{\card{\vec{x}_j}} \right) & \Leftrightarrow \\[5pt]
    
    \forall f\in F \exists j \in [n] \,. \, \mu\sys(p_j) \subseteq 
    \bigcup_{i \in [k], f(\overline{\mathcal{Q}}_i) = j}\mu\sys(q^i_j) & \Leftrightarrow \\[5pt]
    
    \bigwedge_{i=1}^{n^k} \bigvee_{j=1}^n \mu\sys(p_j) \subseteq \bigcup 
    \{\mu\sys(q^\ell_j) \mid \ell \in [k], f_i(\overline{\mathcal{Q}}_\ell) = j\} & \Leftrightarrow \\[5pt]
    
    \bigvee_{\bar{\imath} \in [n]^{n^k}} \bigwedge_{j=1}^{n^k}\mu\sys(p_{\bar{\imath}_j})
    \subseteq \bigcup \{\mu\sys(q^\ell_{\bar{\imath}_j}) \mid \ell \in [k], 
    f_j(\overline{\mathcal{Q}}_\ell) = \bar{\imath}_j\} & \text{\qed}
    \end{array}\]}

\subsection{Proof of Lemma \ref{lemma:sl-local-soundness}}
\proof{For $\AXsl$, soundness follows from the side condition and its consequent admits no counterexamples, thus the lemma is trivially true in this case. We analyse $\LU$, $\RUsl$, $\RDsl$, $\RI$ and $\SPsl$ individually, in a similar fashion as we did for the local soundness of $\RInd$ (Lemma \ref{lemma:local-soundness}).
    
    \vspace{10pt}
    \paragraph{$(\LU)$} Let $p(\vec{x}) \in \lseq$ be a predicate atom, where $p(\vec{x}) \leftarrow_\sys R_1(\vec{x}) \mid \ldots \mid R_n(\vec{x})$. The antecedents of $\lseq \vdash \rseq$ are $\lseq_i \vdash \rseq_i \equiv R_i(\vec{x}, \vec{y}_i),\lseq \setminus p(\vec{x}) \vdash \rseq$, where $i \in [n]$ and each $\vec{y}_i$ is a tuple of fresh variables. In this case, the least solution of $\lseq$ is
    
    \begin{align*}
    \mu\sys^\tinyseplog(\Asterisk \lseq) & = \mu\sys^\tinyseplog\left(p(\vec{x}) * \Asterisk \left(\lseq \setminus p(\vec{x}) \right)\right) = \mu\sys^\tinyseplog(p(\vec{x})) \uplus \mu\sys^\tinyseplog\left(\Asterisk \left(\lseq \setminus p(\vec{x}) \right)\right) \notag \\
    & = \left(\bigcup_{i=1}^n\mu\sys^\tinyseplog\left(\Asterisk R_i(\vec{x}) \right)\right) \uplus \mu\sys^\tinyseplog\left(\Asterisk \left(\lseq \setminus p(\vec{x})\right)\right) \notag \\
    & = \bigcup_{i=1}^n\left(\mu\sys^\tinyseplog\left(\Asterisk R_i(\vec{x}) \right) \uplus \mu\sys^\tinyseplog\left(\Asterisk \left(\lseq \setminus p(\vec{x})\right)\right)\right) \notag \\
    & = \bigcup_{i=1}^n\mu\sys^\tinyseplog\left(\Asterisk R_i(\vec{x}) * \Asterisk \left(\lseq \setminus p(\vec{x})\right)\right)
    \end{align*}
    If there exists $(\nu, h) \in \mu\sys^\tinyseplog(\Asterisk\lseq) \setminus \mu\sys^\tinyseplog(\Asterisk\rseq)$, then $(\nu,h) \in \mu\sys^\tinyseplog(\Asterisk R_i(\vec{x}) \wedge \Asterisk (\lseq \setminus p(\vec{x})))$ for some $i \in [n]$. In consequence, there also exists $(\nu_i,h_i) \in \mu\sys^\tinyseplog(\Asterisk R_i(\vec{x},\vec{y}_i) \wedge \Asterisk (\lseq \setminus p(\vec{x}))) \setminus \mu\sys^\tinyseplog(\Asterisk\rseq)$ such that $h_i = h$ and for every $x \in \fv{\lseq}$ we have $\nu_i(x) = \nu(x)$. Then $h_i \unlhd h$ holds trivially.
    
    \vspace{10pt}
    \paragraph{$(\RUsl)$} Let $p(\vec{x}) \in \rseq$ be a predicate atom, where $p(\vec{x}) \leftarrow_\sys R_1(\vec{x}) \mid \ldots \mid R_n(\vec{x})$. Then $\lseq \vdash \rseq$ has only one antecedent $\lseq_1 \vdash \rseq_1 \equiv \lseq \vdash \exists \vec{y}_1 . \Asterisk R_1(\vec{x},\vec{y}_1), \ldots, \\ \exists \vec{y}_n. \Asterisk R_n(\vec{x},\vec{y}_n), \rseq \setminus p(\vec{x})$, where each $\vec{y}_i$ with $i \in [n]$ is a tuple of fresh variables. In this case, the least solution of $\rseq$ is
    \begin{align*}
    \mu\sys^\tinyseplog\left(\bigvee \rseq\right) & = \mu\sys^\tinyseplog\left(p(\vec{x}) \lor \bigvee \left(\rseq \setminus p(\vec{x}) \right)\right) = \mu\sys^\tinyseplog(p(\vec{x})) \cup \mu\sys^\tinyseplog\left(\bigvee \left(\rseq \setminus p(\vec{x}) \right)\right) \\
    & = \left(\bigcup_{i=1}^n\mu\sys^\tinyseplog\left(\Asterisk R_i(\vec{x}) \right)\right) \cup \mu\sys^\tinyseplog\left(\bigvee \left(\rseq \setminus p(\vec{x})\right)\right) \\
    & = \left(\bigcup_{i=1}^n\mu\sys^\tinyseplog\left(\exists \vec{y}_i . \Asterisk R_i(\vec{x}, \vec{y}_i) \right)\right) \cup \mu\sys^\tinyseplog\left(\bigvee \left(\rseq \setminus p(\vec{x})\right)\right) \notag \\
    & = \mu\sys^\tinyseplog\left(\bigvee_{i=1}^n\exists \vec{y}_i . \Asterisk R_i(\vec{x}, \vec{y}_i) \right) \cup \mu\sys^\tinyseplog\left(\bigvee \left(\rseq \setminus p(\vec{x})\right)\right) \\
    & = \mu\sys^\tinyseplog\left(\bigvee_{i=1}^n\exists \vec{y}_i . \Asterisk R_i(\vec{x}, \vec{y}_i) \lor \bigvee \left(\rseq \setminus p(\vec{x})\right)\right) = \mu\sys^\I(\bigvee \rseq_1)
    \end{align*}
    If there exists $(\nu,h) \in \mu\sys^\tinyseplog\left(\Asterisk \lseq\right) \setminus \mu\sys^\tinyseplog\left(\bigvee\rseq\right)$, then it is also the case that $(\nu,h) \in \mu\sys^\tinyseplog(\Asterisk \lseq_1) \setminus \mu\sys^\tinyseplog\left(\bigvee\rseq_1\right)$. Therefore, the counterexample for the antecedent is $(\nu_1, h_1) = (\nu, h)$ and $h_1 \unlhd h$ holds trivially.
    
    \vspace{10pt}
    \paragraph{$(\RDsl)$} Then the sequent $\lseq \vdash \rseq \equiv \phi(\vec{x}, \vec{x}_1, \ldots, \vec{x}_n), p_1(\vec{x}_1),\ldots,p_n(\vec{x}_n) \vdash \{\exists \vec{y}_j . \\ \psi_j (\vec{x}, \vec{y}_j) * \mathcal{Q}_j(\vec{y}_j)\}_{j=1}^k$ has only one antecedent $\lseq_1 \vdash \rseq_1 \equiv p_1(\vec{x}_1),\ldots,p_n(\vec{x}_n) \vdash \{\mathcal{Q}_j\theta \mid \theta \in S_j\}_{j=1}^i$. By the side condition of $\RD$, $\phi \models^\tinyseplog \Asterisk_{j=1}^i \exists \vec{y}_j . \psi_j$.  Also, by Definition \ref{def:fol-fvi}, we have $\mu\sys^\tinyseplog(\phi) \subseteq \mu\sys^\tinyseplog(\psi_j\theta)$ for each $\theta \in \subst{\phi,\psi_j}$ and $j \in [i]$. In this case, the least solution of $\rseq$ is
    
    \begin{align*}
    \mu\sys^\tinyseplog\left(\bigvee\rseq\right) & = \mu\sys^\tinyseplog\left(\bigvee_{j=1}^k\exists \vec{y}_j . \psi_j * \mathcal{Q}_j\right) \supseteq \bigcup_{j=1}^k \mu\sys^\tinyseplog\left(\exists \vec{y}_j . \psi_j * \mathcal{Q}_j\right)\\
    &\supseteq \bigcup_{j=1}^i \mu\sys^\tinyseplog\left(\exists \vec{y}_j . \psi_j * \mathcal{Q}_j\right) = \bigcup_{j=1}^i \bigcup_{\theta \in \subst{\phi,\psi_j}} \mu\sys^\tinyseplog\left((\psi_j * \mathcal{Q}_j)\theta\right) \\
    &= \bigcup_{j=1}^i \bigcup_{\theta \in \subst{\phi,\psi_j}} \mu\sys^\tinyseplog\left(\psi_j\theta * \mathcal{Q}_j\theta\right) \supseteq \bigcup_{j=1}^i \bigcup_{\theta \in \subst{\phi,\psi_j}} \left( \mu\sys^\tinyseplog\left(\psi_j\theta\right) \uplus \mu\sys^\tinyseplog(\mathcal{Q}_j\theta)\right) \\
    &= \bigcup_{j=1}^i \bigcup_{\theta \in \subst{\phi,\psi_j}} \mu\sys^\tinyseplog\left(\psi_j\theta\right) \uplus \bigcup_{j=1}^i \bigcup_{\theta \in \subst{\phi,\psi_j}}\mu\sys^\tinyseplog(\mathcal{Q}_j\theta)
    \end{align*}
    
    \noindent Note that also \(\mu\sys^\tinyseplog(\Asterisk \lseq)= \mu\sys^\tinyseplog\left(\phi\right) \uplus \mu\sys^\tinyseplog\left(p_1(\vec{x}_1) * \ldots * p_n(\vec{x}_n)\right)\).
    If there exists $(\nu, h) \in \mu\sys^\tinyseplog(\Asterisk \lseq) \setminus \mu\sys^\tinyseplog\left(\bigvee\rseq\right)$, then $h = h' \uplus h''$ such that $(\nu, h') \in \mu\sys^\tinyseplog\left(\phi\right)$ and $(\nu, h'') \in \mu\sys^\tinyseplog(p_1(\vec{x}_1) * \ldots * p_n(\vec{x}_n))$. It follows that we have $(\nu, h' \uplus h'') \in (\mu\sys^\tinyseplog\left(\phi\right) \uplus \mu\sys^\tinyseplog(p_1(\vec{x}_1) * \ldots * p_n(\vec{x}_n))) \setminus (\bigcup_{j=1}^i \bigcup_{\theta \in \subst{\phi,\psi_j}} \mu\sys^\tinyseplog\left(\psi_j\theta\right) \uplus \bigcup_{j=1}^i \bigcup_{\theta \in \subst{\phi,\psi_j}}\mu\sys^\tinyseplog(\mathcal{Q}_j\theta))$ and, since as previously stated $\mu\sys^\tinyseplog(\phi) \subseteq  \bigcup_{j=1}^i \bigcup_{\theta \in \subst{\phi,\psi_j}} \mu\sys^\tinyseplog\left(\psi_j\theta\right)$, we obtain $(\nu, h'') \in \mu\sys^\tinyseplog\left(p_1(\vec{x}_1) * \ldots * p_n(\vec{x}_n)\right) \setminus \bigcup_{j=1}^i \bigcup_{\theta \in \subst{\phi,\psi_j}}\mu\sys^\tinyseplog(\mathcal{Q}_j\theta) = \mu\sys^\tinyseplog(\Asterisk \lseq_1) \setminus \mu\sys^\tinyseplog(\bigwedge \rseq_1)$. Therefore, the counterexample for the antecedent is $(\nu_1, h_1) = (\nu, h'')$. Because $\phi$ is introduced to the left-hand side by left unfolding and $\sys$ is ranked, we have $h' \neq \emptyset$ and, thus, $h_1 \lhd h$.
    
    \vspace{10pt}
    \paragraph{$(\RI)$} As in the $\RI$ case from the proof of Theorem \ref{thm:ta-soundness}, we have that, for any counterexample $(\nu, h) \in \mu\sys^\tinyseplog(\Asterisk \lseq) \setminus \mu\sys^\tinyseplog\left(\bigvee\rseq\right)$, it is also the case that $(\nu, h) \in (\mu\sys^\tinyseplog(\Asterisk \lseq) \setminus \mu\sys^\tinyseplog(\bigvee \rseq_1)) \cup (\mu\sys^\tinyseplog(\Asterisk \lseq) \setminus \mu\sys^\tinyseplog(\bigvee \rseq_2))$. Therefore, $\nu \in \mu\sys^\tinyseplog(\Asterisk \lseq_i) \setminus \mu\sys^\tinyseplog(\bigvee \rseq_i)$ for some $i \in [2]$ and the counterexample for $\lseq_{i} \vdash \rseq_{i}$ is $(\nu_i, h_i) = (\nu, h)$. Then $h_i \unlhd h$ holds trivially.

    \vspace{10pt}
    \paragraph{$(\SPsl)$} Then $\lseq \vdash \rseq \equiv p_1(\vec{x}_1), \ldots, p_n(\vec{x}_n) \vdash \{\Asterisk_{i=1}^n q_i^j(\vec{x}_i) \}_{j=1}^k$. For each $\bar{\imath} \in [n]^{n^k}$, the antecedents of $\lseq \vdash \rseq$ are $\lseq^{\bar{\imath}}_{j} \vdash \rseq^{\bar{\imath}}_{j} \equiv  p_{\bar{\imath}_j}(\vec{x}_{\bar{\imath}_j}) \vdash \{q_{\bar{\imath}_j}^\ell(\vec{x}_{\bar{\imath}_j}) \mid \ell \in [k],~ f_j(\overline{\mathcal{Q}}_\ell) = \bar{\imath}_j\}, j \in [n^k]$.
    
    If there exists $(\nu,h) \in \mu\sys^\tinyseplog\left(\Asterisk_{i=1}^n p_i(\vec{x}_i)\right) \setminus \mu\sys^\tinyseplog\left(\bigvee_{j=1}^k \Asterisk_{i=1}^n q_i^j(\vec{x}_i)\right)$, then by the proof of Lemma \ref{lemma:spsl}, there exists $j \in [n^k]$ and pairs $(\nu_1, h_1), \ldots, (\nu_n, h_n)$ such that $(\nu_i, h_i) \in \mu\sys^\tinyseplog(p_i(\vec{x}_i)) \setminus \bigcup\{\mu\sys^\tinyseplog(q_i^\ell) \mid \ell \in [k],~ f_j(\overline{\mathcal{Q}}_\ell(\vec{x}_i)) )=i\}$ for all $i \in [n]$, where $h = h_1 \uplus \ldots \uplus h_n$ and $\nu(\vec{x}_i) = \nu_i(\vec{x}_i)$ for each $i \in [n]$. In other words, for all tuples $\bar{\imath} \in [n]^{n^k}$ we have \((\nu_{\bar{\imath}_j}, h_{\bar{\imath}_j}) \in \mu\sys^\tinyseplog(p_{\bar{\imath}_j}(\vec{x}_{\bar{\imath}_j})) \setminus \bigcup \{\mu\sys^\tinyseplog(q_{\bar{\imath}_j}^\ell(\vec{x}_{\bar{\imath}_j})) \mid \ell\in[k], f_j(\overline{\mathcal{Q}}_\ell) = \bar{\imath}_j\} = \mu\sys^\tinyseplog(\Asterisk \lseq^{\bar{\imath}}_j) \setminus \mu\sys^\tinyseplog(\bigvee \rseq^{\bar{\imath}}_j)\). Therefore, given such $j \in [n^k]$, the counterexample for each antecedent $\lseq^{\bar{\imath}}_{j} \vdash \rseq^{\bar{\imath}}_{j}$ is $(\nu_{\bar{\imath}_j}, h_{\bar{\imath}_j})$. Since $h = h_1 \uplus \ldots \uplus h_n$, we have $h_{\bar{\imath}_j} \unlhd h$, for each $\bar{\imath} \in [n]^{n^k}$.\qed}

\subsection{Proof of Lemma \ref{lemma:sl-direct-counterexample-path}}
\proof{Through a similar reasoning as the one in the proof of Lemma \ref{lemma:direct-counterexample-path} and using Lemma \ref{lemma:sl-local-soundness} to support the fact that $h_1 \unrhd \ldots \unrhd h_k$, we obtain that $\RDsl$ is required to occur in $\typelabel(\pi)$, leading to $h_i \rhd h_{i+1}$ for some $i \in [k-1]$. Then $h_1 \rhd h_k$.\qed}

\subsection{Proof of Theorem \ref{thm:sl-soundness}}
\proof{This proof by contradiction closely follows the one of Theorem \ref{thm:ta-soundness}. We suppose that $\Asterisk\lseq \models_\sys^\tinyseplog \bigvee\rseq$ does not hold and, by Lemma \ref{lemma:sl-local-soundness}, obtain a path from $v_0$ to a leaf $v_k \in V$ and an associated sequence of counterexamples $(\nu_1, h_1) \nextval \ldots \nextval(\nu_k, h_k)$.
    
Clearly, $R(v_k) = \ID$. Let $(v_k, v_{k+1})$ be a backlink, $S(v_k) = \lseq_k \vdash \rseq_k$ and $S(v_{k+1}) = \lseq_{k+1} \vdash \rseq_{k+1}$ such that $\lseq_k = \lseq_{k+1}\theta$,  $\rseq_{k} = \rseq'_{k+1} \theta$ and $\rseq_{k+1} \subseteq \rseq'_{k+1}$, for some injective substitution $\theta : \fv{\lseq_{k+1} \cup \rseq_{k+1}} \rightarrow \fv{\lseq_k\cup \rseq_k}$. Again, we can assume w.l.o.g. that $\theta$ is surjective, by defining $\theta(x) = x$ for each $x \in \fv{\lseq_k \cup \rseq_k} \setminus \theta(\fv{\lseq_{k+1} \cup \rseq_{k+1}})$. Since $\theta$ is also injective, by the side condition of $\ID$, its inverse exists and $(\nu_k \theta^{-1}, h_k)$ is a counterexample for $\lseq_{k+1} \vdash \rseq_{k+1}$. Therefore, we can extend the relation $\nextval$ with the pair $((\nu_k, h_k),(\nu_k \theta^{-1}, h_k))$. 

We obtain an infinite trace $\tau = v_0,v_1, \ldots$ in $D$ together with an infinite sequence $(\nu_0, h_0) \nextval (\nu_1, h_1) \nextval \ldots$. By Lemma \ref{lemma:sl-local-soundness}, we have $h_i \unrhd h_{i+1}$, for all $i\geq0$. By Proposition \ref{prop:direct-path}, $\tau$ contains infinitely many direct paths $\pi_j = v_{k_j}, \ldots, v_{\ell_j}$, where $\set{k_j}_{j\geq0}$ and $\set{\ell_j}_{j\geq0}$ are infinite strictly increasing sequences of integers such that $k_j < \ell_j \leq k_{j+1} < \ell_{j+1}$, for all $j\geq0$. By Lemma \ref{lemma:sl-direct-counterexample-path}, we obtain that $h_{k_j} \rhd h_{\ell_j}$, for all $j \geq 0$, which leads to a strictly decreasing sequence $h_{k_0} \rhd h_{k_1} \rhd \ldots$, contradicting the fact that $\unlhd$ is a wfqo. We can conclude that our initial assumption was false, thus the entailment $\Asterisk\lseq \models_\sys^\tinyseplog \bigvee\rseq$ holds. \qed}

\subsection{Proof of Lemma \ref{lemma:sl-u-cex}}

\proof{Let $(\DSSL)^*(\lseq \vdash \rseq) = \{d \mid d \in \DSSL(\lseq \vdash \rseq) \text{ or } \exists d' \in (\DSSL)^*(\lseq \vdash \rseq) \,.\, d \subtree d'\}$. Because, by Lemma \ref{lemma:sl-max-irred-finite}, any derivation in $\DSSL(\lseq \vdash\rseq)$ is finite and $\DSSL(\lseq \vdash\rseq)$ itself is finite, then $(\DSSL)^*(\lseq \vdash \rseq)$ is also finite. Since $((\DSSL)^*(\lseq \vdash \rseq),\sqsubseteq)$ is a wqo, by Lemma \ref{lemma:finite-sets-wqo} $(\finpow{(\DSSL)^*(\lseq \vdash \rseq)},\sqsubseteq^{\forall\exists})$ is also a wqo and we can prove this lemma by induction on it.
    
    Let $\DSSL(\lseq \vdash \rseq) = \bigcup_{\IR \in \RIndsl} \DSSL_\IR(\lseq \vdash \rseq)$, where $\DSSL_\IR(\lseq \vdash \rseq)$ denotes the subset of $\DSSL(\lseq \vdash \rseq)$ consisting of derivations starting with an inference rule of type $\IR$. We do not consider the cases $\IR \in \{\AXsl, \ID\}$ because these inference rules do not lead to a leaf of the required form. Given $D \in \DSSL_\IR(\lseq \vdash \rseq)$ with $\IR \in \RIndsl \setminus \{\AXsl, \ID\}$, we assume that, for each antecedent $\lseq' \vdash \rseq'$ of $\lseq \vdash \rseq$, the lemma holds for any $D' \in \bigcup_{\mathit{IR'} \in \RInd} \DSSL_\mathit{IR'}(\lseq' \vdash \rseq')$ and show that it extends to $D$.
    
    \vspace{10pt}
    \paragraph{$(\LU)$} Let $r(\vec{y}) \in \lseq$ be the predicate atom chosen for replacement and $r(\vec{y}) \leftarrow_\sys R_1(\vec{y}) \mid \ldots \mid R_m(\vec{y})$. Then $R_1(\vec{y},\vec{y}_1),\lseq \setminus r(\vec{y}) \vdash \rseq, \ldots, R_m(\vec{y},\vec{y}_m),\lseq \setminus r(\vec{y}) \vdash \rseq$ are the antecedents for the application of $\IR = \LU$ on $\lseq \vdash \rseq$. If $\rseq = \emptyset$, then, similarly to the $\LU$ case in Lemma \ref{lemma:cex}, there exists a counterexample $\nu \in \mu\sys^\utree(\bigwedge \lseq) \setminus \mu\sys^\utree(\bigvee\rseq) = \mu\sys^\utree(\bigwedge \lseq)$.
    
    If $\rseq \neq \emptyset$, suppose that, for each $i \in [m]$, there exists $D_i \in \DSSL(R_i(\vec{y},\vec{y}_i),\lseq \setminus r(\vec{y}) \vdash \rseq)$ not containing any leaf $\lseq' \vdash \emptyset$. Then we can choose these derivations and create one for $\lseq \vdash \rseq$ with the same property, which contradicts the hypothesis of the lemma. Therefore, there must exist $i \in [n]$ such that every derivation in $\DSSL(R_i(\vec{y},\vec{y}_i),\lseq \setminus r(\vec{y}) \vdash \rseq)$ contains a leaf $\lseq' \vdash \emptyset$.
    
    Because $p(\vec{x}) \vdash q_1(\vec{x}),\ldots, q_n(\vec{x}) \leadsto^\tinyseplog \lseq \vdash \rseq$, $\lseq$ is a tree-shaped set and let $T_1, \ldots, T_k$ be the singly-tree shaped sets, represented as trees labelled with formulae, such that $\lseq = \bigcup_{i=1}^k T_i$. Then there exists a tree $T_j$, $j \in [k]$ and a frontier position $\alpha_r \in \fr(T_j)$ such that $T_j(\alpha_r) = r(\vec{y})$. Then the tree-shaped set $R_i,\lseq \setminus r(\vec{y})$ is represented by the singly-tree shaped sets $T'_1,\ldots,T'_k$, where $T'_\ell = T_\ell$ for all $\ell \in [k] \setminus \set{j}$ and $T'_j = \context{T_j}{\alpha_r} \circ R_i$.
    
    By the induction hypothesis, there exists a counterexample $(\nu,h_1 \uplus \ldots \uplus h_k,\{t_1, \ldots,t_k\}) \in \mu\sys^\utree(\Asterisk (R_i(\vec{y},\vec{y}_i) \cup \lseq \setminus r(\vec{y}))) \setminus \mu\sys^\utree( \bigvee \rseq)$, where $(\nu,h_\ell,t_\ell) \in \mu\sys^\utree(\Asterisk T'_\ell)$, for all $\ell \in [k]$. Because $T'_\ell = T_\ell$ for all $\ell \in [k] \setminus \set{j}$, we only need to show that $(\nu,h_j,t_j) \in \mu\sys^\utree(\Asterisk T_j)$. Since $(\nu,h_j,t_j) \in \mu\sys^\utree(\Asterisk (\context{T_j}{\alpha_r} \circ R_i(\vec{y},\vec{y}_i)))$, there exist some disjoint heaps $h'_j$ and $h''_j$, a context $\context{t'_j}{\alpha_r}$ and a cover $t''_j$ such that: \begin{enumerate*}[label=(\roman*)] \item $h_j = h'_j \uplus h''_j$ and $t_j = \context{t'_j}{\alpha_r} \circ t''_j$, \item $\context{t'_j}{\alpha_r}$ covers $h'_j$ and $(\nu,t''_j,h''_j) \in \mu\sys^\utree(\Asterisk R_i(\vec{y},\vec{y}_i))$.\end{enumerate*} Since $\tuple{r(\vec{y}), R_i(\vec{y})} \in \sys$ is a predicate rule, we have $\mu\sys^\utree(\Asterisk R_i(\vec{y},\vec{y}_i)) \subseteq \mu\sys^\utree(r(\vec{y}))$. This leads to $(\nu,t''_j,h''_j) \in \mu\sys^\utree(r(\vec{y}))$ and $(\nu,h_j,t_j) \in \mu\sys^\utree(\Asterisk (\context{T_j}{\alpha_r} \circ R_i(\vec{y},\vec{y}_i))) \subseteq \mu\sys^\utree(\Asterisk (\context{T_j}{\alpha_r} \circ r(\vec{y}))) = \mu\sys^\utree(\Asterisk T_j)$. Therefore, $(\nu,h_\ell,t_\ell) \in \mu\sys^\utree(\Asterisk T_\ell)$ for all $\ell \in [k]$ and $(\nu,h_1 \uplus \ldots \uplus h_k,\set{t_1,\ldots,t_k}) \in \mu\sys^\utree(\Asterisk\lseq) \setminus \mu\sys^\utree(\bigvee\rseq)$.
    
    \vspace{10pt}
    \paragraph{$(\RUsl)$} Let $r(\vec{y}) \in \rseq$ be the predicate atom chosen for replacement and $r(\vec{y}) \leftarrow_\sys R_1(\vec{y}) \mid \ldots \mid R_m(\vec{y})$. Then \(\lseq \vdash \exists \vec{z}_1 . \Asterisk R_1(\vec{y}, \vec{z}_1), \ldots, \exists \vec{z}_m . \Asterisk R_m(\vec{y}, \vec{z}_m), \rseq \setminus r(\vec{y})\) is the antecedent for the application of $\IR = \RUsl$ on $\lseq \vdash \rseq$. Since $\RU$ is applicable, $\rseq \neq \emptyset$. Also, the right-hand side of the antecedent cannot be $\emptyset$ because all predicates in $\sys$ must be goals for at least one predicate rule of $\sys$.
    
    Suppose that there exists $D \in \DSSL(\lseq \vdash \exists \vec{z}_1 . \Asterisk R_1(\vec{y}, \vec{z}_1), \ldots, \exists \vec{z}_m . \Asterisk R_m(\vec{y}, \vec{z}_m), \\ \rseq \setminus r(\vec{y}))$ not containing any leaf $\lseq' \vdash \emptyset$. Then we can choose $D$ and create a derivation for $\lseq \vdash \rseq$ with the same property, which contradicts the hypothesis of the lemma. Therefore, it must be the case that every derivation in $\DSSL(\lseq \vdash \exists \vec{z}_1 . \Asterisk R_1(\vec{y}, \vec{z}_1), \ldots, \exists \vec{z}_m . \Asterisk R_m(\vec{y}, \vec{z}_m), \rseq \setminus r(\vec{y}))$ contains a leaf $\lseq' \vdash \emptyset$.
    
    By the induction hypothesis, there exists $(\nu, h, U) \in \mu\sys^\utree(\Asterisk \lseq) \setminus \mu\sys^\utree(\bigvee_{i=1}^m \exists \vec{z}_i . \\ \Asterisk R_i(\vec{y}, \vec{z}_i) \lor \bigvee (\rseq \setminus r(\vec{y})))$. Then it is also the case that $\nu \in \mu\sys^\utree(\Asterisk\lseq) \setminus \mu\sys^\utree(\bigvee\rseq)$, because $\mu\sys^\utree(\bigvee\rseq) = \mu\sys^\utree(\bigvee_{i=1}^m \exists \vec{z}_i . \Asterisk R_i(\vec{y}, \vec{z}_i) \lor \bigvee (\rseq \setminus r(\vec{y})))$ by a similar argument as in the ($\RUsl$) case of the proof for Theorem \ref{lemma:sl-local-soundness}.
    
    \vspace{10pt}
    \paragraph{$(\RDsl)$} Let $\lseq = \{\phi(\vec{y},\vec{y}_1,\ldots,\vec{y}_n),r_1(\vec{y}_1),\ldots, r_m(\vec{y}_m)\}$ and $\rseq = \{\exists \vec{z}_1.\psi_1(\vec{y}, \vec{z}_1) * \mathcal{Q}_1(\vec{z}_1), \ldots, \\ \exists \vec{z}_k . \psi_k(\vec{y},\vec{z}_k) * \mathcal{Q}_k(\vec{y}_k)\}$, where $\mathcal{Q}_1, \ldots, \mathcal{Q}_k$ are separating conjunctions of
    predicate atoms. Also, let $S_j = \subst{\phi,\psi_j}$, for all $j \in [k]$. Each $S_j$ is finite because $\sys$ has the fvi property. 
    
    If $m = 0$, the proof is similar to the one in the ($\RD$) case of Lemma \ref{lemma:cex}.
    
    If $m > 0$, then $r_1(\vec{y}_1),\ldots,r_m(\vec{y}_m) \vdash \{\mathcal{Q}_j\theta \mid \theta \in S_j\}_{j=1}^i$ is the antecedent for the application of $\IR = \RDsl$ on $\lseq \vdash \rseq$, where, by the side condition, $\phi \models^\utree \bigwedge_{j=1}^i \exists \vec{y}_j.\psi_j$ and $\phi {\not\models}^\utree \bigvee_{j=i+1}^k \exists \vec{y}_j . \psi_j$ (with a possible reordering of $\rseq$). If $k = 0$, because $\sys$ is non-filtering, $\vec{y}_1, \ldots, \vec{y}_m$ are distinct and there are no predicates in $\sys$ with empty least solutions, it follows that $\mu\sys^\utree(\Asterisk \lseq)$ is not empty. Therefore there exists a counterexample $(\nu, h, t) \in \mu\sys^\utree(\Asterisk \lseq) \setminus \mu\sys^\utree(\Asterisk \rseq) = \mu\sys^\utree(\bigwedge \lseq)$. Similarly, if $i = 0$, we have that $\mu\sys^\utree(r_1(\vec{y}_1) * \ldots 
    * r_m(\vec{y}_m))$ is not empty and it contains counterexamples for the antecedent. Given $(\nu, h, \{t_1, \ldots, t_m\}) \in \mu\sys^\utree(r_1(\vec{y}_1) * \ldots * r_m(\vec{y}_m))$, because the system is non-filtering, there exist a pair $(\overline{\ell},h_0) \in \locsi^{\card{\vec{y}}} \times \heaps$ such that $h_0$ is disjoint from $h$ and $\nu[\vec{y} \leftarrow \overline{\ell}],h_0 \models^\tinyseplog \phi$. By the side condition of $\RD$ and the fact that $\sys$ is non-overlapping, we have that $\nu[\vec{y} \leftarrow \overline{\ell}], h {\not\models}^\tinyseplog \exists \vec{z}_j . \psi_j(\vec{z}_j), \forall j \in [k]$. We obtain $(\nu[\vec{y} \leftarrow \overline{\ell}], h_0 \uplus h, \tau_m (h_0, t_1, \ldots, t_m)) \not\in \mu\sys^\utree(\exists \vec{z}_j.\psi_j(\vec{y}, \vec{z}_j) * \mathcal{Q}_j(\vec{z}_j)), \forall j \in [k]$, thus $(\nu[\vec{y} \leftarrow \overline{\ell}], h_0 \uplus h, \tau_m (h_0, t_1, \ldots, t_m)) \in \mu\sys^\utree(\Asterisk \lseq) \setminus \mu\sys^\utree(\Asterisk \rseq)$.

    If both $k > 0$ and $i > 0$, suppose that there exists $D \in \DSSL(r_1(\vec{y}_1), \ldots,r_m(\vec{y}_m) \vdash \{\mathcal{Q}_j\theta \mid \theta \in S_j\}_{j=1}^i)$ not containing any leaf $\lseq' \vdash \emptyset$. Then we can choose $D$ and create a derivation for $\lseq \vdash \rseq$ with the same property, which contradicts the hypothesis of the lemma. Thus, it must be the case that every derivation $\DSSL(p_1(\vec{x}_1),\ldots,p_n(\vec{x}_n) \vdash \{\mathcal{Q}_j\theta \mid \theta \in S_j\}_{j=1}^i$ contains a leaf $\lseq' \vdash \emptyset$.
    
    By the induction hypothesis, there exists $(\nu, h, \{t_1, \ldots, t_m\}) \in\mu\sys^\utree(r_1(\vec{y}_1) * \ldots * r_m(\vec{y}_m)) \setminus \mu\sys^\utree\left(\bigvee_{j=1}^i \{\mathcal{Q}_j\theta \mid \theta \in S_j\}\right)$, and also, for every $j \in [i]$ and $\theta \in S_j$, $(\nu, h, \{t_1, \ldots, t_m\}) \in\mu\sys^\utree\left(r_1(\vec{y}_1) * \ldots * r_m(\vec{y}_m)\right) \setminus \mu\sys^\utree(\mathcal{Q}_j\theta)$. Then $h = \biguplus_{j=1}^m h_j$, where $(\nu(\vec{y}_j),h_j,t_j) \in \mu\sys^\utree(r_j)$, for all $j \in [m]$.
    Because $\sys$ is non-filtering, there exists a pair $(\overline{\ell},h_0) \in \locsi^{\card{\vec{y}}} \times \heaps$ such that $h_0$ is disjoint from $h$ and $\nu[\vec{y} \leftarrow \overline{\ell}],h_0 \models^\tinyseplog \phi$.
    
    Because $p(\vec{x}) \vdash q_1(\vec{x}), \ldots, q_n(\vec{x}) \leadsto^\tinyseplog \lseq \vdash \rseq$, by Lemma \ref{lemma:tree-shaped} the set $\lseq = \{\phi(\vec{y}, \vec{y}_1, \ldots, \vec{y}_m), \\ r_1(\vec{y}_1),\ldots,r_m(\vec{r}_m)\}$ is tree-shaped, and because the number of predicate atoms equals the number of tuples of subgoal variables, it must be a singly-tree shaped set. Then $(\nu[\vec{y} \leftarrow \overline{\ell}], h_0 \uplus h, \tau_n(h_0,t_1,\ldots,t_n)) \in \mu\sys^\utree(\Asterisk\lseq)$.
    
    Suppose that \((\nu[\vec{x} \leftarrow \overline{\ell}], h_0 \uplus h, \tau_n(h_0,t_1,\ldots,t_m))$ is a model of $\exists \vec{y}_j . \psi_j(\vec{x},\vec{y}_j) * \mathcal{Q}_j(\vec{y}_j))\) for some $j \in [k]$. Then $\nu[\vec{x}\leftarrow\overline\ell], h_0 \models^\tinyseplog \exists \vec{y}_j . \psi_j(\vec{x},\vec{y}_j)$ and, because $\sys$ is non-overlapping, we have that $\phi \models^\tinyseplog \exists \vec{y}_j . \psi_j$. It follows that $j \not\in [i+1,k]$, otherwise it would contradict the side condition $\phi \not\models^\tinyseplog \exists \vec{y}_{i+1} . \psi_{i+1} \lor \ldots \lor \exists \vec{y}_k . \psi_k$. Therefore, $j \in [i]$ and, because $\sys$ has the fvi property, $\nu[\vec{x}\leftarrow\overline{\ell}],h_0 \models^\tinyseplog \psi_j\theta$, for all $\theta \in \subst{\phi,\psi_j}$. Moreover, there is no other Skolem function that witnesses this entailment, besides the ones in $S_j$. Then also $(\nu[\vec{x} \leftarrow \overline{\ell}], h, \{t_1,\ldots,t_n\}) \in \mu\sys^\utree(\mathcal{Q}_j\theta)$, for all $\theta \in S_j$ and only for these substitutions. Because the range of each $\theta \in S_j$ is $\vec{y}_1 \cup \ldots \cup \vec{y}_m$, it must be the case that $(\nu, h, \{t_1,\ldots,t_m\}) \in \mu\sys^\utree(\mathcal{Q}_j\theta)$ for all $\theta \in S_j$, which contradicts our assumption that this is a counterexample of the antecedent. It follows that $(\nu[\vec{x} \leftarrow \overline{\ell}], h_0 \uplus h, \tau_n(h_0,t_1,\ldots,t_m))$ cannot be a model of $\exists \vec{y}_j . \psi_j(\vec{x},\vec{y}_j) \land\mathcal{Q}_j(\vec{y}_j)$ for any $j \in [k]$ and $(\nu[\vec{x} \leftarrow \overline{\ell}], h_0 \uplus h, \tau_n(h_0,t_1,\ldots,t_n)) \in \mu\sys^\utree(\Asterisk\lseq) \setminus \mu\sys^\utree(\bigvee\rseq)$, as required.

    \vspace{10pt}
    \paragraph{$(\RI)$} Similar to the ($\RI$) case in the proof of Lemma \ref{lemma:cex}.
    
    \vspace{10pt}
    \paragraph{$(\SPsl)$} Similar to the ($\SP$) case in the proof of Lemma \ref{lemma:cex}, using a variation of Lemma \ref{lemma:spsl} in which $\mathcal{U}_k = \locsi^k \times \heaps \times \trees$ and $\biguplus_{i=1}^n \mu\sys^\utree(p_i) \overset{\text{def}}{=} \{(\overline{\ell}_1 \cdot \ldots \cdot \overline{\ell}_n, h_1 \uplus \ldots \uplus h_n, \{t_1, \ldots, t_n\}) \mid (\overline{\ell}_i, h_i, t_i) \in \mu\sys^\utree(p_i), i \in [n]\}$, for some predicates $p_1,\ldots, p_n$ in $\sys$. \qed}

\end{document}